\documentclass[11pt]{article}
\usepackage{amsfonts}
\usepackage{amsmath}
\usepackage{amssymb}
\usepackage{bbold}
\usepackage[margin=2.2cm]{geometry}
\usepackage{longtable}
\usepackage{lscape,graphicx} 
\usepackage{graphics}
\usepackage{color}
\usepackage{cite}
\usepackage{array}
\usepackage{makecell}

\usepackage{sidecap}
\usepackage{caption}
\usepackage{subcaption}

\newcommand{\ba}{\begin{array}{c}}
\newcommand{\ea}{\end{array}}

\newcommand{\be}{\beta}

\def\be{\begin{equation}}
\def\ee{\end{equation}}
\def\beq{\begin{equation}}
\def\eeq{\end{equation}}
\def\bc{\begin{center}}
\def\ec{\end{center}}
\def\bea{\begin{eqnarray}}
\def\eea{\end{eqnarray}}

\definecolor{darkgreen}{rgb}{.1,0.4,0.3}

\begin{document}
\begin{titlepage}
\vspace*{-1cm}
\phantom{hep-ph/***}
\flushright
\hfil{CP3-Origins-2016-005 DNRF90}
\hfil{DIAS-2016-5}

\vskip 1.5cm
\begin{center}
\mathversion{bold}
{\LARGE\bf 
Flavor and CP symmetries for
leptogenesis\\[0.1in] and $0\nu\beta\beta$ decay
}\\[3mm]
\mathversion{normal}
\vskip .3cm
\end{center}
\vskip 0.5  cm
\begin{center}
{\large Claudia Hagedorn}
{\large and Emiliano Molinaro}
\\
\vskip .7cm
{\footnotesize
CP$^3$-Origins and Danish Institute for Advanced Study, University of Southern Denmark,\\
Campusvej 55, DK-5230 Odense M, Denmark
\vskip .5cm
\begin{minipage}[l]{.9\textwidth}
\begin{center} 
\textit{E-mail:} 
\tt{hagedorn@cp3.sdu.dk}, \tt{molinaro@cp3.sdu.dk}
\end{center}
\end{minipage}
}
\end{center}
\vskip 1cm
\begin{abstract}
We perform a comprehensive analysis of the phenomenology of leptonic low and high energy CP phases 
in a scenario with three heavy right-handed neutrinos in which
a flavor and a CP symmetry are non-trivially broken. All CP phases as well as lepton mixing angles 
are determined by the properties of the flavor and CP symmetry and one free real parameter.
We focus on the generation of the baryon asymmetry $Y_B$
of the Universe via unflavored leptogenesis and the predictions of $m_{ee}$, the quantity measurable in neutrinoless double beta decay. 
We show that the sign of $Y_B$ can be fixed and the allowed parameter range of $m_{ee}$ can be strongly constrained. 
We argue on general grounds that the CP asymmetries $\epsilon_i$ are dominated by the contribution associated with one
Majorana phase and that in cases in which only the Dirac phase is non-trivial the sign of $Y_B$ depends on further 
parameters. In addition, we comment on the case of flavored leptogenesis where in general the knowledge of the CP phases and light neutrino mass spectrum is also not sufficient 
in order to fix  the sign of the CP asymmetries. As examples we discuss the series of flavor groups $\Delta (3 \, n^2)$ and $\Delta (6 \, n^2)$, $n \geq 2$ integer,
and several classes of CP transformations.
\end{abstract}
\end{titlepage}
\setcounter{footnote}{0}

\section{Introduction}
\label{intro}

The baryon asymmetry $Y_B$ of the Universe has been precisely measured \cite{YBexp}
\begin{equation}
\label{YBnumber}
Y_B = \frac{n_B-n_{\overline{B}}}{s} \Bigg|_0= \left( 8.65 \pm 0.09 \right) \times 10^{-11} \, .
\end{equation}
The error is given at the $1 \, \sigma$ level and the subscript ``0" refers to the present epoch. The generation of $Y_B$ requires
the fulfillment of the three Sakharov conditions \cite{Sakharov}: C and CP violation, departure from thermal equilibrium
and baryon number violation. All three of them are met by the mechanism of unflavored and flavored (thermal) leptogenesis \cite{leptogenesisFY}. 
In fact the decay of right-handed (RH) neutrinos $N_i$ (we always assume the existence of three such states and thus $i=1,2,3$) leads to a lepton 
asymmetry that is partially converted into a baryonic one via sphaleron processes \cite{Kuzmin:1985mm}. Departure from thermal equilibrium
arises, since the rate of the Yukawa interactions of RH neutrinos is small compared to the Hubble rate, while 
complex Yukawa couplings $Y_D$ are responsible for C and CP violation. The relevant quantities for computing $Y_B$ are:
the yield of RH neutrinos at high temperatures and the sphaleron conversion factor giving together rise to $10^{-3}$, the efficiency
factors, taking into account washout and decoherence effects and usually of order of $10^{-3} \div 10^{-1}$, and the CP asymmetries 
$\epsilon^{(\alpha)}_i$, generated in the decays of the $i^{\mathrm{th}}$ RH neutrino $N_i$ (and to the flavor $\alpha=e, \mu, \tau$, if
flavored leptogenesis is studied).
As has been shown in \cite{phasesnorelation}, even in a scenario in which light neutrinos acquire
masses via the ordinary type 1 seesaw mechanism \cite{type1seesaw} the CP phases present in the Yukawa couplings
are in general unrelated to the CP phases potentially measurable at low energies: the Dirac phase $\delta$ in neutrino
oscillation experiments \cite{T2K,MINOS,NOvA,DUNE,HyperK} and a combination of the two Majorana phases $\alpha$
and $\beta$ in neutrinoless double beta ($0\nu\beta\beta$) decay \cite{0nubbreview}. However, it is also well-known 
that low energy CP phases may be crucial for having successful leptogenesis in case flavor effects are relevant \cite{PPRpiu}.

In theories with flavor $G_f$ and CP 
symmetries all CP phases can be related.
Such symmetries are usually introduced in order to explain the mixing pattern(s) observed in the lepton (and quark) sector.
Since two of the lepton mixing angles (and thus most of the entries of the Pontecorvo-Maki-Nakagawa-Sakata (PMNS) mixing matrix $U_{PMNS}$) are large \cite{nufit} 
\begin{equation}
\label{thetaij_3sigma}
0.270 \leq \sin^2 \theta_{12} \leq 0.344 \; , \; 0.382[9] \leq \sin^2 \theta_{23} \leq 0.643[4] \;\; \mbox{and} \;\; 
0.0186[8] \leq \sin^2 \theta_{13} \leq 0.0250[1] 
\end{equation}
($3 \, \sigma$ ranges for light neutrino masses following normal ordering (NO) and
in square brackets for inverted ordering (IO), see also \cite{otherglobal_1,otherglobal_2}), 
it is natural to assign the three generations of left-handed (LH) lepton doublets $l$
to an irreducible three-dimensional representation ${\bf 3}$ of the flavor symmetry. If the setup contains RH neutrinos, it is reasonable to also
assign these to such a representation. Indeed, it is convenient to use 
the same representation ${\bf 3}$ as for $l$ in a non-supersymmetric (non-SUSY) context, while in SUSY models $\nu^c$ are 
usually put in the complex conjugated representation with respect to $l$.
The most promising choice of $G_f$ turns out to be a discrete, finite, non-abelian group \cite{reviews,review_math} that is broken to different residual symmetries
$G_e$ and $G_\nu$ in the charged lepton and neutrino sectors, respectively \cite{Gfresidual}. These symmetries are in general abelian subgroups of $G_f$ with three different elements at least
and, in the particular case of three Majorana neutrinos, that we consider, $G_\nu=Z_2 \times Z_2$.
As is known, with this approach only the Dirac phase can be
predicted and the analyses in \cite{analysesmixingGf} have shown that it always turns out to be trivial, if mixing angles are required to be in good agreement with
experimental data.\footnote{Considering smaller residual symmetries $G_e$ or $G_\nu$ or taking into account corrections 
obviously changes this conclusion,  but also reduces the predictive power of the approach.} Thus, in order to fix the Majorana phases via symmetries as well as to achieve $\delta$ different from $0$ or $\pi$,
this approach has to be modified. An extension that has been proven very powerful \cite{S4CPgeneral,A4S4CPmodels,D48CPmodel,D96CPmixing,HMM,DeltaCPothers,A5CP} is to amend the flavor with a CP symmetry \cite{S4CPgeneral} 
(see also \cite{mutaureflection}). The latter acts in general also in a non-trivial way on the flavor space \cite{CPnontrivial}, requiring certain conditions
to be fulfilled for having a consistent theory \cite{GrimusRebelo,S4CPgeneral,GfCPothers}. The choice of $G_e$ remains the same, whereas $G_\nu$
is taken to be the direct product of a $Z_2$ symmetry, contained in $G_f$, and the CP symmetry: $G_\nu=Z_2 \times CP$ \cite{S4CPgeneral}.
The main feature of such a setting is predicting the mixing angles as well as all three CP phases in terms of only one free real parameter $\theta$
that can be chosen without loss of generality to lie in the interval $0 \leq \theta < \pi$. This parameter is present and not fixed in this approach,
simply because the residual flavor symmetry in the neutrino sector is only a $Z_2$ group.
 On the other hand, involving CP
leads to the clear advantage to determine all three CP phases and not only the Dirac phase. Thus, such a scenario is in particular suitable for 
studying the phenomenology related to CP phases, like leptogenesis and $0\nu\beta\beta$ decay.

Leptogenesis, unflavored as well as flavored, has already been studied in scenarios with the flavor symmetries $A_4$ and $S_4$ only
\cite{JM_08,Bertuzzo_09,HMP1,ASierra_09,AM_09,Bazzocchi_09,Adhikary_08}. One striking feature of these scenarios is the fact that CP asymmetries vanish in case one only considers the leading order (LO) terms in the theory,
i.e.~those which preserve $G_e=Z_3$ and $G_\nu=Z_2 \times Z_2$ in the charged lepton and neutrino sectors, respectively. Thus, taking into account next-to-leading order (NLO)
corrections (in the neutrino sector, in particular, to $Y_D$) is mandatory. These are in general proportional to the (small) symmetry breaking parameter
$\kappa \sim 10^{-(2 \div 3)}$. As shown in \cite{JM_08,Bertuzzo_09}, the CP asymmetries $\epsilon_i$ are proportional to $\kappa^2$ for unflavored leptogenesis, whereas in the
case of flavored leptogenesis the suppression is less and $\epsilon_i^{\alpha}$ are proportional to $\kappa$ only. 

$0\nu\beta\beta$ decay has already been discussed in contexts with a flavor and a CP symmetry, see first reference in \cite{D48CPmodel},
 last reference in \cite{A4S4CPmodels},  \cite{D6n2CPZ2Z2}, first reference in \cite{A5CP} and  \cite{DeltaCPothers}.
In  \cite{D6n2CPZ2Z2} the authors have considered the series of groups $\Delta (6 \, n^2)$, but they have assumed that the residual symmetry in the neutrino sector is larger $G_\nu=Z_2 \times Z_2 \times CP$. Thus, the Dirac phase is fixed to be trivial as well as one of the Majorana phases which clearly affects the predictions for $m_{ee}$. In \cite{DeltaCPothers} the authors instead
study $G_\nu=Z_2 \times CP$ as residual group like in our analysis. However, the presented results are mostly in the limit in which the group theoretical parameter $n$ is taken to
be very large and no particular choice of the CP symmetry is made.

In the present paper we study leptogenesis and $0\nu\beta\beta$ decay in a scenario with the flavor symmetry $\Delta (3 \, n^2)$ or $\Delta (6 \, n^2)$ and a CP symmetry that are broken
in a non-trivial way. 
As discussed in \cite{HMM,DeltaCPothers}, for several choices of flavor groups $\Delta (3 \, n^2)$ or $\Delta (6 \, n^2)$, CP symmetries, residual groups $G_e$ and $G_\nu$ and the free parameter $\theta$ 
it is possible to obtain lepton mixing angles in agreement with the experimental data and in turn to predict the CP phases $\delta$ and $\alpha$, $\beta$.
We base our study on these results and we assume throughout that the charged lepton mass matrix is governed by the residual symmetry $G_e=Z_3$, while the RH neutrino mass matrix is taken to be of the most general form compatible
with $G_\nu=Z_2 \times CP$.  
Since LH lepton doublets and RH neutrinos transform according to the same three-dimensional representation ${\bf 3}$ of the flavor group in a non-SUSY framework
(or in a SUSY context $l$ and $\nu^c$ as ${\bf \bar{3}}$ and ${\bf 3}$), the Yukawa coupling $Y_D$ of neutrinos with trivial flavor structure 
 is invariant under the entire flavor group and CP. We note that RH neutrinos are not strongly hierarchical in our scenario and hence all three of them are expected 
 to be relevant for leptogenesis.
 We mainly focus on the case of unflavored leptogenesis and only highlight the main differences occurring, if the baryon asymmetry of the Universe is instead generated 
via flavored leptogenesis. Similarly, most of our results for leptogenesis are obtained in a non-SUSY framework, however, we emphasize the changes that have to be
implemented in order to apply these to a SUSY model.
 We find that the CP asymmetries $\epsilon_i$ vanish for unflavored leptogenesis, if we only consider LO terms. Thus, we have to rely on NLO terms. In a generic model these can be arbitrary,
but frequently it turns out that one type dominates and that the latter is still invariant under one of the residual symmetries. Here we consider a case in which the dominant NLO contribution arises in the neutrino sector and only corrects $Y_D$.
Furthermore, we assume that this correction $\delta Y_D$ stems from the charged lepton sector and is thus constrained by the residual group $G_e$. As expansion parameter we use $\kappa$ 
and hence $\delta Y_D \propto \kappa$. It induces an appropriate suppression of the CP asymmetries $\epsilon_i \propto \kappa^2$, similar to what has been already 
observed in scenarios with a flavor symmetry only. We show that the sign of $\epsilon_i$ depends on the low energy CP phases
and on the loop function, whose sign can be traced back to the light neutrino mass spectrum in our scenario. Light neutrino masses   
can be hierarchical with NO, IO or quasi-degenerate (QD). In particular,
the phases, contained in $\delta Y_D$, are irrelevant for the LO result of $\epsilon_i$. The sign of $Y_B$ is then determined as well and it is given by $\epsilon_i$.
We do not only perform a comprehensive analytical study in which we consider the different choices of residual symmetries, found in \cite{HMM},
 but we also take in account the case in which the mixing in the RH neutrino sector is given by the general form of the PMNS mixing matrix.
 The latter study reveals three interesting features: the CP asymmetries $\epsilon_i$ can be expressed in a certain limit
in terms of the CP invariants $I_i$ that are proportional to $\sin\alpha$, $\sin\beta$ and $\sin (\alpha-\beta)$ (for definition see appendix \ref{app11}). The dominant contribution is expected to arise from terms proportional to $\sin\alpha$,
as long as this Majorana phase does not take a value close to $0$ or $\pi$ or the loop function suppresses these terms for particular values of the light neutrino masses. In case $\sin\delta$ 
dominates the CP asymmetries $\epsilon_i$ their sign cannot be predicted and can depend on e.g.~the relative size of the parameters appearing in $\delta Y_D$ as well as on the 
octant of the atmospheric mixing angle $\theta_{23}$. 
We exemplify these findings with several cases and detail instances in which the sign of $Y_B$ is correctly determined. 
Furthermore, we comment on the case of flavored leptogenesis. The CP asymmetries $\epsilon_i^{\alpha}$ vanish as well, if we only take into account the LO terms, since the Yukawa couplings of neutrinos are taken to be 
invariant under the flavor symmetry $G_f$ and CP in our analysis. Also in this case corrections $\delta Y_D \propto \kappa$ induce non-zero $\epsilon_i^{\alpha}$. However, the sign of the latter depends in general 
on the parameters contained in $\delta Y_D$. A way to generate $\epsilon_i^{\alpha} \neq 0$ without corrections $\delta Y_D$ is to assume
 the Yukawa couplings of neutrinos to be of the most general form compatible with the residual symmetry $G_\nu=Z_2 \times CP$. Again, the sign of $\epsilon_i^{\alpha}$ depends on parameters that are 
 not directly related to the low energy CP phases (and the light neutrino mass spectrum). So, we arrive at the conclusion that the sign of the CP asymmetries cannot be univocally predicted in the flavored 
regime. Regarding $0\nu\beta\beta$ decay we carefully study it analytically for different choices of symmetries $G_f$, CP, $G_e$ and $G_\nu$ and present several numerical examples, 
pointing out the following interesting features: for light neutrino masses following IO
it is possible to achieve only values close to the very upper limit ($m_{ee} \gtrsim 0.05 \, \mathrm{eV}$) of the parameter space, generally compatible with experimental data, and thus enhancing the prospects for a discovery of $0\nu\beta\beta$ decay
 in the not-too-far future; for NO light neutrino masses we observe that the well-known cancellation of the different terms contributing to $m_{ee}$ can be avoided
  and thus a lower limit of $m_{ee} \gtrsim 10^{-3} \, \mathrm{eV}$ can be set or at least the interval of the lightest neutrino mass $m_0$ for which a noticeable cancellation in $m_{ee}$ occurs
  to be considerably shrunk.

The paper is structured as follows: in section \ref{sec2} we describe our scenario in which a flavor group $G_f=\Delta (3 \, n^2)$ or $G_f=\Delta (6 \, n^2)$
combined with a CP symmetry is broken to $G_e=Z_3$ and $G_\nu=Z_2 \times CP$ and corrections from the charged lepton sector to the neutrino one
are crucial for the generation of non-vanishing CP asymmetries. We also briefly repeat the results obtained in \cite{HMM} for lepton mixing,
in particular for the CP invariants, in section \ref{sec2}. Section \ref{sec3} contains the analysis of unflavored leptogenesis in our scenario, first mentioning some
general properties of the underlying leptogenesis framework, then studying the dependence of $Y_B$ on the low energy CP phases,
discussing the case in which the mixing matrix $U_R$, diagonalizing the RH neutrino mass matrix,
is taken to be of the general form of the PMNS mixing matrix, and afterwards studying analytically the properties of the scenarios in which $U_R$ is determined by 
the breaking of $G_f$ and the CP symmetry (distinguishing the different cases case 1) through case 3 b.1) and case 3 a) that give rise to different results for
lepton mixing) and eventually also numerically. We emphasize the cases in which the sign of the CP asymmetries (and thus of the baryon asymmetry of the Universe) that is intimately related
with the CP phases $\delta$, $\alpha$ and $\beta$ and the light neutrino mass spectrum can be predicted and also point out in which cases this is in general impossible. In section~\ref{sec40} we discuss flavored leptogenesis and
 show with several examples that the sign of the CP asymmetries $\epsilon_i^{\alpha}$ can in general not be predicted just from the knowledge of the low 
 energy CP phases and the light neutrino mass spectrum. We also demonstrate that for $Y_D$ invariant under $G_\nu$ (and not $G_f$ and CP)
non-zero CP asymmetries can be achieved without corrections $\delta Y_D$. 
 We comment on the changes occurring, if our scenario is realized in a SUSY context, in section \ref{sec4} and point out the similarities and crucial differences. Section \ref{sec5} is dedicated
to an analytical and numerical study of $0\nu\beta\beta$ decay in which the effects of constraining the lepton mixing parameters, in particular the two Majorana
phases $\alpha$ and $\beta$, become apparent. The summary of our results can be found in section \ref{concl}. Our conventions for mixing angles $\theta_{ij}$
and the CP phases $\delta$, $\alpha$ and $\beta$ together with the results of the global fit \cite{nufit} are given in appendix \ref{app1}. The choice of generators of the flavor
groups $\Delta (3 \, n^2)$ and $\Delta (6 \, n^2)$ is presented in appendix \ref{app2}. For completeness, in appendix \ref{app2a} some results for $\epsilon_i$
are shown for the case in which $U_R$ is identified with the general form of the PMNS
mixing matrix. Appendix \ref{app3} contains sketches of explicit models which realize the envisaged scenario in a non-SUSY as well as
a SUSY context, choosing $G_f=\Delta (6 \, n^2)$ with $n=4$ and for the parameter $s$ characterizing the CP symmetry $s=1$.

\mathversion{bold}
\section{Approach and lepton mixing resulting from $G_f=\Delta (3 \,  n^2)$ and $G_f=\Delta (6 \, n^2)$ and CP}
\mathversion{normal}
\label{sec2}

Here we present the approach we follow and the results obtained for lepton mixing from the groups $G_f=\Delta (3 \, n^2)$ and $\Delta (6 \, n^2)$
discussed in \cite{HMM}.

\subsection{Approach}
\label{sec21}

We consider in the following a scenario with three RH Majorana neutrinos $N_i$. These form a faithful and irreducible 
representation ${\bf 3}$ of the flavor group $G_f$. To the very same three-dimensional representation we also assign the
three generations of LH leptons $l$.\footnote{In the SUSY model we choose $l$ and $\nu^c$ to transform in two complex conjugated three-dimensional representations of the 
group $G_f$, see section \ref{sec4}.} The RH charged leptons $\alpha_R$, $\alpha= e$, $\mu$, $\tau$ are all assumed to transform trivially
under $G_f$, i.e.  $\alpha_R \sim {\bf 1}$ under $G_f$. In order to distinguish them we employ an auxiliary symmetry $Z_3^{(\mathrm{aux})}$ under which they carry the charges
$1$, $\omega$ and $\omega^2$. The phase $\omega$ is defined as the third root of unity: $\omega= e^{2 \, \pi \, i/3}$. LH leptons and RH neutrinos are instead neutral under $Z_3^{(\mathrm{aux})}$. 
In addition, the theory is invariant under a CP symmetry whose action is represented by the CP transformation $X_{\mathrm{\bf r}}$
in the different representations ${\mathrm{\bf r}}$ of $G_f$. In general, this CP symmetry acts also non-trivially on the flavor space
\cite{CPnontrivial} (see also \cite{GrimusRebelo}). The matrix $X_{\mathrm{\bf r}}$ is unitary and symmetric in flavor space (for details see \cite{S4CPgeneral})
\begin{equation}
\label{Xcond}
X_{\mathrm{\bf r}} X_{\mathrm{\bf r}}^\dagger = X_{\mathrm{\bf r}} X_{\mathrm{\bf r}} ^\star = 1 \; .
\end{equation}
As shown in \cite{S4CPgeneral}, the consistent combination of a flavor group $G_f$ and a CP symmetry, represented by $X_{\mathrm{\bf r}}$,
requires that the condition 
\begin{equation} 
\label{XGfcon}
(X_{\mathrm{\bf r}}^{-1} g_{\mathrm{\bf r}} X_{\mathrm{\bf r}})^\star = g^\prime_{\mathrm{\bf r}}
\end{equation}
is fulfilled for all elements $g$ of $G_f$ with $g^\prime$ being also an element of $G_f$ that is in general different from $g$.
Here $g_{\mathrm{\bf r}}$ and $g^\prime_{\mathrm{\bf r}}$ indicate that both elements, $g$ and $g^\prime$, are given in the representation ${\mathrm{\bf r}}$.
Since we only make explicit use of the trivial representation and the representation ${\bf 3}$ (and its complex conjugate), we only need 
to specify the form of the CP transformation in these two representations. Without loss of generality we can choose $X_{\bf 1}=1$,
while the form of $X \equiv X_{\bf 3}$ is in general non-trivial. Its particular form is given explicitly in the different cases, see (\ref{possibleCPnu}).
Similarly, we only need to consider the representation matrices $g_{\mathrm{\bf r}}$ of the abstract elements $g$ of the group $G_f$
in the trivial representation and in ${\bf 3}$. Those belonging to the former representation always equal the identity,  $g_{\bf 1}=1$, while
we denote those of the latter, for simplicity, with the same symbol as the abstract elements themselves, i.e. $g_{\bf 3}\equiv g$.\footnote{In appendix \ref{app3}
we discuss sketches of models. This requires the knowledge of the representation matrices in further representations of the flavor group as well as of the
form of the CP transformation $X_{\mathrm{\bf r}}$ in the other representations ${\mathrm{\bf r}}$. However, these pieces of information are either available in the literature or can be straightforwardly 
calculated.}

We focus on a non-SUSY framework (and comment on the implementation in a SUSY context in section \ref{sec4} and appendix \ref{app3}) in the present paper and thus
the form of the relevant Lagrangian is
\begin{equation}
\label{Ll}
{\cal L}_l= -Y_l \, \bar{l} \, H \, \alpha_R - Y_D \, \bar{l} \,  H^c  N - \frac{1}{2} \overline{N^c} M_R N + \mathrm{h.c.}  
\end{equation}
with $H$ being the Higgs $SU(2)_L$ doublet and $H^c=\epsilon \, H^\star$. Note that the Higgs field does neither transform under the flavor nor the CP symmetry of the theory.

The residual symmetry in the 
charged lepton sector is chosen as
\begin{equation}
G_e=Z_3^{(D)} \; ,
\end{equation}
where $Z_3^{(D)}$ is the diagonal subgroup of the $Z_3$ symmetry contained in $G_f$ and the additional $Z_3$ 
group $Z_3^{\mathrm{(aux)}}$. The requirement to preserve $G_e$ in the charged lepton sector is equivalent to 
requiring that the Yukawa matrix $Y_l$ is invariant under this symmetry.
For convenience, we choose a basis for all groups $G_f$ we discuss in such a way that the generator $Q$ of the 
subgroup $Z_3$ contained in $G_f$ is diagonal. Thus, the charged lepton mass matrix $m_l$ is diagonal and it contains three independent parameters that can be identified with the charged lepton masses
\begin{equation} 
\label{ml0}
m_l= Y_l \, v_\text{EW} = \mbox{diag} \, (m_e, m_\mu, m_\tau) \; .
\end{equation}
The value of the Higgs vacuum expectation value (VEV) is fixed to
$v_\text{EW}=\langle H \rangle \approx 174 \, \mathrm{GeV}$ in our convention.
If the ordering of $m_e$, $m_\mu$ and $m_\tau$ is not the standard one, we apply a permutation matrix $P$. 
Thus, the contribution of the charged leptons to the PMNS mixing matrix is in the case at hand given by $U_l=P$ (up to unphysical phases).
In an explicit model the matrix $m_l$ can be generated when 
appropriate flavor symmetry breaking fields take a  VEV leaving invariant $Z_3^{(D)}$, see appendix \ref{app3}. In such a realization also
the mass hierarchy of the charged leptons is usually explained either with the help of an additional Froggatt-Nielsen symmetry $U(1)_{\footnotesize\mbox{FN}}$ \cite{FN}
 under which RH charged leptons carry different charges, see e.g. \cite{AF2}, or it is generated through operators with multiple insertions of flavor symmetry breaking fields, see e.g. \cite{multiflavon}.
Notice the additional symmetry  $Z_3^{\mathrm{(aux)}}$, not present in the model-independent approach \cite{HMM}, 
 does not have a direct impact on our results on lepton mixing and thus the results in subsection \ref{sec22} hold without loss of generality. 

Since LH leptons and RH neutrinos transform in the same way under all flavor and CP symmetries, the second term in (\ref{Ll}) is automatically invariant 
under these symmetries and the Yukawa coupling $Y_D$ is proportional to the identity matrix in flavor space.
The Dirac mass matrix of the neutrinos hence reads
\begin{equation}
\label{mD0}
m_D= Y_D \, v_\text{EW} = y_0 \, v_\text{EW} \; 1 \; .
\end{equation}
It is clear that the form of $m_D$ preserves $G_f$ and the phase of the parameter $y_0$ 
can always be chosen in such a way that the imposed CP symmetry remains intact. 
Throughout this paper we take $y_0$ to be positive, since the phase of $y_0$ can be absorbed in a re-definition of fields and is thus unphysical.
Clearly, $m_D$ is, as a consequence, also invariant under all subgroups of $G_f$ and CP.

The presence of the non-trivial residual symmetry
\begin{equation}
G_\nu=Z_2 \times CP 
\end{equation}
that is the direct product of a $Z_2$ subgroup of the flavor symmetry $G_f$ and the CP symmetry instead manifests 
itself in the form of the Majorana mass matrix $M_R$ of RH neutrinos. Before presenting its explicit form we mention
that the generator of the $Z_2$ group, denoted by $Z$ in ${\bf 3}$, and the CP transformation $X$, have to fulfill\footnote{This equation is trivially fulfilled, if we consider
the trivial representation of $G_f$ instead of ${\bf 3}$.}
\begin{equation}
\label{XZ} 
 X Z^\star - Z X =0 \; ,
 \end{equation}
 since the product of these two symmetries is required to be a direct one.
 Furthermore, we note that also $Z X$ is a CP symmetry present in the neutrino sector, satisfying  the condition in (\ref{XZ}) \cite{S4CPgeneral}.
  The Majorana mass matrix $M_R$ of RH neutrinos is assumed to be invariant under the residual group $G_\nu$, i.e.
 \begin{equation}
 \label{MR0}
 U_R^T \, M_R \, U_R =  \mbox{diag} \, (M_1, M_2, M_3)
 \end{equation}
 with $M_i$ being the three RH neutrino masses and the unitary matrix $U_R$ is of the form
 \begin{equation}
 \label{UR0}
 U_R=\Omega \, R_{ij} (\theta) \, K_\nu \; .
 \end{equation}
 The matrix $\Omega$ is determined by $X$ and $Z$, $R_{ij} (\theta)$ is a rotation matrix in the $(ij)$-plane through the real parameter
 $\theta$, $0 \leq \theta < \pi$, and $K_\nu$ is a diagonal
 matrix with elements $\pm 1$ and $\pm i$. The latter is necessary for making the masses of RH neutrinos real and positive and we parametrize it 
 as 
\begin{equation}
\label{Knu}
K_\nu = \left( \begin{array}{ccc}
 1 & 0 & 0 \\
 0 & i^{k_1} & 0\\
 0 & 0 & i^{k_2}
\end{array}
\right) \;\;\; \mbox{with} \;\;\; k_{1,2}=0,1,2,3 \; .
\end{equation}
 Thus, the matrix $M_R$ contains in general four independent real parameters: the RH neutrino masses $M_i$ and $\theta$.
 In an explicit model $M_R$ is generated, like the charged lepton mass matrix $m_l$, via the couplings of RH neutrinos to flavor symmetry breaking fields. Obviously, 
 the VEVs of the latter should leave the residual symmetry $G_\nu$ invariant. 
  The light neutrino masses originate from the type 1 seesaw mechanism \cite{type1seesaw}
 \begin{equation}
 \label{mnuMR}
m_\nu =-m_D \, M_R^{-1} \, m_D^T =-y_0^2 \, v_\text{EW}^2 \, M_R^{-1} \; .
\end{equation}
This matrix is diagonalized by $U_\nu$
\begin{equation}
\label{mnuUnu}
U_\nu^\dagger \, m_\nu \, U_\nu^\star = \mbox{diag} \, (m_1, m_2, m_3) \;\;\; \mbox{with} \;\;\; U_\nu=U_R \; .
 \end{equation}
The light neutrino masses $m_i$ are inversely proportional to the RH neutrino masses $M_i$
 \begin{equation}
 \label{miMi}
 m_i =- \frac{y_0^2 \, v_\text{EW}^2}{M_i} \; .
 \end{equation}
 Note that we omit the minus sign appearing in (\ref{miMi}) in the following.
 The matrix $U_\nu^\star$ appears in (\ref{mnuUnu}), since the light neutrino mass matrix is given in the basis $\overline{\nu_L} \, m_\nu \, \nu^c_L$
 and we identify $U_\nu$ with the matrix that brings the fields $\nu_L$ into their mass basis so that the lepton mixing matrix $U_{PMNS}$
 is given by
 \begin{equation}
 U_{PMNS} = U^\dagger_l \, U_\nu = P^T U_R = P^T \Omega \, R_{ij} (\theta) \, K_\nu \; .
\end{equation}
In accordance with what has been shown in \cite{S4CPgeneral}, in such an approach the form of the PMNS mixing matrix is fixed by the symmetries $G_f$, CP, $G_e$ and $G_\nu$ 
up to the free real parameter $\theta$, possible permutations of rows and columns, since the masses of charged leptons and neutrinos are not predicted, and possible, but
unphysical, phases. Consequently,
all mixing angles and CP phases are strongly correlated, because they all only depend on $\theta$. The latter has to be fixed
to a particular value in order to accommodate the data on lepton mixing angles well and thus it has to be explained by some mechanism in an explicit 
model, see e.g.~\cite{A4S4CPmodels}. Since all lepton mixing angles $\theta_{ij}$ have been measured with a certain accuracy by now \cite{nufit}, the admitted values of $\theta$
are usually constrained to a rather narrow range, even if the experimentally preferred $3 \, \sigma$ ranges for $\sin^2 \theta_{ij}$, given in (\ref{thetaij_3sigma}), are taken into account. In addition, we note
that due to symmetries of the formulae for $\sin^2 \theta_{ij}$ two intervals of values of $\theta$ lead to good agreement with experimental data in several occasions.

 As mentioned, masses are generally
 not predicted in this approach, unless a particular model is considered, see e.g. second reference in \cite{A4S4CPmodels}. Thus, we parametrize the three light neutrino
 masses in terms of the two measured mass squared differences 
 \begin{equation}
 \Delta m_{\mathrm{sol}}^2=m_2^2 -m_1^2 \;\;\; \mbox{and} \;\;\; \Delta m_{\mathrm{atm}}^2 = \left\{
 \begin{array}{cc}
m_3^2-m_1^2 & \mbox{for NO}\\
m_3^2-m_2^2 & \mbox{for IO}
 \end{array} \right.
 \end{equation}
and the lightest neutrino mass $m_0$. For NO the light neutrino mass spectrum is thus of the form 
\begin{equation}
\label{massesNO}
m_1= m_0 \;\; , \;\;\; m_2= \sqrt{m_0^2 + \Delta m_{\mathrm{sol}}^2} \;\; , \;\;\; m_3= \sqrt{m_0^2 + \Delta m_{\mathrm{atm}}^2} \; ,
\end{equation}
while we find for IO
\begin{equation}
\label{massesIO}
m_1= \sqrt{m_0^2 + |\Delta m_{\mathrm{atm}}^2| - \Delta m_{\mathrm{sol}}^2 } \, , \;\; m_2= \sqrt{m_0^2 + |\Delta m_{\mathrm{atm}}^2| } \, , 
\;\; m_3= m_0 \; .
\end{equation}
If $m_0 \gtrsim 0.1 \, \mathrm{eV} > \sqrt{|\Delta m_{\mathrm{atm}}^2|}$, the light neutrino mass spectrum is QD and $m_1 \approx m_2 \approx m_3$.
The best fit values obtained from the global fit analysis in \cite{nufit} are 
\begin{eqnarray}
\label{dmbf}
&\Delta m_{\mathrm{sol}}^2 = 7.50 \times 10^{-5} \; \mathrm{eV}^2 \; ,&
\\ \nonumber
&\Delta m_{\mathrm{atm}}^2 = 2.457 \times 10^{-3} \;\; \mathrm{eV}^2  \;\; \mbox{(NO)} \;\;\; \mbox{and} \;\;\;  \Delta m_{\mathrm{atm}}^2 =-2.449 \times 10^{-3} \;\mathrm{eV}^2 \;\; \mbox{(IO)} \; .&
\end{eqnarray}
For the $3 \, \sigma$ intervals of the mass squared differences see appendix \ref{app12}.
The sum of the three light neutrino masses is constrained by cosmology and an upper bound is given by the Planck  Collaboration (using TT, TE, EE and lowP constraints)  \cite{YBexp}
 \begin{equation}
 \label{sumBOUND}
 \sum\limits_{k=1}^3 m_k <  0.492\;\text{eV} \;\;\; \text{at 95\% C.L.} \; .
 \end{equation}
 As a consequence the lightest neutrino mass $m_0$ has to be smaller than
 \begin{equation}
 \label{m0BOUND}
m_0 \lesssim  0.164 \;\text{eV}\; .
\end{equation}
Using (\ref{miMi}) we see that the masses of the RH neutrinos are determined by the light neutrino mass ordering and by the lightest neutrino mass $m_0$ and thus fixed once the latter are fixed.

 The LO results are in general perturbed in an explicit model, e.g. the Dirac mass matrix of 
the neutrinos can receive contributions from flavor symmetry breaking fields dominantly coupling to charged leptons.
Thus, we have in general 
\begin{eqnarray}
 \label{corrgeneral}
 &&Y_l=Y_l^0 + \delta Y_l +  \dots \; , \;\;
\\ \nonumber
 &&Y_D=Y_D^0 + \delta Y_D +  \dots \; , \;\; M_R = M_R^0 + \delta M_R + \dots
 \end{eqnarray}
with $Y_l^0$, $Y_D^0$, $M_R^0$ denoting the LO results, given in (\ref{ml0}), (\ref{mD0}) and (\ref{MR0}). The corrections 
$\delta Y_l$, $\delta Y_D$ and $\delta M_R$
are suppressed  with respect to the LO results by (powers of) the small (real, positive) parameter $\kappa$, $\kappa \ll 1$. 
For simplicity, we include this suppression factor in the definition of the corrections.
The latter are responsible for changes in the matrices that diagonalize $m_l$, $M_R$ and $m_\nu$, i.e. the mixing matrices read
\begin{equation}  
 U_l=P \, \delta U_l \; , \;\;  U_R=U_R^0 \, \delta U_R  \;\;\; \mbox{and} \;\;\; U_\nu=U_\nu^0 \, \delta U_\nu \;\;\; \mbox{with} \;\;\; \delta U_l \approx 1 \; , \;\;
 \delta U_R \approx 1 \; , \;\;  \delta U_\nu \approx 1 
\end{equation} 
leading to a PMNS mixing matrix of the form
\begin{equation}
U_{PMNS}= U_l^\dagger \, U_\nu = (\delta U_l)^\dagger\, P^T  U_\nu^0 \, \delta U_\nu \approx P^T  U_\nu^0 = U_{PMNS}^0 \; .
\end{equation}
In the discussion of leptogenesis in our scenario we are particularly interested in the corrections $\delta Y_D$ to the Yukawa coupling of the neutrinos.
There are three principle possibilities: either the dominant correction $\delta Y_D$ arises from fields coupling dominantly
to RH neutrinos, then $\delta Y_D$ is also invariant under $G_\nu$, or $\delta Y_D$ arises from fields coupling dominantly
to charged leptons, then $\delta Y_D$ is expected to possess as residual symmetry $G_e=Z_3^{(D)}$ or $\delta Y_D$
respects none of these residual symmetries, since combinations of both types of flavor symmetry breaking fields give
the dominant correction. A particularly interesting case is the second one. We can parametrize the correction $\delta Y_D$
in this case as
\begin{equation}
\label{deltaYD}
\delta Y_D = \left( \begin{array}{ccc} 
  \frac{2}{\sqrt{3}} z_1 & 0 & 0\\
  0 & -\frac{1}{\sqrt{3}} z_1 - z_2 & 0\\
  0 & 0 & -\frac{1}{\sqrt{3}} z_1 + z_2 
\end{array} \right) \, \kappa
\end{equation}
with $z_{1,2}$ being complex numbers with absolute values of order one. These parameters are in general complex, because we do 
not require a residual CP symmetry to be preserved in the charged lepton sector.
Note that the trace of $\delta Y_D$ vanishes, since
the correction proportional to the trace can be absorbed in the LO term. Note further that the corrections are normalized in 
such a way that the trace of the square of the matrices that accompany $z_1$ and $z_2$, respectively, is the same, while the trace of the product of 
these two matrices vanishes. 
In section \ref{sec3}  we focus on leptogenesis in a scenario with $\delta Y_D$ as in (\ref{deltaYD}).

We present sketches of a non-SUSY and a SUSY model in appendix \ref{app3}, in particular, in order to motivate the size of the parameter $\kappa$
that we assume later in our phenomenological study, see (\ref{estimatekappa}).

\mathversion{bold}
\subsection{Lepton mixing from $G_f=\Delta (3 \,  n^2)$ and $G_f=\Delta (6 \, n^2)$ and CP}
\mathversion{normal}
\label{sec22}

In \cite{HMM} the mixing patterns that can -- for certain values of the parameters of the theory -- fit the experimental data on the 
lepton mixing angles well \cite{nufit} have been found in a scenario with $G_f=\Delta (3 \,  n^2)$ or $G_f=\Delta (6 \, n^2)$, $n$ being an integer, and CP.
Like in \cite{HMM}, we require in the following that three does not divide $n$ as well as for case 1) and case 2) also that $n$ is even. 
The residual symmetries in the charged lepton and neutrino sectors are $G_e=Z_3$ and $G_\nu=Z_2 \times CP$. 
As regards the mixing there is no difference in 
considering as residual symmetry $G_e=Z_3$ generated by $Q=a$ (see definition of $a$ in appendix \ref{app2}) or the residual symmetry $G_e=Z_3^{(D)}$ where $Z_3^{(D)}$ is the
diagonal subgroup of the $Z_3$ symmetry generated by $Q=a$ and the additional $Z_3$ group $Z_3^{(\mathrm{aux})}$, since the relevant property, namely the fact that the charged lepton mass
matrix is diagonal, is not altered. In \cite{HMM} all
possible choices of $Z_3$ and $Z_2$ groups and a certain set of CP transformations $X$ have been studied.\footnote{These results have been confirmed in \cite{DeltaCPothers} and there also extended to
the case in which $G_e$ is not a $Z_3$ symmetry.}
 In particular,
the following representative cases that lead to different mixing patterns have been singled out
 \begin{equation}
 \label{possibleCPnu}
 \begin{array}{llll}
\Delta (3 \, n^2) \, , \; \Delta (6 \, n^2)&\mbox{case 1)}&Z=c^{n/2}&X= a \, b \, c^s \, d^{2 s} \, P_{23} \; ,\\
\Delta (3 \, n^2) \, , \; \Delta (6 \, n^2)&\mbox{case 2)}&Z=c^{n/2}&X= c^s \, d^t \, P_{23}\; ,\\
\Delta (6 \, n^2)&\mbox{case 3 a) and case 3 b.1)}&Z=b \, c^m \, d^m&X= b \, c^s \, d^{n-s} \, P_{23}\; ,
\end{array}
 \end{equation}
 with $s, t, m$ taking integer values in the interval $0 \leq s, t, m \leq n-1$, $a$, $c$ and $d$ (and $b$) being the generators of the group $\Delta (3 \, n^2)$ (and $\Delta (6 \, n^2)$), see appendix \ref{app2} 
 and \cite{HMM}, and the matrix $P_{23}$ reading
 \begin{equation}
 \label{P23}
  P_{23}=\left(
 \begin{array} {ccc}
  1 & 0 & 0\\
  0 & 0 & 1\\
  0 & 1 & 0
  \end{array}
  \right) \; .
 \end{equation} 
We note that the CP transformation $X=P_{23}$ in the representation ${\bf 3}$ corresponds to the following automorphism acting on the generators $a$, $b$, $c$ and $d$ as
 \begin{equation}
 \label{XP23aut}
 a \;\; \rightarrow \;\; a \;\; , \; b \;\; \rightarrow \;\; b \;\; , \; c \;\; \rightarrow \;\; c^{-1} \;\;\; \mbox{and} \;\;\; d \;\; \rightarrow \;\; d^{-1} \; .
 \end{equation}
 Since in our scenario only the Majorana mass matrix of the RH neutrinos is non-diagonal, mixing solely originates from this sector.
 As shown in the preceding subsection,  the diagonalization matrix $U_R$ and thus also the lepton mixing matrix $U_{PMNS}$
 are given at LO by the matrix in (\ref{UR0}),
 up to permutations of rows and columns. 
 
 The case $X=P_{23}$ in the charged lepton mass basis is well-known in the literature \cite{mutaureflection}
and is called $\mu\tau$ reflection symmetry. Its predictions are maximal atmospheric mixing and maximal Dirac phase
 as well as trivial Majorana phases, while the reactor and the solar mixing angle remain in general unconstrained without additional symmetries.
  
In the following we repeat the form of the CP invariants $J_{CP}$, $I_1$ and $I_2$ for the cases in (\ref{possibleCPnu}), as these are relevant for our analysis 
 of leptogenesis. We note that we add 
 the results of a third Majorana CP invariant, called $I_3$ and defined in (\ref{I3def}) in appendix \ref{app11}. Clearly, this third invariant is not
 independent of the two other ones $I_1$ and $I_2$. However, it proves useful to employ $I_3$ in the discussion of leptogenesis, see e.g. formulae
 in (\ref{eps1genz20}-\ref{eps3genz20}). For the formulae of the mixing angles we refer the reader  to \cite{HMM}.
 
In case 1) the PMNS mixing matrix is of the form
\begin{equation}
\label{URcase1}
U_{PMNS,1} = U_{R , 1}= \Omega_1 \, R_{13} (\theta) \, K_\nu \; ,
\end{equation}
where
\begin{equation}
\label{Omegacase1}  
\Omega_1 = e^{i \, \phi_s} \, U_{TB} \, \left( \begin{array}{ccc}
1 & 0 & 0\\
0 & e^{-3 \, i \, \phi_s} & 0\\
0 & 0 & -1
\end{array}
\right) 
\;\;\; \mbox{with} \;\;\;
\phi_s = \frac{\pi s}{n} \; ,
\end{equation}
\begin{equation}
\label{UTB} 
U_{TB}=\left( \begin{array}{ccc}
\sqrt{\frac{2}{3}} & \frac{1}{\sqrt{3}} &0\\
-\frac{1}{\sqrt{6}} & \frac{1}{\sqrt{3}} & \frac{1}{\sqrt{2}}\\
-\frac{1}{\sqrt{6}} & \frac{1}{\sqrt{3}} & -\frac{1}{\sqrt{2}}
\end{array}
\right) \;\;\;  \mbox{and} \;\;\;
R_{13} (\theta) = \left( \begin{array}{ccc}
\cos \theta & 0 &  \sin \theta\\
0 & 1 & 0\\
-\sin \theta & 0 & \cos \theta
\end{array} \right) \; .
\end{equation}
The results for the CP invariants are 
\begin{eqnarray}
\nonumber
&&J_{CP}=0 \;, \;\;\; I_1 = \frac 29 \, (-1)^{k_1+1} \, \cos^2 \theta \, \sin 6 \, \phi_s  \; , \;\;\; I_2=0\;,  
\\ \label{CPinvcase1}
 &&\mbox{and} \;\;\; I_3 = \frac 29 \, (-1)^{k_1+k_2} \, \sin^2 \theta \, \sin 6 \, \phi_s \; .
\end{eqnarray}
In this case also the form of the Majorana phase $\alpha$ is particularly simple
\begin{equation}
\label{alphacase1}
\sin \alpha = (-1)^{k_1+1} \, \sin 6 \, \phi_s  
\end{equation}
and we see immediately that for the choices $s=0$ and $s=n/2$ (i.e. if the CP transformations $X=a \, b \, P_{23}$ and $X=a \, b \, c^{n/2} \, P_{23}$ are imposed, respectively)
also the Majorana phase $\alpha$ is trivial, meaning that CP is conserved in this case, see also \cite{HMM}.
We note that replacing $\theta$ with $\pi-\theta$ leads to the same results for the solar and reactor mixing angle as well as for the in general non-trivial Majorana phase $\alpha$,
whereas the atmospheric mixing angle changes its octant.

In case 2) the matrix $U_{PMNS}$ reads
\begin{equation}
\label{URcase2}
U_{PMNS, 2}=U_{R , 2}= \Omega_2 \, R_{13} (\theta) \, K_\nu 
\end{equation}
with
\begin{equation}
\Omega_2= e^{i \, \phi_v/6} \, U_{TB} \, R_{13} \left( - \frac{\phi_u}{2} \right) \, \left( \begin{array}{ccc}
 1 & 0 & 0\\
0 & e^{-i \, \phi_v/2} & 0\\
0 & 0 & -i
\end{array} 
\right) \;\;\;
\mbox{with} \;\;\;
\phi_u = \frac{\pi u}{n} \;\;\; \mbox{and} \;\;\; \phi_v = \frac{\pi v}{n} \; .
\end{equation}
The integer parameters $u$ and $v$ depend on the choice of the CP transformation $X= c^s d^t \, P_{23}$ and are related 
to the exponents $s$ and $t$ as follows
\begin{equation}
\label{defuv}
 u= 2 \, s -t \;\;\;\;\; \mbox{and} \;\;\;\;\; v = 3 \, t \; .
\end{equation}
Since $0 \leq s, t \leq n-1$, we get for $u$ and $v$ as admitted intervals: $- (n-1) \leq u \leq 2 \, (n-1)$ and $0 \leq v \leq 3 \, (n-1)$.
According to \cite{HMM} the CP invariants are of the form
 \begin{eqnarray}\nonumber
 && J_{CP}= -\frac{\sin 2 \theta}{6 \sqrt{3}} \;\; , \;\;  I_2 = \frac{1}{9} (-1)^{k_2} \,  \sin 2 \phi_u \sin 2 \theta \;\; ,
 \\  \label{CPinvcase2}
 && I_1 = \frac{1}{9} (-1)^{k_1+1} \, \left( \left[\cos\phi_u + \cos 2 \theta \right] \sin\phi_v-  \sin \phi_u \cos \phi_v \sin 2 \theta \right) \; ,
 \end{eqnarray}
and the third Majorana CP invariant  is given by
\begin{equation}
\label{I3case2}
 I_3 = \frac{1}{9} (-1)^{k_1+k_2} \, \left( \left[\cos\phi_u - \cos 2 \theta \right] \sin\phi_v+  \sin \phi_u \cos \phi_v \sin 2 \theta \right) \; .
  \end{equation}
We can derive approximate formulae for the CP phases \cite{HMM}, if we take into account the constraints
on the group theoretical quantities $u$ and $n$ as well as on the free parameter $\theta$ arising from the requirement to describe the experimentally
measured values of the lepton mixing angles well. In particular, we see that the sine of the Majorana
phase $\alpha$ can be expressed as
\begin{equation}
\label{sinaapproxcase2}
\sin\alpha \approx (-1)^{k_1+1} \, \sin \phi_v \; .
\end{equation}
The study of several explicit cases in \cite{HMM} has demonstrated that this formula holds to very good accuracy
for all values of the parameter $v$. Thus, we see that choices of $v\approx 0, n, 2 \, n$ lead to a very small phase $\alpha$, while
values of $v\approx n/2, 3 \, n/2, 5\, n/2$ entail (almost) maximal $\alpha$, see also table \ref{tab:case2} in our numerical discussion in subsection \ref{sec34}.

In addition, we note that by permuting the rows of the mixing matrix $U_{PMNS, 2}=U_{R,2}$, defined in (\ref{URcase2}), i.e. we consider either
 \begin{equation}
 \label{shiftPcase2}
  P_1 \, \Omega_2 \, R_{13} (\theta) \, K_\nu 
 \;\;\; \mbox{or} \;\;\;
 P_1^2 \, \Omega_2 \, R_{13} (\theta) \, K_\nu 
 \end{equation}
using the permutation matrix $P_1$
\begin{equation}
\label{P1}
P_1=\left( \begin{array}{ccc}
0&1&0\\
0&0&1\\
1&0&0
\end{array}
\right) \; ,
 \end{equation}
two slightly different mixing patterns can be obtained whose results for the CP invariants can be deduced from those
given in (\ref{CPinvcase2}-\ref{I3case2}), if we apply either the transformations
\begin{equation}
\label{shift1case2}  
u \;\;\; \rightarrow \;\;\; u - \frac{n}{3} \;\; , \;\; \theta \;\;\; \rightarrow \;\;\; \frac{\pi}{2} - \theta \;\;\; \mbox{and} \;\;\; k_1 \;\;\; \rightarrow \;\;\; k_1 +1 \; ,
\end{equation}
or
\begin{equation}
\label{shift2case2}
u \;\;\; \rightarrow \;\;\; u + \frac{n}{3} \;\; , \;\; \theta \;\;\; \rightarrow \;\;\; \frac{\pi}{2} - \theta \;\;\; \mbox{and} \;\;\; k_1 \;\;\; \rightarrow \;\;\; k_1 +1 \; .
\end{equation}
For details see \cite{HMM}. We only notice here that applying any of the two sets of transformations also changes the sign in the approximate
formula for the sine of the Majorana phase $\alpha$ given in (\ref{sinaapproxcase2}).
In case 2) replacing $\theta$ with $\pi-\theta$  does not affect the mixing angles, whereas the sign of $J_{CP}$ and $I_2$ changes and the CP invariants
$I_1$ and $I_3$ do not transform in a definite way under this replacement. Nevertheless, using the approximate relation in (\ref{sinaapproxcase2}) for $\sin\alpha$, we see
that it does not change, if we only replace $\theta$ with $\pi-\theta$. The two CP invariants $I_1$ and $I_3$ can be made transforming in a definite way
by additionally sending $k_1$ into $k_1+1$ and $v$ into $k \, n-v$ (with $k$ being
an odd integer chosen in such a way that $k \, n-v$ is still in the admitted interval for the parameter $v$). Then, both, $I_1$ and $I_3$,  
change sign so that all CP invariants change sign. This statement is in agreement with the observation that $\sin\alpha$ in (\ref{sinaapproxcase2}) is not affected by
the replacement of $\theta$ by $\pi-\theta$ alone, but it changes sign, if we change $k_1$ into $k_1+1$.

The third case can be divided into two subcases, case 3 a) and case 3 b.1). Since we show numerical results only for case 3 b.1) in our analysis of leptogenesis, 
we first present the findings of \cite{HMM} for this case. For case 3 b.1) the matrix $U_{PMNS}$ takes the form
\begin{equation}
\label{URcase3}
U_{PMNS, 3b1}=U_{R, 3b1} = \Omega_3 \, R_{12} (\theta) \, K_\nu \, P_1
\end{equation}
with
\begin{equation}
\Omega_3 =\left(
\begin{array}{ccc}
1&0&0\\
0&\omega&0\\
0&0&\omega^2
\end{array}
\right) \, \Omega_1 \, R_{13} (\phi_m) \;\;\; \mbox{and} \;\;\; \phi_m = \frac{\pi m}{n}  \; ,
\end{equation}
\begin{equation}
R_{12} (\theta) = \left( \begin{array}{ccc}
\cos \theta & \sin \theta & 0\\
-\sin \theta & \cos \theta & 0\\
0 & 0 & 1
\end{array} \right) 
\end{equation}
and $P_1$ as defined in (\ref{P1}).
The CP invariants are
\begin{eqnarray}
&& J_{CP,3b1}= -\frac{1}{6 \sqrt{6}} \, \sin 3 \, \phi_m \sin 3 \, \phi_s \sin 2 \theta \;\; ,
 \nonumber \\
&& I_{1,3b1}=\frac 49 \, (-1)^{k_2+1} \, \sin^2 \phi_m \sin 3 \, \phi_s \sin\theta \, \left( \cos 3 \, \phi_s \sin \theta - \sqrt{2} \cos \phi_m \cos \theta \right)  \;\; ,
\nonumber \\  \label{CPinvcase3b1}
&& I_{2,3b1}= \frac 49 \, (-1)^{k_1+k_2+1} \, \sin^2 \phi_m \sin 3 \, \phi_s \cos\theta \left( \cos 3 \, \phi_s \cos \theta +\sqrt{2} \cos \phi_m \sin \theta \right) 
\end{eqnarray}
and
\begin{equation}
\label{I3case3b1}
\!\! I_{3,3b1}=\frac{1}{9} \, (-1)^{k_1+1} \, \cos \phi_m \, \sin 3 \, \phi_s \, \left( 4 \cos \phi_m \cos 3 \, \phi_s \cos 2 \theta + \sqrt{2}  \, \cos 2 \, \phi_m \, \sin 2 \theta \right) \; .
\end{equation}
The expressions of the CP invariants  in terms of $\phi_s$ and $\theta$ are considerably simplified in the  case
$m=n/2$ ($\phi_m=\pi/2$). Thus we can easily extract the Majorana and Dirac phases
\begin{equation}
\label{phaseseasy3b1}
 \sin\alpha = (-1)^{k_2+1}\,\sin 6\,\phi_s\;\; , \;\; \sin\beta = (-1)^{k_1+k_2+1}\,\sin 6\,\phi_s \;\; , \;\;
  \sin\delta
	\approx \pm\,\sin 3\,\phi_s \; .
\end{equation}
The last approximation holds, because $\theta\approx\pi/2$ (see \cite{HMM} for details), and the plus (minus) sign refers to $\theta$  
smaller (larger) than $\pi/2$. For $s=n/2$ in addition, $\sin\delta$ is actually independent of $\theta$ \cite{HMM} and the approximation in (\ref{phaseseasy3b1})
is exact. So, we find in this case $\sin\delta= \mp 1$, i.e. the Dirac phase is maximal, while the Majorana phases $\alpha$ and $\beta$ are trivial for $s=n/2$.

For case 3 a) we consider instead of $U_{PMNS, 3b1}=U_{R,3b1}$ in (\ref{URcase3}) the matrix without the permutation $P_1$
\begin{equation}
\label{URcase3a}
U_{PMNS, 3a}=U_{R, 3a} = \Omega_3 \, R_{12} (\theta) \, K_\nu \; .
\end{equation}
We find the same result for the Jarlskog invariant derived from $U_{PMNS, 3b1}$ and $U_{PMNS, 3a}$, while the CP invariants $I_i$
are permuted among each other
\begin{eqnarray}\nonumber
&&J_{CP, 3a}=J_{CP, 3b1} \; ,
\\
&& I_{1, 3a}=I_{3, 3b1} \;\; , \;\; I_{2, 3a}=-I_{1, 3b1} \;\;\; \mbox{and} \;\;\; I_{3, 3a}=-I_{2, 3b1}  \; .
\label{CPinvcase3a}
\end{eqnarray}
Obviously, the results for the mixing angles in case 3 a) are different from those obtained in case 3 b.1). Thus, the sets of parameters $n$, $m$, $s$ and $\theta$
that lead to lepton mixing angles in accordance with the experimental data are different in the two cases, see the extensive analysis in \cite{HMM}  for details.
We remark that the formulae of mixing angles and CP invariants in case 3 b.1) and case 3 a) have certain symmetry properties. Applying the set of transformations
\begin{equation}
\label{trafocase3}
s \;\;\; \rightarrow \;\;\; n-s \;\; (\phi_s \; \rightarrow \; \pi-\phi_s) \;\;\; \mbox{and} \;\;\; \theta \;\;\; \rightarrow \;\;\; \pi - \theta 
\end{equation}
the mixing angles remain invariant, whereas the CP invariants change sign and thus also the sines of all three CP phases. For $s=n/2$, thus, sending $\theta$ into $\pi-\theta$
leaves the mixing angles untouched, while all CP phases change their sign.
Using instead the transformations
\begin{equation}
\label{trafo2case3}
m \;\;\; \rightarrow \;\;\; n-m  \;\; (\phi_m \; \rightarrow \; \pi-\phi_m) \;\;\; \mbox{and} \;\;\; \theta \;\;\; \rightarrow \;\;\; \pi - \theta 
\end{equation}
the solar and reactor mixing angle as well as the CP invariants $I_i$ remain unchanged, while $\sin^2 \theta_{23}$ becomes $\cos^2\theta_{23}$ and $J_{CP}$
changes sign. In the particular case $m=n/2$, we can conclude that the replacement of $\theta$ with $\pi-\theta$
does not affect the solar and reactor mixing angle as well as the CP invariants $I_i$, while the atmospheric mixing angle changes its octant and $J_{CP}$ its sign.
If we set $m=n/2$ and $s=n/2$ and apply the transformation $\theta \, \rightarrow \, \pi-\theta$, we see from (\ref{trafocase3}) and (\ref{trafo2case3}) that the solar and reactor mixing angle are 
unchanged, the atmospheric mixing angle must be maximal, the CP invariants $I_i$ must vanish and $J_{CP}$ changes sign and is in general non-zero.
For more symmetry transformations of this type see table 6 in \cite{HMM}.

As in case 2), mixing matrices arising from permutations of the rows of the PMNS mixing matrices derived from $U_{R, 3b1}$ and $U_{R, 3a}$ do also admit
good agreement with the experimental data on lepton mixing angles for certain choices of the parameters. The formulae for mixing angles and CP
invariants can be obtained from those found for the PMNS mixing matrices with non-permuted rows by taking into account shifts in the parameters $m$ and $\theta$.
Again, details can be found in \cite{HMM}.

\section{Leptogenesis}
\label{sec3}

We first collect several pieces of information regarding leptogenesis in general. We continue with the discussion of leptogenesis
 in our framework. In particular, we determine conditions on the form of $Y_D$ under which the CP asymmetries $\epsilon_i$ can 
 be directly related to the low energy CP phases. We also study the results for $\epsilon_i$, assuming a generic form of the PMNS mixing matrix. 
 Subsequently, we turn to the detailed analysis of leptogenesis
in the different scenarios of mixing, case 1) through case 3 b.1) and case 3 a). In doing so, we separate our discussion in an analytical and a numerical part.
In order to not expand the latter too much we concentrate on case 3 b.1) when studying the third class of mixing.

\subsection{Preliminaries}
\label{sec31}

As already mentioned, we focus on unflavored leptogenesis as mechanism to generate correctly the measured value of the baryon asymmetry of the Universe
$Y_B = \left( 8.65 \pm 0.09 \right) \times 10^{-11}$ \cite{YBexp}. Thus, we assume RH neutrino masses to be larger than $10^{12} \, \mathrm{GeV}$ \cite{review_lepto_2}.
Since the RH neutrino mass spectrum is not expected to be strongly hierarchical, we consider the contributions of all three RH 
neutrinos to the generation of the baryon asymmetry of the Universe 
\begin{equation}
Y_B = \sum\limits_{i=1}^3 Y_{B i} \; .\label{Ytot}
\end{equation}
The quantities $Y_{B i}$  correspond to the part of the baryon asymmetry produced by the $i^{\mathrm{th}}$ RH neutrino. 
The latter can be expressed in the following way  \cite{leptasymindep} 
\begin{equation}
\label{YBi}
Y_{B i} \approx  1.38  \times 10^{-3} \, \epsilon_i \sum\limits_{j=1}^3 \eta_{ij}\, .
\end{equation}
The numerical value in (\ref{YBi}) depends on the yield of RH neutrinos at high temperatures and on the sphaleron conversion factor,
while $\epsilon_i$ is the CP asymmetry, generated in decays of the RH neutrino $N_i$.  The efficiency factor $\eta_{ij}$ 
parametrizes the washout and decoherence effects of the lepton charge asymmetry generated in $N_i$ decays
 due to lepton number violating interactions which involve the states $N_j$ present in the thermal bath at temperatures $T\sim M_i$. 
The CP asymmetry $\epsilon_i$, arising from the decay of the RH neutrino $N_i$ with the mass $M_i$, is defined as 
\begin{equation}
 \label{epsi}  
\!\!\!\epsilon_i =  -\frac{\sum_\alpha [\Gamma (N_i \rightarrow H l_\alpha) - \Gamma (N_i \rightarrow H^\star \bar{l}_\alpha)]}{\sum_\alpha [\Gamma (N_i \rightarrow H l_\alpha) + \Gamma (N_i \rightarrow H^\star \bar{l}_\alpha)]}
= -\frac{1}{8 \pi} \sum_{j \neq i}  \frac{\mathrm{Im} \left( (\hat{Y}_D^\dagger \hat{Y}_D)_{ij}^2 \right)}{(\hat{Y}_D^\dagger \hat{Y}_D)_{ii}} \, f(M_j/M_i)\,,
\end{equation}
and the loop function $f (x)$ in the Standard Model (SM) reads \cite{Covi:1996wh}
\begin{equation}
\label{floop}
f(x) = x \left[ \frac{1}{1-x^2} +1 - (1+x^2) \ln \left( 1+ \frac{1}{x^2} \right) \right] \, .
\end{equation}
Notice that the  quantities $\epsilon_i$ are specified as the CP asymmetries for anti-leptons such that the sign of $\epsilon_i$ is the same as for $Y_{B i}$ in (\ref{YBi}). 

The matrix $\hat{Y}_D$ is given by the neutrino Yukawa coupling $Y_D$ in the RH neutrino mass basis and thus reads with our conventions, see (\ref{Ll}) and (\ref{MR0}),
\begin{equation}
\label{defYDhat}
\hat{Y}_D= Y_D \, U_R \; .
\end{equation}
The efficiency factors $\eta_{ij}$ are expected to lie in the interval $10^{-3} \lesssim \eta_{ij} \lesssim 1$, see comments at the end of subsection \ref{sec32} and figure \ref{Fig:1b}. 
Their computation requires in general the numerical 
solution of Boltzmann equations. However, in our scenario several instances allow
for a simplified treatment. The three different RH neutrinos $N_i$ couple at LO to orthogonal linear combinations of the lepton flavors, since we find that the Yukawa
coupling $\hat{Y}_D$ is proportional to the unitary matrix $U_R$, if we take into account that $Y_D$ is proportional to the identity matrix at LO, see (\ref{mD0}).
The RH neutrino masses are taken to be larger than $10^{12} \, \mathrm{GeV}$
so that lepton flavor dynamics is not resolved at the temperatures relevant for leptogenesis and the Boltzmann equations of the three orthogonal lepton charge asymmetries 
generated by the decays of each RH neutrino are almost independent  \cite{leptasymindep}.
In the RH neutrino mass basis the efficiency factors $\eta_{ij}$ thus reduce to 
 \begin{equation}\label{etaij}
\eta_{ij}\approx \eta_{ii} \,\delta_{ij}\;. 
 \end{equation}
 Furthermore, we constrain RH neutrino masses to be smaller than $10^{14} \, \mathrm{GeV}$. Thus, possible washout effects due to  scattering processes which violate lepton number 
 by two units are out of equilibrium
 and the efficiency factors $\eta_{ii}$ in (\ref{etaij}) can be approximated well as  \cite{review_lepto_1}
 \begin{equation}
 \label{eta}
 \eta_{ii} = \left(  \frac{3.3 \times 10^{-3} \, \mathrm{eV}}{\tilde{m}_i}  + \left( \frac{\tilde{m}_i}{0.55 \times 10^{-3} \, \mathrm{eV}} \right)^{1.16}  \right)^{-1}
 \end{equation}
 with $\tilde{m}_i$ being defined as
\begin{equation}
\tilde{m}_i = v_\text{EW}^2\,\frac{(\hat{Y}^\dagger_D \hat{Y}_D)_{ii}}{M_i}\,.
\end{equation}
Using that the RH neutrino masses $M_i$ are directly related to the light neutrino masses $m_i$ at LO, see (\ref{miMi}), and are, in particular, not degenerate, 
we can show that  the relative mass splitting of any pair $N_i$ and $N_j$ ($i\neq j$) satisfies the bound  
 \begin{equation}
 	\frac{\left| M_i-M_j\right|}{M_i}\; \approx\;\left|1-\frac{m_i}{m_j}\right|\gg\frac{\left|(\hat{Y}^\dagger_D \hat{Y}_D)_{ij}\right|}{16\,\pi^2}\;\approx 3\times10^{-(6\div7)}\left(\frac{\kappa}{10^{-3}}\right)\,,
 \end{equation}
 for $y_0$ given in (\ref{rangey0}) and typical values of the expansion parameter $\kappa$, see (\ref{estimatekappa}). Thus, the perturbative expansion of the CP asymmetries in $Y_D$ in (\ref{epsi}) is still valid, see  
 e.g.~\cite{Buchmuller:1997yu, Blanchet:2006dq}. 
  Similarly, thanks to the relation in (\ref{miMi}) 
we can express the argument of the loop function $f(x)$ in (\ref{epsi}) in terms of $m_i$
\begin{equation}
 f (M_j/M_i) \approx f (m_i/m_j)\;.
 \end{equation}
In addition, using that $Y_D$ is proportional to the identity matrix at LO, see (\ref{mD0}), and that (\ref{miMi}) holds, we find 
\begin{equation}
\tilde{m}_i \approx m_i \; .
\end{equation}
Thus, also the efficiency factors $\eta_{ii}$ are functions of the light neutrino masses
\begin{equation}
\eta_{ii} \approx \eta (m_i) \; .
\end{equation}
We can distinguish two regimes: for $\tilde m_i\approx m_i \lesssim 1.1 \times 10^{-3} \, \mathrm{eV}$ we are in the weak washout regime, while for larger values of $m_i$ 
in the strong washout regime, see \cite{review_lepto_2}. 
In the first case the neutrino Yukawa interactions are so small that  the number density of RH neutrinos
does not  reach thermal equilibrium. Consequently, there is a partial cancellation between the anti-lepton 
asymmetry generated during the production of the RH neutrinos and the lepton
asymmetry produced by their subsequent decays. In the second case the abundance of RH neutrinos matches the equilibrium density and a sufficiently large
 lepton asymmetry can be entirely realized from the out-of-equilibrium decays of RH neutrinos. 

Given the strong correlation between light and RH neutrino masses, focussing on the range $10^{12} \, \mathrm{GeV} \lesssim M_i \lesssim 10^{14} \, \mathrm{GeV}$
also constrains the masses $m_i$. In particular, we use as range for the 
 lightest neutrino mass $m_0$  in our numerical analysis
 \begin{equation}
\label{rangem0}
 5 \times 10^{-4} \, \mathrm{eV} \lesssim m_0 \lesssim 0.1 \, \mathrm{eV} \; .
 \end{equation}
 This corresponds, if we fix the mass of the heaviest 
 RH neutrino to $10^{14} \, \mathrm{GeV}$, to the following interval of the Yukawa coupling $y_0$
\begin{equation}
\label{rangey0}
0.04 \lesssim \, y_0 \, \lesssim 0.6 \; .  
\end{equation}  
In this case the masses of the two lighter RH neutrinos automatically lie in the interval $10^{12} \, \mathrm{GeV} \lesssim M_i \lesssim 10^{14} \, \mathrm{GeV}$.
 As one can see, the coupling $y_0$ is (slightly) smaller than an order one number.\footnote{Its suppressed value can be easily achieved 
by an additional $Z_2$ symmetry under which, for example, only RH neutrinos transform. Then the Yukawa coupling $Y_D$ of the neutrinos, see (\ref{Ll}),
does not originate anymore from (a) renormalizable operator(s), but requires an insertion of a field whose VEV spontaneously breaks the $Z_2$ symmetry. 
This idea has been applied in e.g. \cite{HMP1}. Note that in this case also the correction $\delta Y_D$ in (\ref{deltaYD}) is not only suppressed by $\kappa$, but also by the expansion parameter 
associated with the VEV of the $Z_2$ group breaking field.}

As has been noticed in \cite{JM_08,Bertuzzo_09}, if RH neutrinos transform in an irreducible three-dimensional representation of a flavor group $G_f$, the CP asymmetries $\epsilon_i$
vanish in the limit of exact symmetry in the neutrino sector, since the combination $\hat{Y}_D^\dagger \hat{Y}_D$ is always proportional to the identity matrix. This also happens in our scenario.
For this reason, corrections $\delta Y_D$ to the Yukawa coupling $Y_D$  of the neutrinos play a crucial role for having successful leptogenesis.
We can expand the matrix  $\hat{Y}_D^\dagger \hat{Y}_D$ which enters in the expression of the CP asymmetries (\ref{epsi}) as
\begin{equation}
\label{YDYDexp}
\hat{Y}_D^\dagger \hat{Y}_D \approx U_R^\dagger \left( (Y_D^0)^\dagger Y_D^0 + (\delta Y_D)^\dagger Y_D^0 + (Y_D^0)^\dagger \delta Y_D   \right) U_R
\end{equation}
at lowest orders in the correction $\delta Y_D$. The first term in this expansion is proportional to the identity matrix and therefore does not contribute to the numerator of $\epsilon_i$. 
However, the other two terms lead to off-diagonal entries in the matrix combination $\hat{Y}_D^\dagger \hat{Y}_D$.
This expansion in $\delta Y_D$ also corresponds to an expansion in the parameter $\kappa$, simply because any correction $\delta U_R$ to the diagonalization
matrix of the RH neutrino mass matrix can only be effective, if at the same time also the correction $\delta Y_D$ is present. Thus, we expect from (\ref{epsi}) and (\ref{YDYDexp})
\begin{equation}
\label{epsikappa}
\epsilon_i \propto \kappa^2 \; .
\end{equation}
This observation is in accordance with the results obtained in models with a flavor symmetry only \cite{JM_08,Bertuzzo_09,HMP1}. Furthermore,
  using (\ref{epsi}) and (\ref{YDYDexp}) we see that the leading term in the CP asymmetries is independent of the parameter $y_0$.
Since this statement does not hold for higher order terms in $\kappa$, we study this issue in our numerical analysis (see discussion of case 1) in subsection \ref{sec34}).
We show that for $y_0$ in the range given in (\ref{rangey0}) and $\kappa$, $\tilde\kappa$, see definition in (\ref{tildekappa}), chosen as in (\ref{estimatekappa}),
the relative difference between the exact expression of the CP asymmetries defined in (\ref{epsi}) and their LO expansion in $\kappa$ 
is typically less than $10\%$.

\mathversion{bold}
\subsection{Dependence of $\epsilon_i$ on the low energy CP phases}
\label{sec31add}
\mathversion{normal}

We determine the conditions under which the dominant source of CP violation in leptogenesis is given by the low energy CP phases, contained in the PMNS mixing matrix. 
We thus study under which conditions the non-vanishing off-diagonal elements of the matrix combination $\hat{Y}^\dagger_D\hat{Y}_D$
depend only on the CP phases encoded in the mixing matrix $U_R=U_{PMNS}$. We perform this study at LO in the expansion parameter $\kappa$.   

From (\ref{YDYDexp}) we see that CP violation relevant for leptogenesis is related to the Dirac phase $\delta$ and the Majorana phases $\alpha$ and $\beta$
 provided that the matrix combination
\begin{equation}
\label{deltaYDYD0comb}
 (Y_D^0)^\dagger \delta Y_D
\end{equation}
fulfils one of the following conditions: $i)$ $(Y_D^0)^\dagger \delta Y_D$ is real, $ii)$ $(Y_D^0)^\dagger \delta Y_D$ is imaginary, $iii)$ $(Y_D^0)^\dagger \delta Y_D$ is complex and symmetric or $iv)$ $(Y_D^0)^\dagger \delta Y_D$ is complex and antisymmetric.
The first two possibilities could be ensured with the help of a CP symmetry.\footnote{In our scenario this is not necessary, since the matrix combination in (\ref{deltaYDYD0comb})
fulfils condition $iii)$. In any case it would require that this additionally imposed CP symmetry is distinct from the one preserved in the neutrino sector, since otherwise the CP asymmetries $\epsilon_i$ would vanish.} Notice that, if  conditions $i)$ and $iv)$ or conditions $ii)$ and $iii)$
are simultaneously realized, the CP asymmetries $\epsilon_i$ become more suppressed, $\epsilon_i \propto \kappa^4$ instead of proportional to $\kappa^2$, see (\ref{epsikappa}). It is thus not possible to reproduce the measured value of $Y_B$ for the expected size of $\kappa$, see (\ref{estimatekappa}) for an estimate.

In our scenario $Y_D^0$ is real and proportional to the identity matrix, see (\ref{mD0}), and the LO correction $\delta Y_D$ has the form of a complex diagonal matrix, see
 (\ref{deltaYD}), which is the most general form under the assumption that the dominant corrections to the neutrino Yukawa matrix arise from the charged lepton sector. The matrix combination $(Y_D^0)^\dagger \delta Y_D$ in (\ref{deltaYDYD0comb}) thus satisfies condition $iii)$. Consequently, all our results for the CP asymmetries are independent of the phases of the parameters $z_{1,2}$ at lowest order. In our numerical analysis we
also study the effect of the latter phases, entering in the subdominant terms, on the results for the CP asymmetries and find it to be typically less than $10\%$, if 
we compute the relative difference between the results using the expression of the CP asymmetries in (\ref{epsi}), which include 
phases of the complex parameters $z_{1,2}$ in (\ref{deltaYD}), and the corresponding LO expressions in $\kappa$ (see discussion of case 1) in subsection \ref{sec34}).

Finally, we note that also in several models with the flavor symmetry $A_4$ only \cite{JM_08,HMP1,Bertuzzo_09,AM_09}
the LO results of the CP asymmetries $\epsilon_i$ turn out to be independent of the phases of the correction $\delta Y_D$
and to only depend on the phases appearing in the diagonalization matrix $U_R$ that are identified with the Majorana phases at low energies. 
This result can be traced back to the fact that the matrix combination $ (Y_D^0)^\dagger \delta Y_D$ in (\ref{deltaYDYD0comb}) fulfils condition $iii)$ also in these models.

\subsection{General results in our framework}
\label{sec32}

Before discussing explicitly the results for the different cases, case 1) through case 3 b.1) and case 3 a), we would like to work out the form of the 
CP asymmetries obtained for $U_R=U_{PMNS}$. In this case mixing angles and CP phases are not assumed to be predicted by any flavor or CP symmetry, 
but only to have values compatible with the experimental data. We thus parametrize $U_R=U_{PMNS}$ using the convention given in (\ref{UPMNSdef}) in appendix \ref{app11}.
The form of the neutrino Yukawa coupling $Y_D$ is taken to be the sum of the LO term $Y_D^0$ in (\ref{mD0}) and the correction $\delta Y_D$ in (\ref{deltaYD}). 

As argued in the preceding subsection, at the dominant order in $\kappa$
only the real parts of the parameters $z_{1,2}$, see (\ref{deltaYD}), enter the expressions of the CP asymmetries. Thus, we define  
\begin{equation}
\label{Rez12}
\mathrm{Re} (z_1) = z \, \cos \zeta \;\;\; \mbox{and} \;\;\; \mathrm{Re} (z_2) = z \, \sin \zeta\, .
\end{equation}
We assume $z \geq 0$ and $\zeta$ to lie between $0$ and $2 \, \pi$. 
The vanishing of one of the parameters $z_1$ and $z_2$ refers to particular choices of $\zeta$, $\zeta=\pi/2, 3 \pi/2$ and $\zeta=0, \pi$, respectively.
As explained in appendix \ref{app3}, in an explicit model such special values can be achieved, for example, with a particular alignment of the VEV of a flavor symmetry breaking field.
Since $\kappa$ is always accompanied by $z$, we furthermore define as (effective) expansion parameter
\begin{equation}
\label{tildekappa}
\tilde\kappa = \kappa \, z \; .
\end{equation}
The analytic expressions of the CP asymmetries at LO in $\tilde\kappa$ in terms of the mixing angles, CP phases and $\zeta$
are in general rather lengthy.
However, they considerably simplify, if we choose $z_2=0$ or equivalently $\zeta=0, \pi$,
\begin{eqnarray}
\nonumber
 \!\!\!\! \!\!\!\!\epsilon_1 &\approx&-\frac{3\,\tilde\kappa^2}{2\,\pi}\,\cos^2\theta_{12}\,\cos^2\theta_{13}\Bigg[\cos^2\theta_{13}\,\sin^2\theta_{12}\,\sin\alpha\,f\left(\frac{m_1}{m_2}\right)
+\,\sin^2\theta_{13}\,\sin\beta\,f\left(\frac{m_1}{m_3}\right)\Bigg]\\
\label{eps1genz20}
  \!\!\!\!\!\!\!\!&=& -\frac{3\,\tilde\kappa^2}{2\,\pi}\, \Bigg[ I_1\,f\left(\frac{m_1}{m_2}\right)\,+\,I_2\,f\left(\frac{m_1}{m_3}\right) \Bigg]
\\ \nonumber
  \!\!\!\!\!\!\!\!\epsilon_2 &\approx& \frac{3\,\tilde\kappa^2}{2\,\pi}\,\sin^2\theta_{12}\,\cos^2\theta_{13} \, \Bigg[\cos^2\theta_{12}\,\cos^2\theta_{13}\,\sin\alpha\,f\left(\frac{m_2}{m_1}\right)
 -\sin^2\theta_{13}\,\sin\left(\beta-\alpha\right)\,f\left(\frac{m_2}{m_3}\right)\Bigg]
\\ \label{eps2genz20}
  \!\!\!\!\!\!\!\!&=& \frac{3\,\tilde\kappa^2}{2\,\pi}\, \Bigg[ I_1\,f\left(\frac{m_2}{m_1}\right)\,-\,I_3\,f\left(\frac{m_2}{m_3}\right) \Bigg]
\\  \nonumber
  \!\!\!\!\!\!\!\!\epsilon_3 &\approx& \frac{3\,\tilde\kappa^2}{8\,\pi}\, \sin^2 2\theta_{13} \Bigg[\cos^2\theta_{12}\,\sin\beta\,f\left(\frac{m_3}{m_1}\right)+\sin^2\theta_{12}\,\sin\left(\beta-\alpha\right)\,f\left(\frac{m_3}{m_2}\right)\Bigg]
\\ \label{eps3genz20}
  \!\!\!\!\!\!\!\!&=&\frac{3\,\tilde\kappa^2}{2\,\pi}\, \Bigg[ I_2\,f\left(\frac{m_3}{m_1}\right)\,+\,I_3\,f\left(\frac{m_3}{m_2}\right) \Bigg] \; .
\end{eqnarray}
The first line of these equations tells us that the sign of all CP asymmetries depends on the sines of the Majorana phases $\alpha$ and $\beta$
as well as on a possible sign coming from the loop function $f(m_i/m_j)$, $i \neq j$. Especially, the lepton mixing angles $\theta_{13}$ and $\theta_{12}$
do not have any influence on the sign. As we see, the CP asymmetries $\epsilon_i$ do neither depend on the Dirac phase $\delta$ nor on the atmospheric mixing angle $\theta_{23}$.
In the second line of these equations we have used the CP invariants $I_{1,2,3}$ given in appendix \ref{app11}. This establishes the connection between
low and high energy CP violation.

\begin{figure}[t!]
\begin{center}
\begin{tabular}{c}
\includegraphics[width=0.8\textwidth]{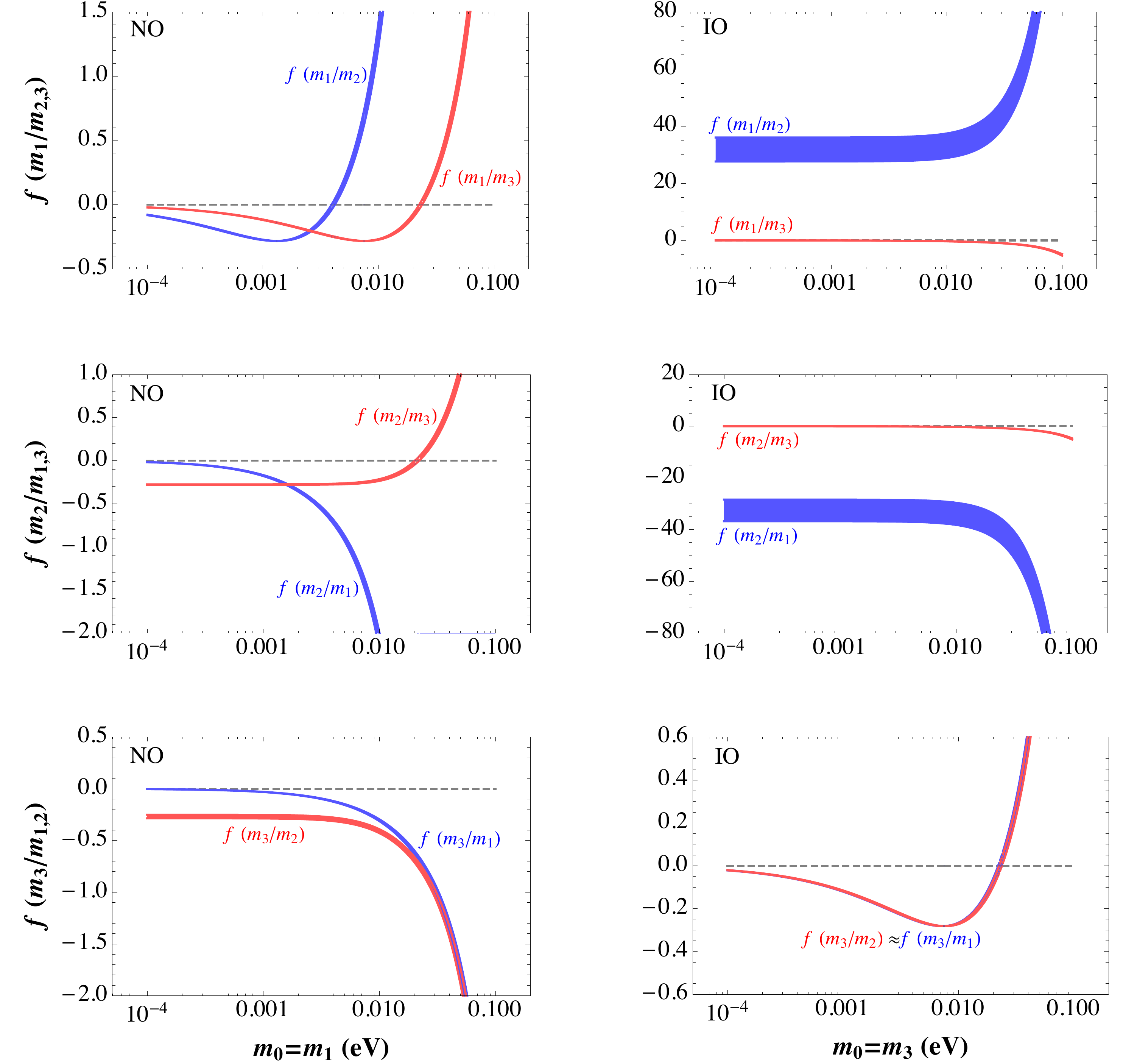}
\end{tabular}
\caption{\small{Behavior of the loop function $f(x)$ shown in (\ref{floop}) for the six different arguments $m_i/m_j$, $i\neq j$, $i,j=1,2,3$, with respect
to the lightest neutrino mass $m_0$. Plots on the left (right) side
correspond to a light neutrino mass spectrum with NO (IO). Note that $m_0$ is given by $m_1$ in the former and by $m_3$ in the latter case.
The mass squared differences $\Delta m_{\mathrm{sol}}^2$ and $\Delta m_{\mathrm{atm}}^2$  are taken to be in their experimentally preferred $3 \, \sigma$ range
given in \cite{nufit} and reported in (\ref{masses3sappNO}) and (\ref{masses3sappIO}) in appendix \ref{app12}, respectively. This variation explains the width of the red and blue 
bands.}}
\label{Fig:1}
\end{center}
\end{figure}

In order to understand the sign of the CP asymmetries and to estimate the size of the expansion parameter $\tilde{\kappa}$, necessary for achieving $\epsilon_i$ that permits a sufficiently large
baryon asymmetry, we first briefly analyze the behavior of the loop function.
The light neutrino mass spectrum, whether it follows NO or IO, affects the value of the loop function $f(x)$. In particular, it affects
also the sign of $f(x)$ as shown in  figure \ref{Fig:1}. We display the behavior of the loop function $f(x)$ for the six different arguments $m_i/m_j$, $i\neq j$, $i,j =1,2,3$,
with respect to the lightest neutrino mass $m_0$. In the panels on the left we assume light neutrino masses with NO, while in the right ones they are supposed to 
follow IO. The solar and atmospheric mass squared differences, $\Delta m_{\mathrm{sol}}^2$ and $\Delta m_{\mathrm{atm}}^2$,  take values in their experimentally preferred $3 \, \sigma$ ranges
given in \cite{nufit} and reported in appendix \ref{app12}, see (\ref{masses3sappNO}) and (\ref{masses3sappIO}). These ranges 
determine the width of the blue and red bands in the six plots of figure \ref{Fig:1}. This is particularly relevant for a light neutrino mass spectrum with IO, since $m_1$ and 
$m_2$ are almost degenerate in this case.  If we assume a hierarchical light neutrino mass spectrum, 
namely for $m_0\lesssim 4 \times 10^{-3}$ eV,\footnote{This value is chosen, since $f(m_1/m_2)$ changes its sign there for NO.} $f(m_i/m_j)$ is always negative apart from $f(m_1/m_{2})$ which is large and positive for IO.
For large $m_0$, $f(m_i/m_j) \approx - f(m_j/m_i)$, $i \neq j$, holds showing that half of the loop functions attain positive and half negative values. 
Notice also that within the whole range of $m_0$ shown in figure \ref{Fig:1}  the sign of the loop function does not change in the case of $f(m_2/m_1)$ and $f(m_3/m_{1,2})$ for NO and in the case of 
$f(m_1/m_{2,3})$ and $f(m_2/m_{1,3})$ for IO (all of them lead to a negative sign apart from $f(m_1/m_2)$ for IO). Relevant is also the 
size of $f(x)$. As is well-known for IO $f(m_1/m_2)\approx-f(m_2/m_1)$ always and very large, thus enhancing both CP asymmetries $\epsilon_1$ and $\epsilon_2$, see 
e.g.~figure \ref{Fig:3} in subsection \ref{sec34}, whereas $f(m_{1,2}/m_3)$ and $f(m_3/m_{1,2})$ are usually subdominant for all values of $m_0$ and for both mass orderings. Thus,
if no special cancellations occur, we expect the contribution from $\epsilon_3$ to the baryon asymmetry of the Universe to be always subdominant and thus unlikely to determine the sign of the
latter.

We also estimate the size of the CP invariants using the experimentally preferred $3 \, \sigma$ intervals of the reactor and the solar mixing angles shown in appendix \ref{app12}
\begin{eqnarray}
\nonumber
&&I_1 \approx \left( 0.18 \div 0.22 \right) \, \sin\alpha \; ,
\\ \label{estimateIi}
&&I_2 \approx \left(  1.2 \div 1.8 \right) \, \times 10^{-2} \, \sin\beta \;\;\; \mbox{and} \;\;\; I_3 \approx \left(  4.9 \div 8.4 \right) \, \times 10^{-3} \, \sin (\beta-\alpha) \; .
\end{eqnarray}
Thus, if the Majorana phase $\alpha$ is not small and the value of $f(m_1/m_2)$ and $f(m_2/m_1)$ is not close to zero for a particular
choice of the lightest neutrino mass $m_0$ (only occurring for NO and for $m_0 \approx 4 \times 10^{-3}$ eV), we expect
 the contributions involving $I_1$ ($\sin\alpha$) to dominate over the others. As a consequence, the absolute values of the CP asymmetries $\epsilon_1$ and $\epsilon_2$ should be (much) larger
than $|\epsilon_3|$. This suppression of $\epsilon_3$ (in addition to the one coming from the loop function $f(x)$) can also be understood by noticing that $\epsilon_3$ is proportional 
to $\sin^2 2 \theta_{13}$, whereas $\epsilon_{1,2}$ are proportional to $\cos^2\theta_{13}$. With (\ref{eps1genz20}) the expected size of the CP asymmetry is found to be: 
for $z$ being of order one and no particular enhancement or suppression of the loop function, i.e. $f(x)$ is also of order one, the expansion parameter has to fulfill
\begin{equation}
\label{estimatekappa}
\kappa \, , \; \tilde\kappa \gtrsim 10^{-3}
\end{equation}
in order to achieve $|\epsilon_{1,2}| \gtrsim 10^{-6}$ which is the typical size of the CP asymmetry necessary to ensure
successful leptogenesis \cite{review_lepto_2}. Our estimate in (\ref{estimatekappa}) for $\kappa$ is naturally reproduced in typical
non-SUSY as well as SUSY models, see our sketches of models in appendix \ref{app3}.

In the other limiting case, $z_1=0$ and thus $\zeta=\pi/2, 3\, \pi/2$, one can show that all three CP phases contained in the PMNS mixing matrix enter the expressions of the CP asymmetries.
In order to simplify the form of the latter, we thus consider cases in which two of the three CP phases are trivial. We focus here on the
case in which the Majorana phase $\beta$ and the Dirac phase $\delta$ are trivial, i.e.
\begin{equation}
\label{bdtrivial}
\beta= k_\beta \, \pi \;\; , \;\; \delta= k_\delta \, \pi \;\;\; \mbox{with} \;\;\; k_{\beta, \delta}=0, 1 \; ,
\end{equation}
since this also happens in case 1), see CP invariants in (\ref{CPinvcase1}).
We find compact formulae for $\epsilon_{1,3}$, if the conditions in (\ref{bdtrivial}) are imposed,
\begin{eqnarray}\nonumber
\epsilon_1 &\approx& -\frac{\tilde\kappa^2}{2 \, \pi} \, \sin\alpha \, \bigg[ \frac 12 \, \left( 1+\sin^2\theta_{13} \right) \, \sin 2\theta_{12} \, \cos 2\theta_{23}
 + (-1)^{k_\delta} \, \sin \theta_{13} \, \cos 2\theta_{12} \, \sin 2\theta_{23} \bigg]^2 \, f \left( \frac{m_1}{m_2} \right) \; ,
\\ \label{e1e3bdtrivial}
\epsilon_3 &\approx& \frac{\tilde\kappa^2}{2 \, \pi} \, (-1)^{k_\beta+1} \, \sin\alpha \, \cos^2 \theta_{13} \, \bigg[ \cos \theta_{12} \, \sin 2\theta_{23}
 + (-1)^{k_\delta} \, \sin \theta_{13} \, \sin\theta_{12} \, \cos 2\theta_{23} \bigg]^2 \, f \left( \frac{m_3}{m_2} \right) 
\end{eqnarray}
while $\epsilon_2$ can be expressed in terms of the other two CP asymmetries
\begin{equation}
\label{e2bdtrivial}
\epsilon_2 = - \epsilon_1 \, f \left( \frac{m_2}{m_1} \right) \left(f \left( \frac{m_1}{m_2} \right)\right)^{-1} 
 - \epsilon_3 \, f \left( \frac{m_2}{m_3} \right) \left(f \left( \frac{m_3}{m_2} \right)\right)^{-1}  \; .
\end{equation}

\begin{figure}[t!]
\begin{center}
\includegraphics[width=0.45\textwidth]{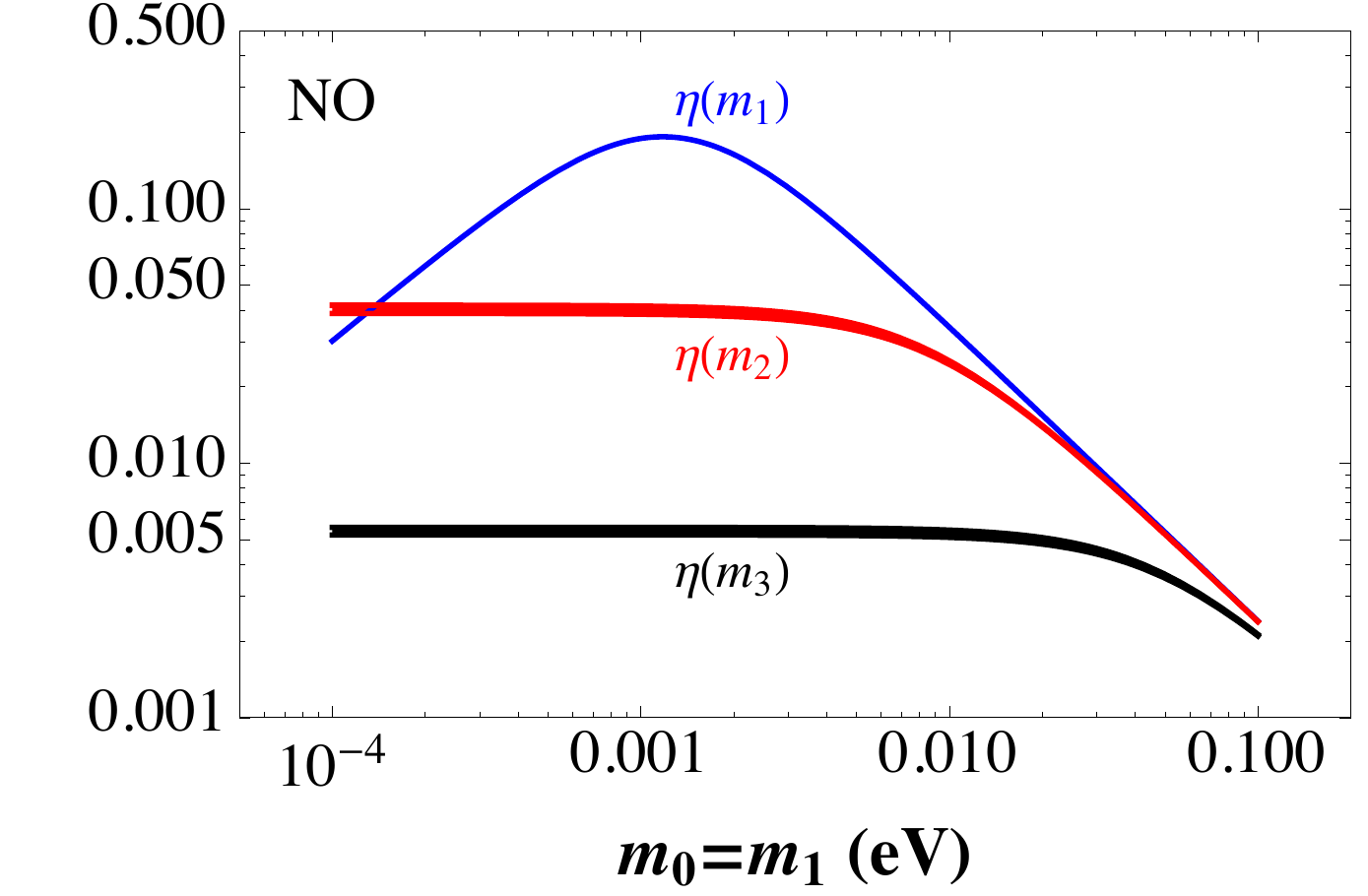} \hspace{0.07\textwidth} \includegraphics[width=0.45\textwidth]{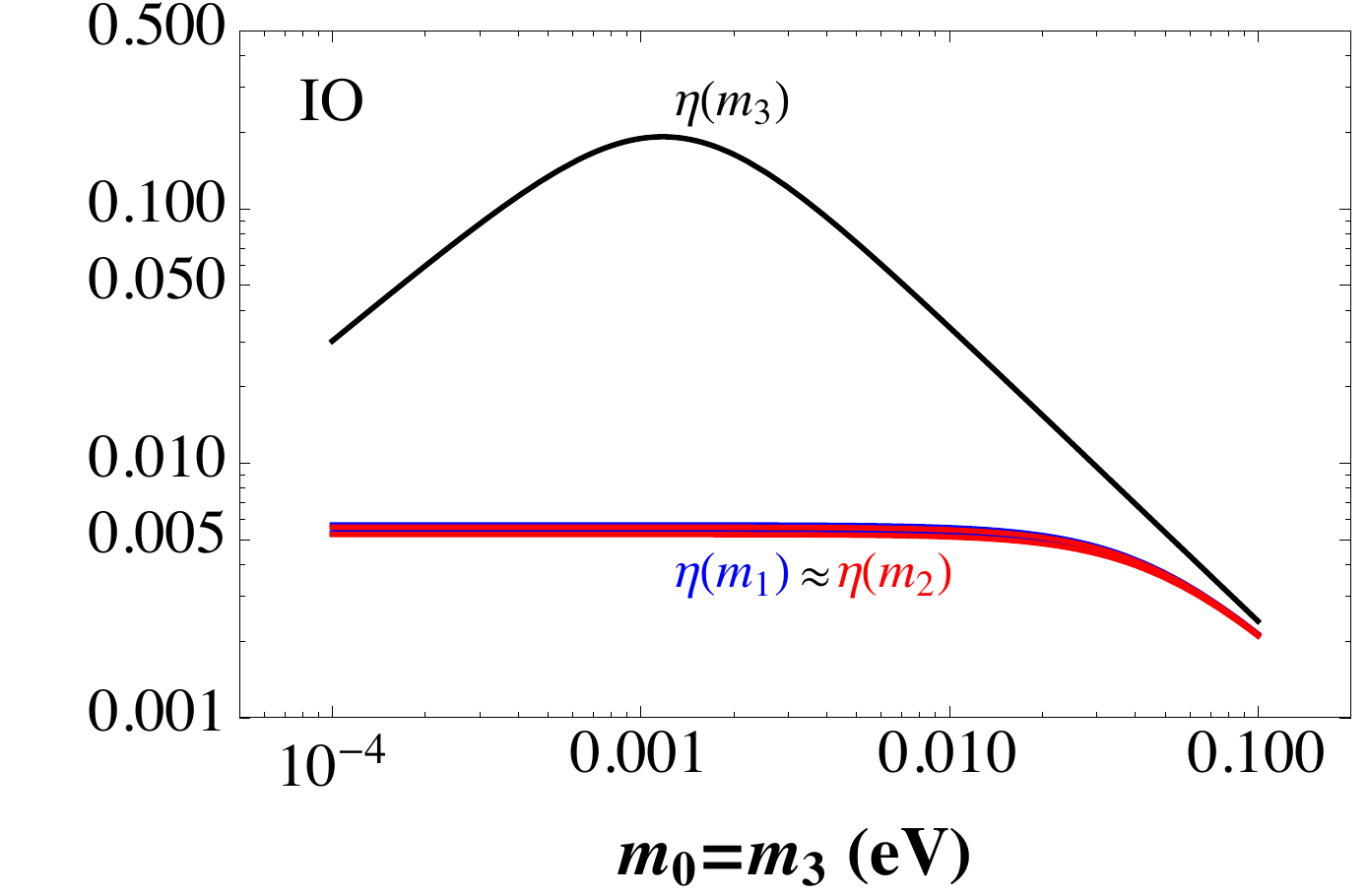}
\caption{\small{Behavior of the efficiency factor $\eta (m_i)$ defined in (\ref{eta}) for the three different neutrino masses $m_i$ with respect
to the lightest neutrino mass $m_0$. Plots on the left (right) side
correspond to a light neutrino mass spectrum with NO (IO). Note that $m_0$ is given by $m_1$ in the former and by $m_3$ in the latter case.
The mass squared differences $\Delta m_{\mathrm{sol}}^2$ and $\Delta m_{\mathrm{atm}}^2$  are taken to be in their experimentally preferred $3 \, \sigma$ ranges
given in \cite{nufit} and reported in (\ref{masses3sappNO}) and (\ref{masses3sappIO}) in appendix \ref{app12}, respectively. This variation explains the width of the curves.}}
\label{Fig:1b}
\end{center}
\end{figure}

\noindent This occurs, because $\mathrm{Im} \left( (\hat{Y}_D^\dagger \hat{Y}_D)_{ij}^2 \right)$ vanishes for $ij=13, 31$  (not only at LO in $\tilde\kappa$, but
exactly) in this case.
At this order in the expansion parameter $\tilde\kappa$ the CP asymmetries $\epsilon_1$ and $\epsilon_3$ depend on three different pieces:
the sine of the non-trivial Majorana phase $\alpha$, the loop function $f(x)$ and a combination of trigonometric functions of the mixing angles 
that forms a square. Thus, the sign of $\epsilon_{1}$ and $\epsilon_3$ depends on the sign of $\sin\alpha$ (and whether $\beta=0$ or $\pi$) and the sign of the loop function.
Like in the case above, we reach the conclusion that the sign of the CP asymmetries can be fixed with the knowledge of the Majorana phase(s).
Results very similar to those in (\ref{e1e3bdtrivial}-\ref{e2bdtrivial}) are obtained, if the Majorana phase $\alpha$ is trivial
instead of $\beta$. For completeness, we mention the formulae belonging to this choice of CP phases
in appendix \ref{app2a}. Moreover, we can analyze the case in which both Majorana phases are trivial, see also appendix \ref{app2a}.  
All three CP asymmetries $\epsilon_i$ are still non-zero in the latter case and all are proportional to $\sin\delta$. However, the sign of the CP asymmetries $\epsilon_i$
does not only depend on the sign of $\sin\delta$ (and the loop function $f(x)$) in this case, but also, for example, on the octant of the atmospheric mixing angle
($\cos 2 \theta_{23} \lessgtr 0$), see (\ref{app2aspecialcasee3}).

Before closing this general part we would like to also comment in a quantitative way on the efficiency factors $\eta_{ii}\approx\eta(m_i)$,  given in (\ref{eta}), that are necessary
for the computation of the baryon asymmetry of the Universe.  We show the variation of $\eta(m_i)$ with respect to
the lightest neutrino mass $m_0$ for NO (left panel) and IO (right panel) in figure \ref{Fig:1b}. Like in the case of the loop function, the width of the curves is given by the experimentally preferred $3 \, \sigma$ intervals of the 
solar and atmospheric mass squared differences. As we can see, the efficiency factors can reach sizable values, up to $0.2$, in the 
 weak washout regime, i.e. \ for $\tilde{m}_i \approx m_i\lesssim 1.1\times 10^{-3}$ eV. Conversely, in case of strong washout
 (for $m_i> 1.1 \times 10^{-3}$ eV), $\eta(m_i)$ is suppressed by one or more orders of magnitude. In particular, 
 for $m_i\gtrsim \sqrt{\Delta m_{\mathrm{sol}}^2 }\approx 9 \times 10^{-3}$ eV ($m_i\gtrsim \sqrt{|\Delta m_{\mathrm{atm}}^2 |}\approx 0.05$ eV) 
 we find $\eta(m_i)\lesssim 0.04$ ($\eta(m_i)\lesssim 5 \times 10^{-3}$). 
In the case of NO light neutrino mass spectrum, the maximum efficiency is obtained in the production of $Y_{B 1}$, since $\eta(m_1)$ is the largest, unless the CP asymmetry $\epsilon_1$
is suppressed. The strongest washout effects are expected for $Y_{B 3}$, which is controlled by the heaviest neutrino mass $m_3$. The opposite happens in the case 
 of a light neutrino mass spectrum with IO, where the lightest neutrino mass is $m_3$, while $m_2 \approx m_1\gtrsim \sqrt{|\Delta m_{\mathrm{atm}}^2 |}$. This does not necessarily mean
 that $\epsilon_{1,2}$ become much suppressed for IO, since the strong suppression of the baryon asymmetry $Y_{B i}$ due to washout effects can be easily compensated by the large
 enhancement of the  loop function $f (m_1/m_2)$ and $f(m_2/m_1)$ in the expression of the CP asymmetries, as shown in figure \ref{Fig:1}. The latter argument has also to be taken into account in the case of a QD light neutrino
 mass spectrum.

\subsection{Analytical discussion}
\label{sec33}

After the general analysis we discuss the features of case 1) through case 3 b.1) and case 3 a) in an analytical way first. Then, we also 
show a numerical study for examples of each of the cases.

\subsubsection*{Leptogenesis in case 1)}
\label{sec331}

We turn now to the predictions of the baryon asymmetry of the Universe for case 1). 
Accordingly, we express the CP asymmetries, defined in (\ref{epsi}), in terms of the relevant parameters which characterize the lepton mixing angles and CP 
phases of the PMNS mixing matrix in this case, namely $\theta$, the group theoretical quantities $n$ and $s$ (in the combination  $\phi_{s}$) as well as $k_{1,2}$ --see~(\ref{URcase1}-\ref{alphacase1}). The expressions 
at LO in the expansion parameter $\tilde\kappa$, given in (\ref{tildekappa}), are 
\begin{eqnarray}
\label{eps1case1}
\epsilon_1 & \approx & (-1)^{k_1} \,\frac{\tilde\kappa^2}{3\,\pi}\, \cos^2\left(\theta+\zeta\right)\,  f \left(\frac{m_1}{m_2}\right) \, \sin 6\,\phi_s\nonumber\\
& = & -\,\frac{3\,\tilde\kappa^2}{2\,\pi}\,I_1\left(\theta\to\theta+\zeta\right)\, f \left(\frac{m_1}{m_2}\right) 
\\
\epsilon_2 & \approx &
(-1)^{k_1+1}\,\frac{\tilde\kappa^2}{3\,\pi}\,\left(\cos^2 \left(\theta+\zeta\right) \, f\left(\frac{m_2}{m_1} \right) + (-1)^{k_2} \sin^2 \left(\theta+\zeta\right) \, f\left(\frac{m_{2}}{m_{3}}\right) \right)\,\sin 6\,\phi_s\nonumber\\
 \label{eps2case1}
&=& \frac{3\,\tilde\kappa^2}{2\,\pi}\, \left[I_1\left(\theta\to\theta+\zeta\right)\, f\left(\frac{m_{2}}{m_{1}}\right) - I_3\left(\theta\to\theta+\zeta\right)\, f\left(\frac{m_{2}}{m_{3}}\right) \right] \,, \\
 \epsilon_3 & \approx & (-1)^{k_1+k_2}\, \frac{\tilde\kappa^2}{3\,\pi}\, \sin^2\left(\theta+\zeta \right)\, f\left(\frac{m_3}{m_2}\right)\,  \sin 6\,\phi_s \nonumber \\
  \label{eps3case1}  
 & = & \frac{3\,\tilde\kappa^2}{2\,\pi}\, I_3\left(\theta\to\theta+\zeta\right)\, f\left(\frac{m_{3}}{m_{2}}\right) \; .                 
\end{eqnarray}
Here $I_{1,3} \left(\theta\to\theta+\zeta\right)$ have to be read as

\small
\begin{equation}
\label{I1I3case1shifted}
\!\!\!\!\!\!	I_1\left(\theta\to\theta+\zeta\right) =	\frac 2 9 (-1)^{k_1+1}\,\cos^2\left(\theta\,+\zeta\right)\,\sin6\,\phi_s \;\;\; \mbox{and} \;\;\;
	I_3\left(\theta\to\theta+\zeta\right) =	\frac 2 9 (-1)^{k_1+k_2}\,\sin^2\left(\theta\,+\zeta\right)\,\sin6\,\phi_s
\end{equation}
\normalsize
meaning that the free parameter $\theta$ is shifted by $\zeta$, which characterizes the correction $\delta Y_D$, see (\ref{deltaYD}) and (\ref{Rez12}).
Thus, the CP asymmetries $\epsilon_i$ can be formally written in terms of the CP invariants $I_1$ and $I_3$ given in (\ref{CPinvcase1}) ($I_2$ is zero anyway).
For $\zeta=0,~\pi$ the expressions in (\ref{I1I3case1shifted}) coincide with the CP invariants $I_{1,3}$ of case 1) and 
thus the formulae in (\ref{eps1case1}-\ref{eps3case1})
represent a special case of the general ones found in (\ref{eps1genz20}-\ref{eps3genz20}). We can also compare the formulae 
in (\ref{eps1case1}-\ref{eps3case1}) with the ones in (\ref{e1e3bdtrivial}-\ref{e2bdtrivial}) that have been obtained for $\zeta=\pi/2, 3 \pi/2$
and a non-zero Majorana phase $\alpha$ only. In doing so, we only have to remember that $\sin6\,\phi_s$ can be identified with  $\sin\alpha$, see (\ref{alphacase1}). 
In particular, we see that case 1) offers an example in which $\epsilon_1$ and $\epsilon_3$ only receive one contribution
and $\epsilon_2$ can be written in terms of the former two ones as in (\ref{e2bdtrivial}). Furthermore, with this explicit case we 
can also confirm several of the observations made in subsection \ref{sec31}, e.g.~no explicit dependence of the CP asymmetries on the
Yukawa coupling $y_0$ at LO, $\epsilon_i$ are proportional to $\tilde{\kappa}^2$ and the phases of the parameters $z_{1,2}$ only enter at higher orders in $\tilde\kappa$, i.e.~these
 terms are suppressed by $(\tilde\kappa/y_0)^\sigma$ with $\sigma\geq 1$ with the respect to the LO terms shown in (\ref{eps1case1}-\ref{eps3case1}).

The sign of the CP asymmetries is only determined by $\phi_s$ and by the one of the loop function $f(x)$ for a given value of $m_0$ and, in particular, is independent of the values of the
parameters $\theta$ and $\zeta$. The identification of $\sin6 \, \phi_s$ with $\sin\alpha$ also shows that in the case of no low energy CP violation, $s=0$ or $s=n/2$, also no
CP violation occurs at high energies. We note that replacing $s$ by $n-s$ reverses the sign of all three CP asymmetries $\epsilon_i$
\begin{equation}
\label{case1relepsisns}
\epsilon_i (n-s) = - \epsilon_i (s) \; .
\end{equation}
This equality holds exactly. 
Furthermore, we note that  at LO in $\tilde\kappa$ the CP asymmetries are periodic functions in 
$\zeta$ with the period $\pi$
\begin{equation}
\label{periodeps}
\epsilon_i (\zeta) \;=\; \epsilon_i (\zeta \;+ \;k\, \pi)\,,\quad \text{for} \; k=0, 1, 2, \dots 
\end{equation}
As can be checked by explicit computation, this does not hold, if we include the higher order terms in  $\tilde\kappa$  in (\ref{eps1case1}-\ref{eps3case1}). 
Similarly, at LO in $\tilde\kappa$ the CP asymmetries $\epsilon_i$ remain invariant, if we replace $\theta$ by $\pi-\theta$ and $\zeta$ by $k \, \pi-\zeta$ where $k$ is an integer chosen
in such a way that $k \, \pi -\zeta$ lies in the fundamental interval $[0, 2 \, \pi)$. Thus, we expect to obtain very similar results (for a different value of the parameter $\zeta$ however)
for $\theta$ in the two different admitted intervals, $0.169 \lesssim \theta \lesssim 0.195$ and $2.95 \lesssim \theta \lesssim 2.97$, see also table \ref{tab:case1}.
Differences, indeed, only arise at order $\tilde\kappa^3$ at most.

\subsubsection*{Leptogenesis in case 2)}
\label{sec332}

In this case the parameters which determine the lepton mixing angles and CP phases are $\phi_u$ and $\phi_v$ that characterize the chosen CP transformation $X$ (remember
$u$ and $v$ are related to $s$ and $t$, see (\ref{defuv})), the free parameter $\theta$ and $k_{1,2}$. Computing the CP asymmetries at the lowest order in $\tilde\kappa$ 
we obtain
\begin{eqnarray}
\epsilon_1 & \approx & 
 \frac{\tilde\kappa^2}{6\,\pi} \, \Bigg[(-1)^{k_1}  f \left(\frac{m_1}{m_2}\right) \Big(\left[\cos\left(\phi_u+2\zeta\right)+\cos2\theta\right]\sin\phi_v-\sin\left(\phi_u+2\zeta\right)\sin2\theta\cos\phi_v \Big) 
 \nonumber\\
&& +(-1)^{k_2+1}  f \left(\frac{m_1}{m_3}\right)\sin2\left(\phi_u-\zeta\right)\sin 2\theta\Bigg] 
\nonumber\\
&=&-\frac{3\,\tilde\kappa^2}{2\,\pi}\, \left[ I_1\left(\phi_u\to\phi_u+2\,\zeta\right) \,  f \left(\frac{m_1}{m_2}\right) +   I_2\left(\phi_u\to\phi_u-\zeta\right) \, f \left(\frac{m_1}{m_3}\right) \right] \, ,
 \label{e1case2}
\end{eqnarray}
\begin{eqnarray}
\epsilon_2 & \approx & 
 -\frac{\tilde\kappa^2}{6\,\pi}\, \Bigg[(-1)^{k_1}  f \left(\frac{m_2}{m_1}\right) \Big(\left[\cos\left(\phi_u+2\zeta\right)+\cos2\theta\right]\sin\phi_v-\sin\left(\phi_u+2\zeta\right)\sin2\theta\cos\phi_v \Big) 
 \nonumber\\
&& +(-1)^{k_1+k_2}  f \left(\frac{m_2}{m_3}\right) \Big(\left[\cos\left(\phi_u+2\zeta\right)-\cos2\theta\right]\sin\phi_v+\sin\left(\phi_u+2\zeta\right)\sin2\theta\cos\phi_v \Big)\Bigg] 
\nonumber\\
 &=&\frac{3\,\tilde\kappa^2}{2\,\pi} \, \left[ I_1\left(\phi_u\to\phi_u+2\,\zeta\right) \, f \left(\frac{m_{2}}{m_{1}}\right) -  I_3\left(\phi_u\to\phi_u+2\,\zeta\right) \, f \left(\frac{m_{2}}{m_{3}}\right) \right] \,,
 \label{e2case2}
\end{eqnarray}
\begin{eqnarray}
   \epsilon_3 & \approx &  \frac{\tilde\kappa^2}{6\,\pi}\, \Bigg[(-1)^{k_2}  f \left(\frac{m_3}{m_1}\right)\sin2\left(\phi_u-\zeta\right)\sin 2\theta 
  \nonumber\\
&& +(-1)^{k_1+k_2}  f \left(\frac{m_3}{m_2}\right) \Big(\left[\cos\left(\phi_u+2\zeta\right)-\cos2\theta\right]\sin\phi_v+\sin\left(\phi_u+2\zeta\right)\sin2\theta\cos\phi_v \Big)\Bigg] 
\nonumber\\
&=&  \frac{3\,\tilde\kappa^2}{2\,\pi}\, \left[ I_2\left(\phi_u\to\phi_u-\zeta\right) \, f \left(\frac{m_{3}}{m_{1}}\right) + I_3\left(\phi_u\to\phi_u+2\,\zeta\right) \, f\left(\frac{m_{3}}{m_{2}}\right)  \right]\,.
\label{e3case2}
\end{eqnarray}
Similar to case 1),  we can express the LO results for the CP asymmetries in terms of the quantities $I_i$, if we shift the group theoretical
parameter $\phi_u$ by appropriate multiples of $\zeta$. Again, only if the latter is taken to be $0$ or $\pi$, we re-cover expressions that depend on the CP invariants $I_i$, as defined
in  (\ref{CPinvcase2}-\ref{I3case2}), and thus on the sines of the Majorana phases $\alpha$ and $\beta$, while the Dirac phase $\delta$ does not appear. In this special case
we can also match to the formulae found in (\ref{eps1genz20}-\ref{eps3genz20}) where the PMNS mixing matrix is of a general form.
In addition, we find that $\epsilon_i$ in (\ref{e1case2}-\ref{e3case2}) fulfill at LO the equality in (\ref{periodeps}), i.e.~the expressions are invariant under the shift of $\zeta$ in $\zeta + k \, \pi$ where $k$ is an integer.
In contrast to case 1), it is not straightforward to determine the sign of $\epsilon_i$ just by looking at (\ref{e1case2}-\ref{e3case2}). However, given that $\theta$ is close to $0$, $\pi/2$ or $\pi$, the terms 
proportional to $\sin 2 \, \theta$ are suppressed compared to those proportional to 
$\cos\left(\phi_u+2\zeta\right) \pm \cos2\theta \approx \cos\left(\phi_u+2\zeta\right) \pm 1$, unless $\sin\phi_v$ is small. An example for the latter case is $v=0$ (if this choice of $v$ is admitted) and
we see that
\begin{equation}
\label{epsisignundetcase2}
\epsilon_i \propto \left( c_1 \, \sin \left( \phi_u + 2 \zeta \right) + c_2 \, \sin 2 \left( \phi_u-\zeta \right) \right) \, \sin 2\theta 
\end{equation}
with $c_{1,2}$ being expressions of the loop function $f (m_i/m_j)$ and $k_{1,2}$. This shows that the
terms in $\epsilon_i$ become proportional to $\sin \left( \phi_u + 2 \zeta \right)$ or $\sin 2 \left( \phi_u-\zeta \right)$ and thus for fixed $u$ ($\phi_u$) crucially depend on
the choice of $\zeta$; for example, changing the latter into $\zeta \pm \pi/2$ changes the overall sign of the CP asymmetries. 
This eventually explains why in our numerical analysis for $v=0$ positive and negative values of the baryon asymmetry of the Universe are equally
likely, see light-blue areas in the upper-left panel of figure \ref{Fig:6}. As a consequence, a prediction of the sign of $Y_B$, depending on the low energy CP phases, is not possible in this case.
The expression in (\ref{epsisignundetcase2}) also shows that the choice of the parameter $\theta$ becomes relevant, in particular whether $\theta \lesssim \pi/2$ or $\theta \gtrsim \pi/2$, and thus 
changing the admitted interval of $\theta$ inevitably changes the sign of the CP asymmetries.\footnote{In general there are two such intervals, one with $\theta \lesssim \pi/2$ and one with $\theta \gtrsim \pi/2$, for each 
admitted value of $\phi_u$, see \cite{HMM} for details.} If $v \neq 0$, however, we expect that our predictions of the CP asymmetries only slightly differ
 for $\theta$ being replaced by $\pi-\theta$, since
the dominant terms in $\epsilon_i$ are then proportional to $\sin\phi_v$, i.e.~$\sin\alpha$ see (\ref{sinaapproxcase2}), and in turn depend on $\cos 2 \, \theta$.

We can also relate the expressions found in case 2) to those in case 1). As has been shown in \cite{HMM}, this is possible, if we set $\theta=0$, identify the parameters of case 2) 
$\phi_u$ and $v$ with $2\, \theta_1$ and $6\, s_1$ of case 1), respectively, assuming that the group index $n$ is the same in both cases, as well as replace $k_2$ by $k_2+1$. Then, $\epsilon_i$ of case 2)
coincide with those of case 1), in particular $I_2$ (for any argument) vanishes. Another way to relate case 2) and case 1) that has been mentioned in \cite{HMM} 
is to set $u=0$ ($\phi_u=0$) and to identify $v$ with $6 \, s_1$. Albeit the solar and reactor mixing angles and the two Majorana phases ($\beta$ is trivial)
are the same in both cases,\footnote{The only differences lie in the values of the atmospheric mixing angle and the Dirac phase, since 
case 2) predicts both of them to be maximal, if $u$ equals zero.} the CP asymmetries are different for case 1) and case 2), since, in particular, $\epsilon_i$ in case 2) include additional 
terms proportional to $I_2\left(\phi_u\to\phi_u-\zeta\right)$. Only for $\zeta=0, \pi$ the CP asymmetries $\epsilon_i$ of case 1) and case 2) do coincide. 

In addition, we can consider leptogenesis in the case in which the PMNS mixing matrix/the matrix $U_R$ is given by one of the two matrices in (\ref{shiftPcase2}).
As explained in subsection \ref{sec22}, the formulae for mixing angles and CP invariants can be derived from those given for case 2), i.e. for $U_{PMNS}$/$U_R$ as in (\ref{URcase2}), 
using the transformations displayed in (\ref{shift1case2}) and (\ref{shift2case2}), respectively. In the same vein, we can obtain the formulae for the CP asymmetries $\epsilon_i$
in these cases. This is relevant for our numerical discussion, since we study an example in which the permutation matrix $P_1$ is applied from the left to
the matrix in (\ref{URcase2}). In doing so, we assume that the additional permutation originates from the Majorana mass matrix of the RH neutrinos and for this reason we have included it
in the matrix $U_R$. However, in principle we can also consider a different situation in which $U_R$ is still given by (\ref{URcase2}), whereas the PMNS mixing matrix is given by one
of the matrices in (\ref{shiftPcase2}). If so, the permutation has to arise from the charged lepton sector due to non-canonically ordered charged lepton masses.
One may wonder whether this difference can lead to new results. Indeed, this is not the case and it is no restriction of our discussion to focus only on the scenario with the permutation originating from 
the RH neutrino sector. The other case can be simply obtained from the latter one by a permutation $P$ of the elements of the (diagonal) correction $\delta Y_D$, i.e.
\begin{equation}
(\delta Y_D)_{P \; \mbox{\footnotesize from} \; m_l} = P \, (\delta Y_D)_{P^T \; \mbox{\footnotesize in} \; U_R} \, P^T \; .
\end{equation}
Thus, we only need to re-define the parameters $z_1$ and $z_2$ appropriately. Clearly, this affects (mainly) the explicit value of the parameter $\zeta$
corresponding to a certain choice of the ratio of the real parts of $z_1$ and $z_2$, see (\ref{Rez12}), but not the general conclusion on the size and the sign of the baryon asymmetry of the Universe, obtained
in the different cases.

\subsubsection*{Leptogenesis in case 3)}
\label{sec333}

We finally consider the predictions of $Y_B$ for the third type of mixing patterns, which are classified as case 3 b.1) and case 3 a). 
As we show, the CP asymmetries predicted in the two cases are closely related.

The analytic approximations of the CP asymmetries in case 3 b.1) are expressed at LO in $\tilde\kappa$ as a function of
the group theoretical quantities $\phi_m$, $\phi_s$, the free parameter $\theta$, $k_{1,2}$ as well as $\zeta$, that is

\small
 \begin{eqnarray}
 \nonumber
&& \epsilon_1  \approx 
  \frac{\tilde\kappa^2}{12\,\pi}\Bigg[(-1)^{k_2}f \left(\frac{m_1}{m_2}\right) \Big(4  \sin^2 (\phi_m +\zeta) \sin^2 \theta \sin 6 \, \phi_s 
+\sqrt{2}\left[\cos3 \, \phi_m-\cos\left(\phi_m-2\zeta\right)\right]\sin3 \, \phi_s\sin2\theta \Big)
\\
  \label{e1case3b1}
 &&+(-1)^{k_1+k_2}  f \left(\frac{m_1}{m_3}\right)\Big(4 \sin^2 (\phi_m +\zeta) \cos^2 \theta  \sin 6 \, \phi_s
 -\sqrt{2}\left[\cos3 \, \phi_m-\cos\left(\phi_m-2\zeta\right)\right]\sin3 \, \phi_s\sin2\theta\Big)  \Bigg]
 \\
 \nonumber 
 &&= -\frac{3\,\tilde\kappa^2}{2\,\pi}\Bigg[  \Big( I_1\left(\phi_m\to\phi_m+\zeta\right)+(-1)^{k_2+1}R_-(\zeta)\Big) \, f \left(\frac{m_1}{m_2}\right)
  +\Big( I_2\left(\phi_m\to\phi_m+\zeta\right)+(-1)^{k_1+k_2}R_-(\zeta)\Big) \, f \left(\frac{m_1}{m_3}\right) \Bigg]
\end{eqnarray}
\begin{eqnarray}
\nonumber
 &&\epsilon_2  \approx 
  -\frac{\tilde\kappa^2}{12\,\pi}\Bigg[ (-1)^{k_2}f \left(\frac{m_2}{m_1}\right) \Big(4 \sin^2 (\phi_m +\zeta) \sin^2 \theta \sin 6 \, \phi_s 
  +\sqrt{2}\left[\cos3 \, \phi_m-\cos\left(\phi_m-2\zeta\right)\right]\sin3\, \phi_s\sin2\theta \Big) 
  \\
 \label{e2case3b1}
 &&+(-1)^{k_1+1}  f \left(\frac{m_2}{m_3}\right)\Big(4 \cos^2 (\phi_m +\zeta) \cos 2 \theta \sin 6 \, \phi_s
 +\sqrt{2}\left[\cos3 \, \phi_m+\cos\left(\phi_m-2\zeta\right)\right]\sin3 \, \phi_s\sin2\theta\Big) \Bigg] 
 \\
 \nonumber
  &&= \frac{3\,\tilde\kappa^2}{2\,\pi}\Bigg[  \Big( I_1\left(\phi_m\to\phi_m+\zeta\right)+(-1)^{k_2+1}R_-(\zeta)\Big) \, f \left(\frac{m_2}{m_1}\right)
 - \Big( I_3\left(\phi_m\to\phi_m+\zeta\right)+(-1)^{k_1}R_+(\zeta)\Big) \, f \left(\frac{m_2}{m_3}\right) \Bigg] 
\end{eqnarray}
 \begin{eqnarray}
\nonumber
 &&\epsilon_3 \approx 
 -\frac{\tilde\kappa^2}{12\,\pi}\Bigg[(-1)^{k_1+k_2}  f \left(\frac{m_3}{m_1}\right)\Big(4 \sin^2 (\phi_m +\zeta)  \cos^2 \theta  \sin 6 \, \phi_s 
 -\sqrt{2}\left[\cos3 \, \phi_m-\cos\left(\phi_m-2\zeta\right)\right]\sin3 \, \phi_s\sin2\theta\Big) 
 \\
 \label{e3case3b1}
 && +(-1)^{k_1}f \left(\frac{m_3}{m_2}\right) \Big(4 \cos^2 (\phi_m +\zeta) \cos2 \theta\sin 6 \, \phi_s 
 +\sqrt{2}\left[\cos3 \, \phi_m+\cos\left(\phi_m-2\zeta\right)\right]\sin3 \, \phi_s\sin2\theta \Big)  \Bigg]  
 \\
 \nonumber
 &&= \frac{3\,\tilde\kappa^2}{2\,\pi}\Bigg[ \Big( I_2\left(\phi_m\to\phi_m+\zeta\right)+(-1)^{k_1+k_2}R_-(\zeta)\Big) \,  f \left(\frac{m_3}{m_1}\right)
 + \Big( I_3\left(\phi_m\to\phi_m+\zeta\right)+(-1)^{k_1}R_+(\zeta)\Big) \, f \left(\frac{m_3}{m_2}\right) \Bigg]
\end{eqnarray}
\normalsize
Like in the other cases, we have arranged the LO expressions of $\epsilon_i$ in terms of the CP invariants $I_1$, $I_2$ and $I_3$, defined in~(\ref{CPinvcase3b1}-\ref{I3case3b1}). 
This requires the group theoretical parameter $\phi_m$ to be shifted into $\phi_m+\zeta$
as well as to add a further piece 
\begin{equation}
\label{Rpm}
R_{\pm} \left(\zeta\right) = -\frac{\sqrt{2}}{9}\sin\frac{3\,\zeta}{2}\left[\sin\left(\phi_m-\frac\zeta2\right)\pm\sin3\left(\phi_m+\frac\zeta2\right)\right]\sin3\,\phi_s\sin 2\theta\; .
\end{equation}
We also confirm in this particular case all statements made in the general part, regarding  the dependence on the Yukawa coupling $y_0$,
the expansion in $\tilde{\kappa}$ and the appearing of the imaginary parts of the parameters $z_{1,2}$ in the subdominant terms only.  
We can furthermore check that the LO expressions of the CP asymmetries in (\ref{e1case3b1}-\ref{e3case3b1}) are periodical in $\zeta$
with periodicity $\pi$. As in the other cases, this does not hold at higher orders in $\tilde{\kappa}$. In addition, we find, like in the general case (see (\ref{eps1genz20}-\ref{eps3genz20})), that 
for $\zeta=0, \, \pi$ the CP asymmetries can be written in terms of the CP invariants $I_i$ in (\ref{CPinvcase3b1}-\ref{I3case3b1}) which shows that there is no explicit dependence
on the Dirac phase. As discussed in \cite{HMM}, the replacement of $\theta$ with $\pi-\theta$ does not give rise to a symmetry transformation in case 3 b.1), unless 
we also replace $m$ ($\phi_m$) with $n-m$ ($\pi-\phi_m$) or $s$ ($\phi_s$) with $n-s$ ($\pi-\phi_s$), see (\ref{trafocase3}-\ref{trafo2case3}). For this reason and because 
$R_\pm (\zeta)$ also change sign for $\theta\; \rightarrow \; \pi-\theta$ and $s \; \rightarrow \; n-s$, we can make the following observation 
\begin{equation}
\label{symmetryepsicase3}
 \epsilon_i \left( n-s, \pi-\theta \right) = -  \epsilon_i \left( s, \theta \right) \; .
\end{equation}
This equality holds for all three asymmetries, at all orders in $\tilde{\kappa}$ and for all choices of the parameters $z_{1,2}$.
We use this fact in our numerical analysis and only display results for $s \leq n/2$.

If $m=n/2$, like in the example studied in subsection \ref{sec343}, we see that changing $\theta$ to $\pi-\theta$ becomes a symmetry transformation, as defined in \cite{HMM}. 
Thus, two admitted intervals of $\theta$ are expected. We, however, only discuss results for one of them in subsection \ref{sec343}, since the CP asymmetries $\epsilon_i$ are likely to be very similar
for both intervals $\theta$.  The reasoning is as follows: replacing $\theta$ with $\pi-\theta$ does not affect the CP invariants $I_i$ and thus the Majorana phases, while it changes 
the sign of $J_{CP}$; at the same time, we know that the CP asymmetries $\epsilon_i$ mostly depend on $\sin\alpha$; in the special case in which only $\delta$ is non-trivial we have already argued in 
subsection \ref{sec32} (see also appendix \ref{app2a}) that fixing the sign of $Y_B$ becomes impossible. We can confirm the latter statement in the case at hand by studying 
the expressions in (\ref{e1case3b1}-\ref{e3case3b1}) for $s=n/2$ in addition to $m=n/2$ (assuming that $n$ is even)
 \begin{eqnarray}
 \label{e1case3b1sim}
 \epsilon_1 & \approx & \frac{\tilde\kappa^2}{6\,\sqrt{2}\,\pi}\, (-1)^{k_2}\Bigg[  f \left(\frac{m_1}{m_2}\right)+ (-1)^{k_1+1} f \left(\frac{m_1}{m_3}\right)\Bigg]\sin2 \zeta\,\sin2 \theta\,,
 \\
  \label{e2case3b1sim}
 \epsilon_2 & \approx & - \frac{\tilde\kappa^2}{6\,\sqrt{2}\,\pi}\, (-1)^{k_2} \, \Bigg[   f \left(\frac{m_2}{m_1}\right)+ (-1)^{k_1+k_2} f \left(\frac{m_2}{m_3}\right)\Bigg]\sin2 \zeta\,\sin2 \theta\,,
 \\
  \label{e3case3b1sim}
 \epsilon_3 &\approx &   \frac{\tilde\kappa^2}{6\,\sqrt{2}\,\pi}\, (-1)^{k_1+k_2}\Bigg[  f \left(\frac{m_3}{m_1}\right)+ (-1)^{k_2} f \left(\frac{m_3}{m_2}\right)\Bigg]\sin2 \zeta\,\sin2 \theta\,.
\end{eqnarray}
Here we clearly see that we cannot predict the sign of the CP asymmetries $\epsilon_i$, since it crucially depends on the choice of the free parameter $\zeta$. 
It also crucially depends on whether $\theta$ is smaller or larger than $\pi/2$. Thus, changing the admitted interval of $\theta$ leads to a change in 
the sign of $\epsilon_i$. Notice the similarity between the formulae in (\ref{e1case3b1sim}-\ref{e3case3b1sim}) and those in (\ref{epsisignundetcase2}) valid for case 2).
In addition, we find that the CP asymmetries are zero at LO in $\tilde\kappa$ for $\zeta=0, \, \pi/2, \, \pi, \, 3\pi/2$,  while they take maximal positive (negative) values for $\zeta=\pi/4, \, 5 \pi/4$ ($\zeta=3\pi/4, \, 7\pi/4$).
If we set either the imaginary part of $z_1$ or $z_2$ to zero, it turns out that $\epsilon_i$ vanish at all orders in $\tilde\kappa$ for $\zeta=0, \, \pi$ and not only at LO. The 
vanishing of $\epsilon_i$ (at LO) for $\zeta=0, \, \pi$ is consistent with the formulae in (\ref{eps1genz20}-\ref{eps3genz20}) obtained in the general case, since case 3 b.1)
entails trivial Majorana phases for $m=n/2$ and $s=n/2$. In the same vein, $\epsilon_i=0$ at LO for $\zeta=\pi/2, \, 3\pi/2$ can be matched to the formulae in (\ref{app2aspecialcasee1}-\ref{app2aspecialcasee3}) in
appendix \ref{app2a}, if we take into account that case 3 b.1) predicts for $m=n/2$ and $s=n/2$ maximal $\theta_{23}$ and $\delta$. 

Moreover, we comment on the results for the CP asymmetries $\epsilon_i$ in case 3 a).
These can be easily obtained from (\ref{e1case3b1}-\ref{e3case3b1}), making the following replacements 
\begin{eqnarray}
\label{epscase3a}
	&& I_{1,3b1}\left(\phi_m\to\phi_m+\zeta\right)+(-1)^{k_2+1}R_-(\zeta)\to I_{1,3a}\left(\phi_m\to\phi_m+\zeta\right)+(-1)^{k_1}R_+(\zeta)\,,\\
	&& I_{2,3b1}\left(\phi_m\to\phi_m+\zeta\right)+(-1)^{k_1+k_2}R_-(\zeta)\to I_{2,3a}\left(\phi_m\to\phi_m+\zeta\right)+(-1)^{k_2}R_-(\zeta)\,,\\
	&& I_{3,3b1}\left(\phi_m\to\phi_m+\zeta\right)+(-1)^{k_1}R_+(\zeta)\to I_{3,3a}\left(\phi_m\to\phi_m+\zeta\right)+(-1)^{k_1+k_2+1}R_-(\zeta)\,.\label{epscase3aa}
\end{eqnarray}
and using the relations among the CP invariants of the two cases, see (\ref{CPinvcase3a}).

Finally, if we consider $U_{PMNS}$/$U_R$ of case 3 b.1) and case 3 a) with rows permuted (similar to what is shown in (\ref{shiftPcase2}) for case 2)), the given formulae for the CP asymmetries $\epsilon_i$ in (\ref{e1case3b1}-\ref{e3case3b1}) and 
(\ref{epscase3a}-\ref{epscase3aa}) can still be applied, as long as
we use the same replacements of the parameters $m$ and $\theta$ that need to be taken into account when computing the lepton mixing angles and CP invariants, see discussion at the end of subsection \ref{sec22}.

\subsection{Numerical discussion}
\label{sec34}

We first summarize our specific choices of several of the involved parameters that we use throughout 
the numerical discussion of the different cases and then show results for
examples of case 1), case 2) as well as case 3 b.1). We also comment on the results expected for case 3 a).
We note that we always use the exact expressions for the CP asymmetries $\epsilon_i$ in our numerical analysis, i.e.~those
that are not expanded in the parameter $\tilde{\kappa}$ and are instead computed using (\ref{epsi}) and the corresponding form of the 
mixing matrix $U_{R, i}$, see (\ref{URcase1}, \ref{URcase2}, \ref{shiftPcase2}, \ref{URcase3}) and (\ref{URcase3a}).

\subsubsection*{Preliminaries}
\label{sec340}

The mixing matrix $U_R$ contains in all cases the two parameters $k_1$ and $k_2$ encoded in the matrix $K_\nu$, see (\ref{Knu}). In the following, 
we set, if not stated otherwise,  
\begin{equation}
\label{k1k2num}
k_1 = 0\quad\text{and}\quad k_2 = 0 \; .
\end{equation}
Furthermore, we fix the heaviest RH neutrino mass to  
\begin{equation}
M_{1(3)} = 10^{14}\,\text{GeV} \;\;\;\;\; \text{for} \;\;\;\;\; \text{NO (IO)} \; .
\end{equation}
Using (\ref{miMi}) we can then express $y_0$ as
\begin{equation}
y_0 \approx 1.82 \,\sqrt{\frac{m_0}{1~\text{eV}}} \; .
\end{equation}
The lightest neutrino mass $m_0$ is chosen in the interval in (\ref{rangem0}) and $y_0$, consequently, falls into the range in (\ref{rangey0}).
The other neutrino masses (light and heavy ones) are then determined by the two experimentally measured mass squared differences, see (\ref{massesNO}-\ref{dmbf}).
In particular, the masses of all three RH neutrinos vary between $10^{12} \, \mathrm{GeV}$ and $10^{14} \, \mathrm{GeV}$ in this setting.

The expansion parameter $\tilde\kappa$ is taken as constant and fixed to 
\begin{equation}
\label{kappafix}
 \tilde\kappa = 4 \times 10^{-3} \; . 
\end{equation}
Thus, when $z$ varies, also the expansion parameter $\kappa$ has to slightly vary.
The value of $\tilde\kappa$ in (\ref{kappafix}) is appropriate in order to achieve large enough CP asymmetries, since it satisfies the bound in (\ref{estimatekappa}).
The parameter $\zeta$ that defines the relative size among the real parts of the two parameters $z_1$ and $z_2$, see (\ref{Rez12}),  
is either fixed to specific values
\begin{equation}
\label{inzeta1}
 \zeta =0,\,\pi \;\;\; \mbox{equivalent to} \;\;\;  \text{Re}(z_2) = 0 \;\;\; \mbox{or} \;\;\;   \zeta = \frac\pi 2,\,\frac{3\,\pi}{2} \;\;\; \mbox{equivalent to} \;\;\; \text{Re}(z_1) = 0
\end{equation}
or is taken to lie in the intervals ${\cal I}_i$ which is equivalent to constraining the real parts of $z_{1,2}$ to assume values in the ranges $[-2,-1/2]$ and/or $[1/2,2]$
\begin{equation}
\label{inzeta2}
\begin{array}{ll}
\mathcal{I}_1= [0.24,1.33] &\mbox{equivalent to} \;\;\; \mathrm{Re}(z_1) > 0 \;\; \mbox{and} \;\; \mathrm{Re}(z_2) > 0 \; ,\\[0in]
\mathcal{I}_2= [1.82,2.90] &\mbox{equivalent to} \;\;\; \mathrm{Re}(z_1) < 0 \;\; \mbox{and} \;\; \mathrm{Re}(z_2) > 0\; ,\\[0in]
\mathcal{I}_3= [3.39,4.47] &\mbox{equivalent to} \;\;\; \mathrm{Re}(z_1) < 0 \;\; \mbox{and} \;\; \mathrm{Re}(z_2) < 0\; ,\\[0in]
\mathcal{I}_4= [4.96,6.04] &\mbox{equivalent to} \;\;\; \mathrm{Re}(z_1) > 0 \;\; \mbox{and} \;\; \mathrm{Re}(z_2) < 0 \; .
\end{array}
\end{equation}
In our numerical discussion these different choices of the parameter $\zeta$ are indicated in different colors in the figures
\begin{equation}
\label{zetacolor}
 \zeta =0,\,\pi \;\; \text{in red} \;\; , \;\;
 \zeta =  \frac\pi 2,\,\frac{3\,\pi}{2}  \;\; \text{in green} \;\; \mbox{and} \;\;
 \zeta \in  {\cal I}_i \;\; \text{in the light-blue areas.}
\end{equation}
If not otherwise stated, the imaginary parts $\mathrm{Im}(z_{1,2}) $ of $z_1$ and $z_2$ are set to zero in the following, since we have argued below (\ref{deltaYDYD0comb})
that these do not enter the expressions of the CP asymmetries $\epsilon_i$ at the dominant order in $\kappa$.

In the discussion of the numerical example for case 1) we not only show figures of the baryon asymmetry $Y_B$ of the Universe, but also of the CP asymmetries $\epsilon_i$.
Moreover, we present for this example results for both, NO as well as IO, light neutrino mass spectra. We also study the effect
of choosing the two parameters $k_1$ and $k_2$ differently from our standard choice in (\ref{k1k2num}). In addition, we analyze the impact and form of subleading terms in the expansion
in $\tilde\kappa$ as well as the effect of non-vanishing imaginary parts of the parameters $z_{1,2}$. In the subsequent discussion of an example for case 2)
and one for case 3 b.1) we focus on a light neutrino mass spectrum with NO and choose $k_1=k_2=0$, since the
relations to and changes coming from the other possible choices become apparent from the study for case 1). We also set the imaginary parts of $z_{1,2}$ to zero for case 2) and case 3 b.1)
after having shown for case 1) that their impact on the final result for $Y_B$ is small.

\subsubsection*{Leptogenesis in case 1)}
\label{sec341}

For all (even) $n$ and all CP transformations with $s=0,\dots,n-1$, the 
values of the free parameter $\theta$, which allow to reproduce the lepton mixing angles in accordance with the experimental data at the $3\, \sigma$ level or better, lie in the two intervals
$0.169\lesssim \theta \lesssim 0.195$ and  $2.95 \lesssim \theta \lesssim 2.97$  \cite{HMM}. 
These two choices are distinguished by the resulting value of the atmospheric mixing angle which is in the second octant for $0.169 \lesssim \theta \lesssim 0.195$ and in the first one 
for $2.95\lesssim \theta \lesssim 2.97$. According to the reasons given in subsection \ref{sec33}, it is sufficient to concentrate on values of  $\theta$ in the first interval. As shown in (\ref{CPinvcase1})
in subsection \ref{sec22}, in case 1) only one of the three CP phases can be non-trivial. The minimal choice of $n$ leading to $\sin\alpha$ different from zero is $n=4$. 
This requires $s$ to be odd. The resulting CP phase $\alpha$ is maximal and for $k_1=k_2=0$ $\sin\alpha$ is positive (negative), if $s=1$ 
($s=3$), according to the formula in (\ref{alphacase1}). In case $s$ is even we find no CP violation at all and consequently vanishing CP asymmetries $\epsilon_i$. 
This information together with the results for the other lepton mixing parameters is gathered in table \ref{tab:case1}.

\begin{figure}[t!]
\begin{center}
\begin{tabular}{c}
\includegraphics[width=\textwidth]{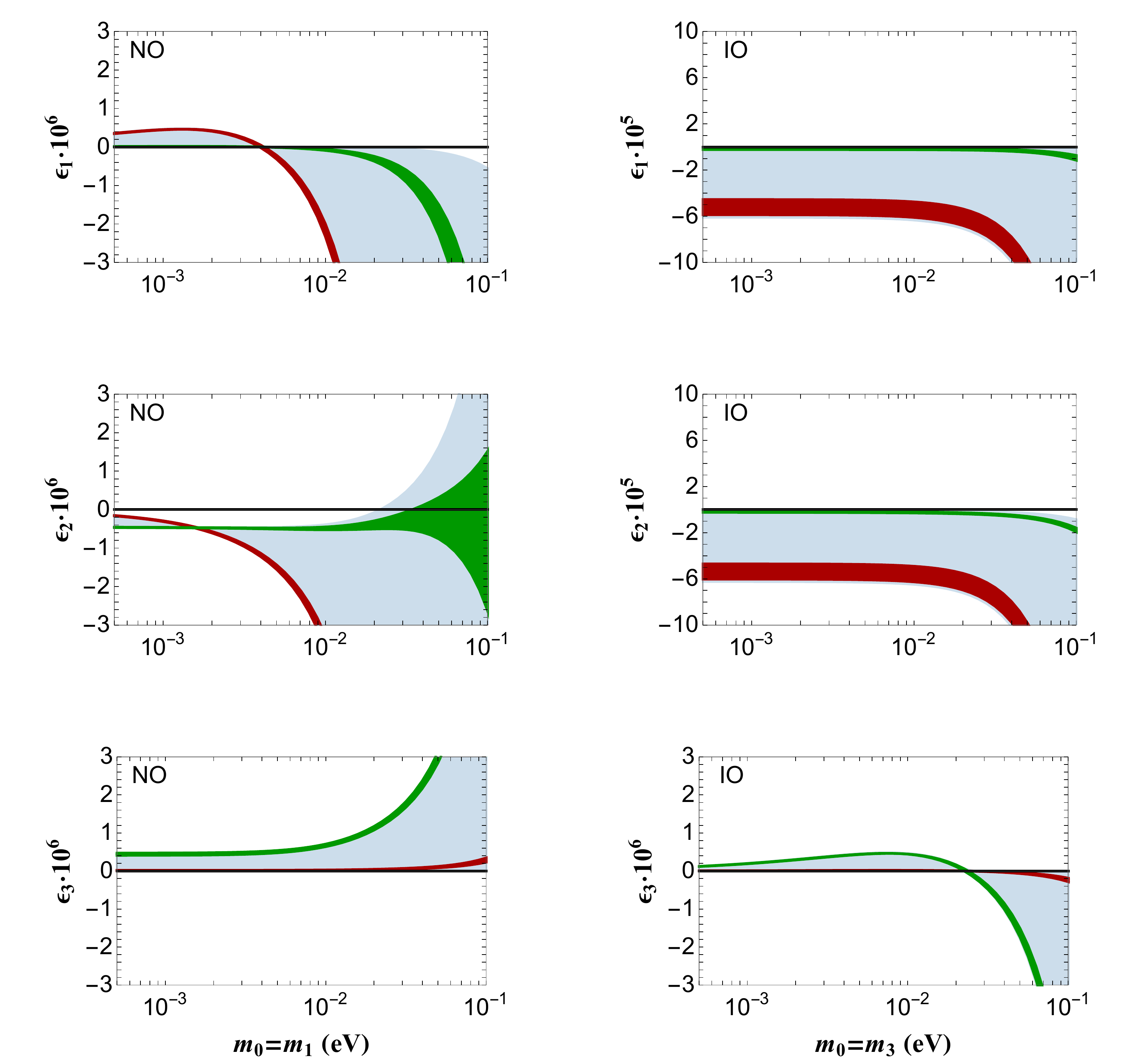}
\end{tabular}
\caption{\small{\textbf{Case 1).} CP asymmetries $\epsilon_i$ for $n=4$, $s=1$, $k_1=k_2=0$ and $\tilde\kappa=4 \times 10^{-3}$ as function of the lightest neutrino mass $m_0$ for
NO (left panel) and IO (right panel) light neutrino masses. The Majorana phase $\alpha$ is  $\alpha=\pi/2$. The parameter $\theta$ is taken in the interval $0.169\lesssim \theta \lesssim 0.195$.
The neutrino mass squared differences are varied within their experimentally preferred $3 \, \sigma$ ranges \cite{nufit}.
Red and green areas correspond to the special choices $\zeta=0,~\pi$ and $\zeta=\pi/2,~3\pi/2$, respectively. The light-blue area is obtained for $\zeta$ lying in one of the four intervals ${\cal I}_i$,
reported in (\ref{inzeta2}). Note the different scale of the vertical axis chosen in the plots for $\epsilon_1$ and $\epsilon_2$ in the case of IO light neutrino masses, showing the enhancement of the CP asymmetries
that arises from the (near) degeneracy of $m_1$ and $m_2$ (compare also the behavior of the loop function $f(x)$ in figure \ref{Fig:1}).}}
\label{Fig:3}
\end{center}
\end{figure}

For $n=4$ and $s=1$ (meaning $\alpha=\pi/2$) we show the predictions of the CP asymmetries $\epsilon_i$ in figure \ref{Fig:3} as a function of the lightest neutrino
mass $m_0$ for a light neutrino mass spectrum with NO (left panel) and IO (right panel). 
We vary the parameter $\theta$ in the interval $0.169\lesssim \theta \lesssim 0.195$ and the 
 solar and atmospheric mass squared differences within their experimentally preferred $3 \, \sigma$ ranges, determined by the global fit analysis \cite{nufit}, see appendix \ref{app12}. 
%
%
\begin{table}[t!]
\centering
\catcode`?=\active \def?{\hphantom{0}}
\begin{tabular}{!{\vrule width 1pt}@{\quad}>{\rule[-2mm]{0pt}{6mm}}l@{\quad}!{\vrule width 2pt}@{\quad\quad}c@{\quad\quad}|@{\quad\quad}c@{\quad\quad}|@{\quad\quad}c@{\quad\quad}|@{\quad\quad}c@{\quad\quad}!{\vrule width 1pt}}
\Xhline{2\arrayrulewidth}
$n$  &   \multicolumn{4}{c@{\quad}!{\vrule width 1pt}}{~4} \\[0.2mm]
\Xhline{3\arrayrulewidth}
 $\theta$  &  \multicolumn{4}{c@{\quad}!{\vrule width 1pt}}{~0.169 $\div$ 0.195}   \\[0.2mm]
\Xhline{3\arrayrulewidth}
 $\sin^{2}\theta_{12}$ &  \multicolumn{4}{c!{\vrule width 1pt}}{0.340 $\div$ 0.342 }   \\[0.2mm]
\hline
 $\sin^{2}\theta_{13}$ &  \multicolumn{4}{c!{\vrule width 1pt}}{0.0188 $\div$ 0.0251} \\[0.2mm]
\hline
 $\sin^{2}\theta_{23}$ & \multicolumn{4}{c!{\vrule width 1pt}}{0.597 $\div$ 0.613} \\[0.2mm]
\Xhline{3\arrayrulewidth}
 $\sin\beta$  & \multicolumn{4}{c@{\quad}!{\vrule width 1pt}}{~0}  \\[0.2mm]
 \hline
 $\sin\delta$  & \multicolumn{4}{c@{\quad}!{\vrule width 1pt}}{~0}  \\[0.2mm]
 \Xhline{3\arrayrulewidth}
  $s$ &  0  &  1 & 2 &  3   \\[0.2mm]
 \hline
 $\sin\alpha$  &  0  &  1 & 0   &  $-1$   \\[0.2mm]
\Xhline{2\arrayrulewidth}
\end{tabular}
\caption{\label{tab:case1}{\small \textbf{Case 1)}.
Results for lepton mixing angles and sines of the CP phases $\alpha$, $\beta$ and $\delta$. The former are within the experimentally preferred $3\, \sigma$ ranges.
The choice $n=4$ is the minimal value of $n$ that can give rise to a non-trivial Majorana phase $\alpha$, if we set $s=1$ or $s=3$. The sign of $\sin\alpha$ depends on $k_1$ and corresponds 
in this table to the choice $k_1=0$ (we also take $k_2=0$), compare to (\ref{alphacase1}). An atmospheric mixing angle in the first 
octant, $0.387\lesssim\sin^2\theta_{23}\lesssim 0.403$, is obtained for $2.95\lesssim\theta\lesssim 2.97$ or if the second and third rows of the PMNS mixing matrix 
are exchanged, while the results for the other two mixing angles and the CP phases are not altered.}}
\end{table}
%
%
%
%
%
\begin{figure}[t!]
\begin{center}
\begin{tabular}{cc}
\includegraphics[width=0.48\textwidth]{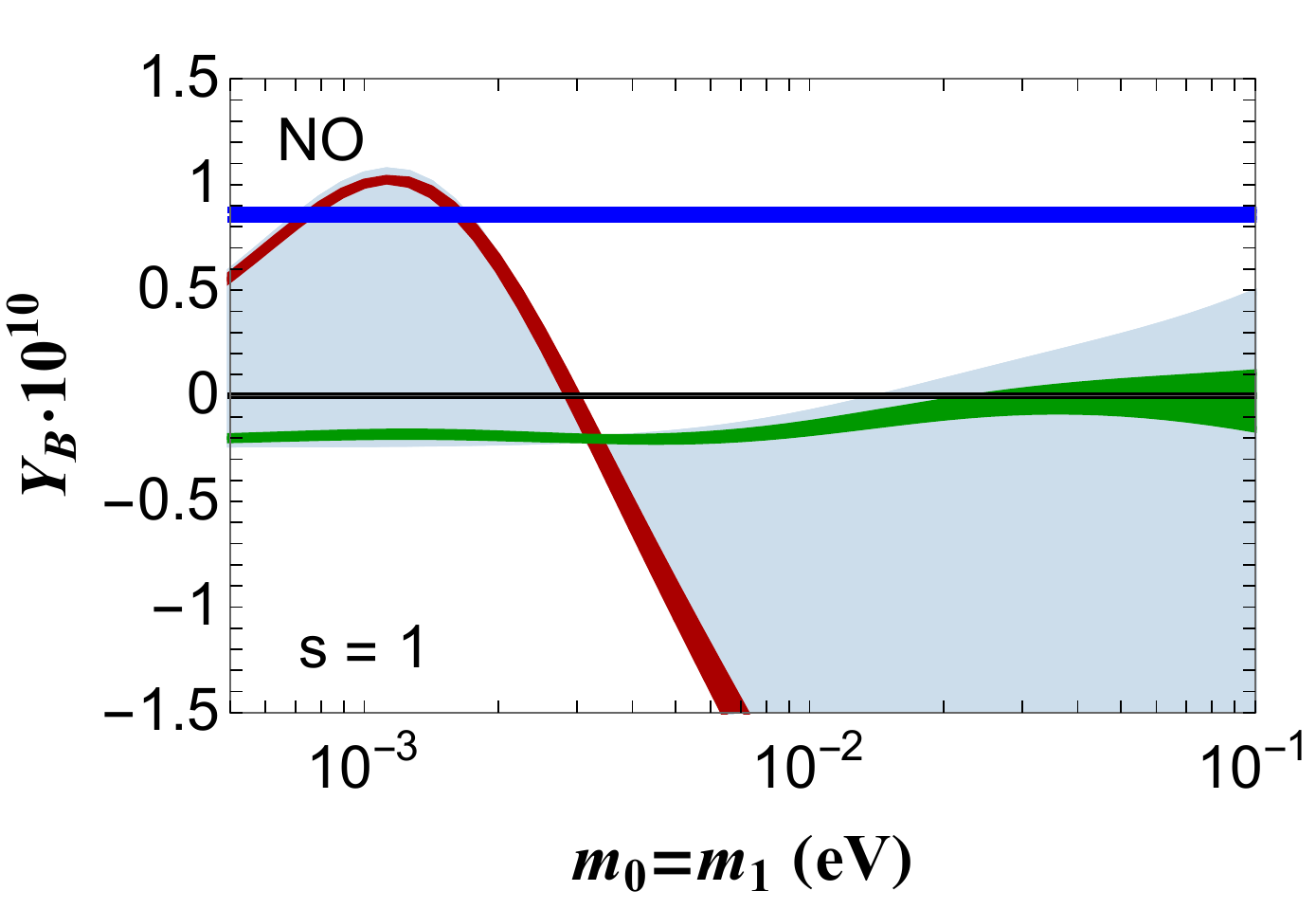} &
\includegraphics[width=0.48\textwidth]{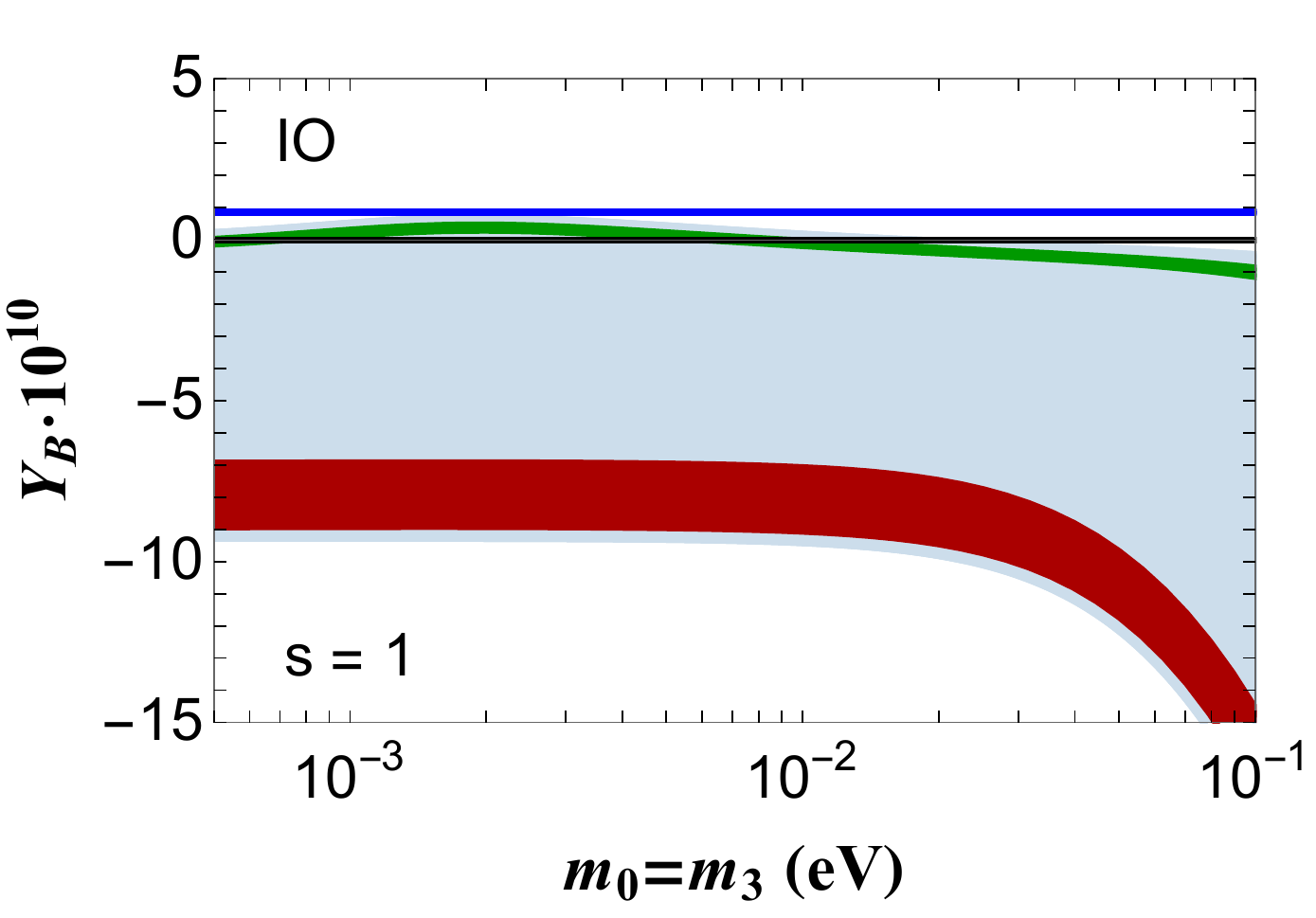} \\
\includegraphics[width=0.48\textwidth]{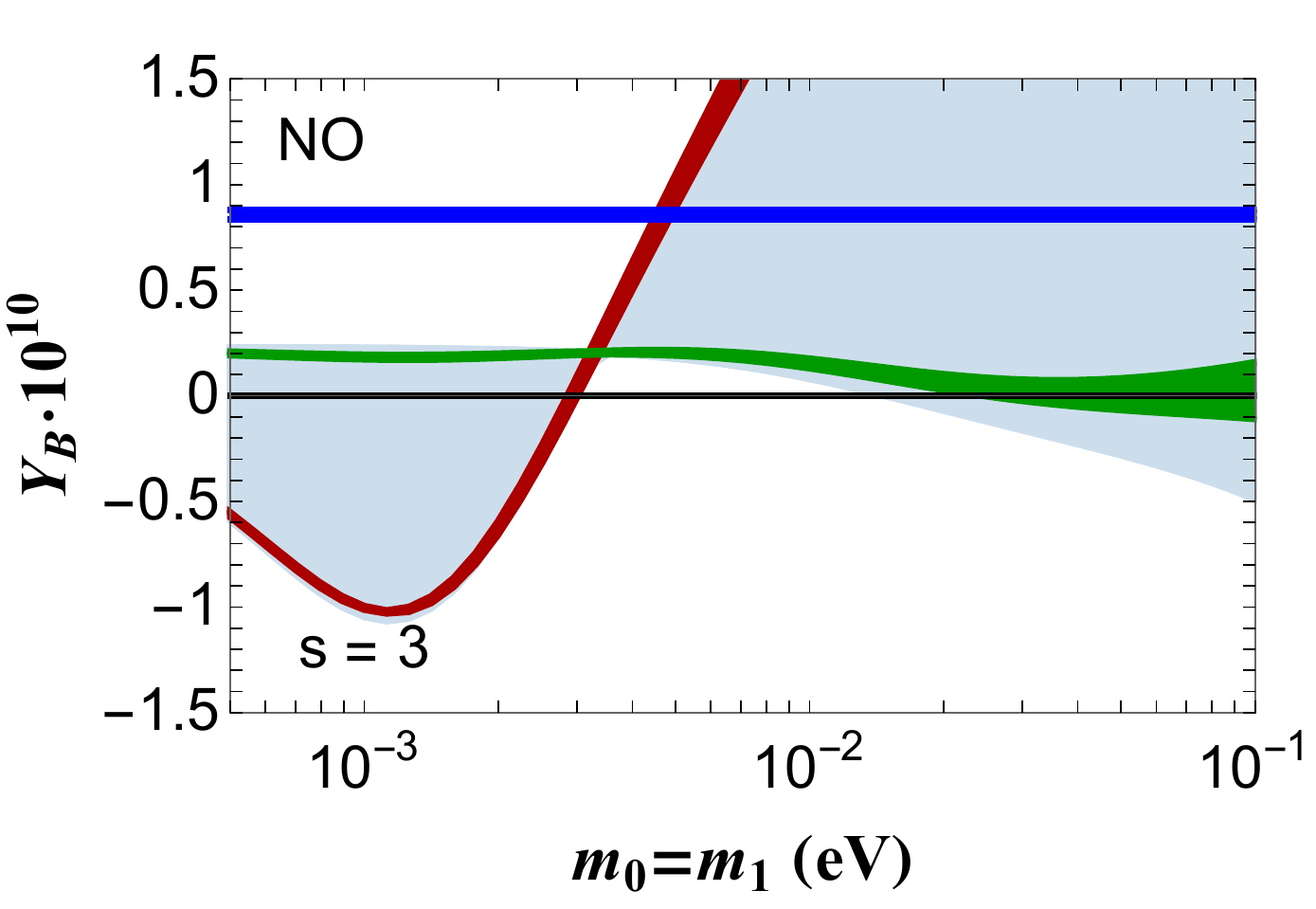} &
\includegraphics[width=0.48\textwidth]{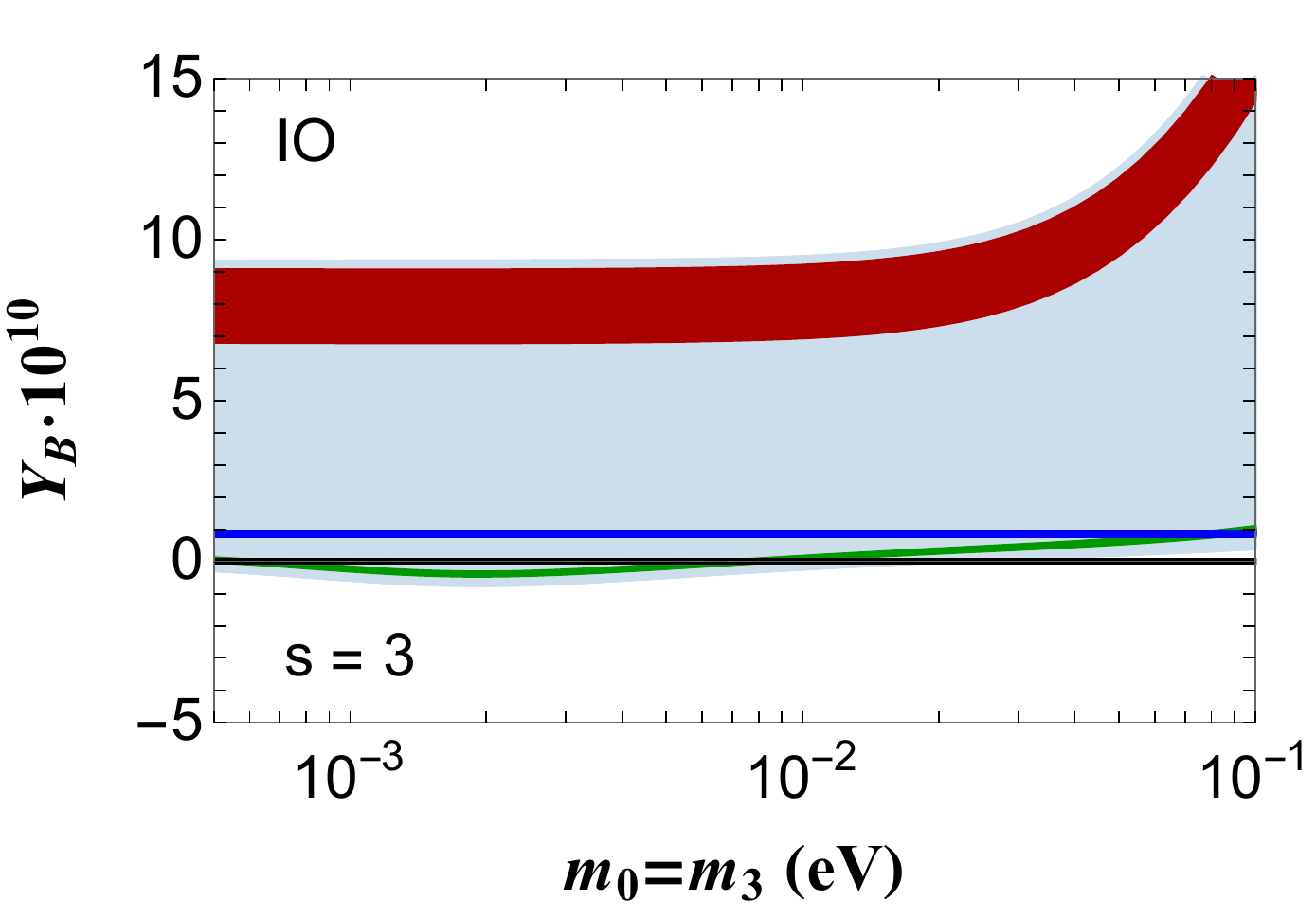}
\end{tabular}
\caption{\small{\textbf{Case 1).} Baryon asymmetry $Y_B$ of the Universe as function of the lightest neutrino mass $m_0$ in the case of a light neutrino mass spectrum with NO (IO) in the left (right) panels. 
The plots at the top (bottom) are realized for $s=1$ ($s=3$) that predicts $\alpha=\pi/2$ ($3\pi/2$). This explains the different sign of $Y_B$ for $s=1$ and for $s=3$.
The other parameters are fixed as in figure \ref{Fig:3}. The horizontal blue band indicates the experimentally preferred $3 \, \sigma$ range of the value of the baryon asymmetry 
$Y_B =(8.65\pm0.27)\times 10^{-11}$ \cite{YBexp}. The differently colored areas correspond to different choices of the parameter $\zeta$ shown in (\ref{inzeta1}-\ref{zetacolor}).}}
\label{Fig:4}
\end{center}
\end{figure}
%
%
First of all, we notice that the CP asymmetry $\epsilon_1$, generated in $N_1$ decays, is suppressed for $\zeta=  \pi/2$ and $\zeta= 3\pi/2$, see green area in the upper panels of figure \ref{Fig:3}, 
since $\epsilon_1$ is proportional to $\cos^2(\theta+\zeta)$ at LO in $\tilde{\kappa}$, see (\ref{eps1case1}), and $\theta$ is small. 
Conversely, the absolute value of $\epsilon_1$ is maximized for $\zeta \approx 0$ and $\zeta \approx\pi$, as can be read off from the red area in the upper panels of figure \ref{Fig:3}. 
Similarly, $\epsilon_3$ is suppressed for $\zeta \approx 0$ and $\zeta \approx\pi$, compare red area in the lower panels of figure \ref{Fig:3}, because it is proportional to $\sin^2(\theta+\zeta)$ at LO 
in $\tilde{\kappa}$, see (\ref{eps3case1}), while this proportionality leads to an enhancement of the CP asymmetry $\epsilon_3$ for $\zeta \approx \pi/2$ and $3 \pi/2$.
The behavior of $\epsilon_2$ is more complex, if light neutrino masses follow NO, since there are two different contributions that can be of similar size, see (\ref{eps2case1}). 
If $m_0$ is small, i.e. $m_0 \lesssim 0.02 \, \mathrm{eV}$, one can see that for the special choices of $\zeta$, see (\ref{inzeta1}), only one of these dominates. For IO light neutrino masses instead $\epsilon_2$ behaves very similar to $\epsilon_1$,
because one of the two contributions clearly dominates.
Regarding the width of the red and green areas in the different plots of figure \ref{Fig:3} we see that these are mostly determined by the variation 
of the solar and atmospheric mass squared
differences in their experimentally preferred $3 \, \sigma$ ranges, like in the case of the loop function $f (x)$, compare to figure \ref{Fig:1}. In the particular case of the CP asymmetry $\epsilon_2$ for NO light neutrino masses
the reason for the broadness of the green band for larger values of $m_0$, see left plot in the middle of figure \ref{Fig:3}, is not this variation, but the fact that there are two different contributions
to $\epsilon_2$, see (\ref{eps2case1}), that are of similar size in this parameter space.
 In the case of IO this does not happen, since the contribution accompanied by the loop function $f (m_1/m_2)$
  dominates over the other one. The sign of the CP asymmetries is determined by $\sin\alpha$ as well as the behavior of the loop function $f(x)$. For this reason, $\epsilon_1$ can only
be positive for small values of $m_0 \lesssim 4 \times 10^{-3} \, \mathrm{eV}$, if light neutrino masses follow NO, while $\epsilon_2$ can only be positive for $m_0$ larger than $0.02 \, \mathrm{eV}$.
In the case of $\epsilon_3$ and NO, the loop function instead does not change sign and thus $\epsilon_3$ is always positive for this choice of $s$. Similarly, the CP asymmetries $\epsilon_{1,2}$
are always negative and can reach large values for light neutrino masses with IO (indeed, the plots for $\epsilon_1$ and $\epsilon_2$ are almost identical in this case), as the relevant 
(dominant) loop function is always positive (negative) for $\epsilon_{1 (2)}$. A suppression of $\epsilon_{1,2}$ occurs in this case for 
$\zeta\approx\pi/2,~3\pi/2$, see (\ref{eps1case1}-\ref{eps3case1}) and green areas in the upper- and middle-right panels of figure~\ref{Fig:3}. In contrast, $\epsilon_3$
is in this case positive for small $m_0 \lesssim 3 \times 10^{-2} \, \mathrm{eV}$ and negative for larger values of $m_0$. It is, however, always strongly suppressed with respect to the other
two CP asymmetries and thus irrelevant for the computation of $Y_B$ (unless $\zeta\approx\pi/2,~3\pi/2$), since such a strong suppression cannot be compensated by the efficiency factor $\eta (m_i)$, see plot on the right of figure \ref{Fig:1b}.

The described behavior of the CP asymmetries $\epsilon_i$ for NO and IO, respectively, explains the results obtained for the baryon asymmetry $Y_B$, shown in the upper panels of figure \ref{Fig:4}.
For NO, upper-left panel of figure \ref{Fig:4}, only values of $m_0 \approx 10^{-3} \, \mathrm{eV}$ allow for the correct sign as well as size of $Y_B$ 
for most of the values of $\zeta$. Indeed, in about $80\%$ of the parameter space $Y_B>0$ is reproduced.\footnote{We compute this percentage as follows: 
we take the interval of $m_0$ with $m_0 \lesssim 3 \times 10^{-3} \, \mathrm{eV}$ and in which for some value of $\zeta$ the size of $Y_B$ can be correctly achieved
and calculate the size of the light-blue area $I_{+ (-)}$ with $Y_B > (<) 0$ for this interval of $m_0$. The ratio $I_+/(I_++I_-)$ then corresponds to the percentage of parameter space
in which $Y_B$ is positive -- in this case about $80\%$. We do so, since by changing the size of $\tilde\kappa$, we can in principle for all choices of $\zeta$ that lead to $Y_B >0$ also achieve
the correct size of $Y_B$, see (\ref{YBnumber}). We use this measure also in the other cases to estimate the predictive power of our approach regarding the sign of $Y_B$.} 
Moreover, $Y_B$ within the experimentally preferred $3 \, \sigma$ range, indicated by the blue band in figure \ref{Fig:4}, is achieved for $6.8\times10^{-4} \, \text{eV}\lesssim m_0 
\lesssim 1.7 \times 10^{-3} \, \text{eV}$. The contribution $Y_{B 1}$,  arising from $N_1$ decays, dominates $Y_B$ and is maximized for this particular range of $m_0$, since the efficiency factor $\eta (m_1)$ takes the largest value, see figure \ref{Fig:1b}.
For larger values of $m_0$, $m_0 \gtrsim 0.02 \, \mathrm{eV}$ instead $Y_B$ can take positive values, but only for a rather small portion of the choices of $\zeta$. In addition,
its size is slightly too small compared to the experimentally preferred $3 \, \sigma$ range for $Y_B$. This can be cured by choosing a value for $\tilde\kappa$
slightly larger than $\tilde\kappa=4 \times 10^{-3}$, see (\ref{kappafix}). Since $Y_B$ is proportional at LO to $\tilde\kappa^2$, an increase of $\tilde\kappa$ by less than a factor of two 
 is in this case sufficient. The situation strongly differs in the case of a light neutrino mass
spectrum with IO, since there $Y_B$ is mostly driven by the CP asymmetries $\epsilon_1$ and $\epsilon_2$ that are both negative for $s=1$, see panels on the right of figure \ref{Fig:3}. Only
for $m_0 \approx 3 \times 10^{-3} \, \mathrm{eV}$ also positive values can be achieved. However, this occurs only in a small portion of the parameter space for $\zeta$ (less than $10\%$). Moreover, the size
of $Y_B$ is below the experimentally preferred lower $3 \, \sigma$ limit (the maximum value achieved is about $6.5 \times 10^{-11}$). Thus, we conclude that for $s=1$ and $k_1=k_2=0$ only for NO light neutrino masses and $m_0 \approx 10^{-3} \, \mathrm{eV}$
we can consider the sign as well as the size of $Y_B$ to be correctly explained. 

The situation is reverse in the case $s=3$, since in this case the sign of $\sin\alpha$ is opposite. This is 
expected, since $s=1$ is related to the choice $s=3$ via the transformation where $s$ is replaced by $n-s$, which changes the sign of all CP asymmetries, see (\ref{case1relepsisns}).
Thus, for $m_0 \gtrsim 3 \times 10^{-3} \, \mathrm{eV}$ and NO of the light neutrino masses the sign of $Y_B$ can be correctly achieved for most of the choices of $\zeta$.
Even more pronounced is the situation for IO, because for $s=3$ the sign of $Y_B$ is for (almost) all values of $\zeta$ correctly accommodated in the whole range of $m_0$ displayed, $5 \times 10^{-4} \, \mathrm{eV} \lesssim m_0 \lesssim 0.1 \, \mathrm{eV}$. Also the size of $Y_B$ can be correctly reproduced for some choice of $\zeta$ for both neutrino mass orderings.

%
\begin{figure}[t!]
    \begin{subfigure}[b]{0.48\textwidth}
      \includegraphics[width=\linewidth]{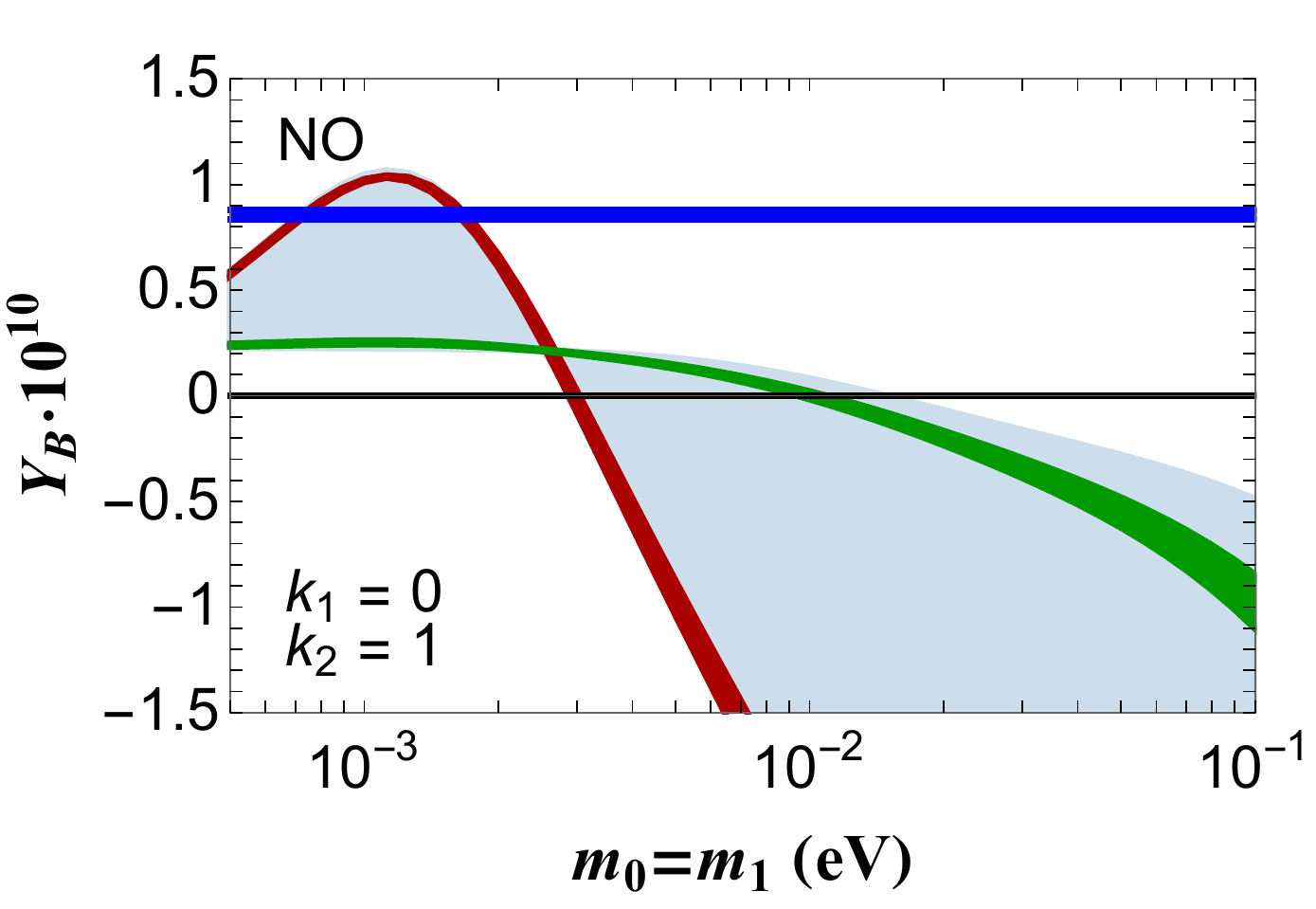}
      \caption*{}
    \end{subfigure}
    \begin{subfigure}[b]{0.48\textwidth}
     \includegraphics[width=\linewidth]{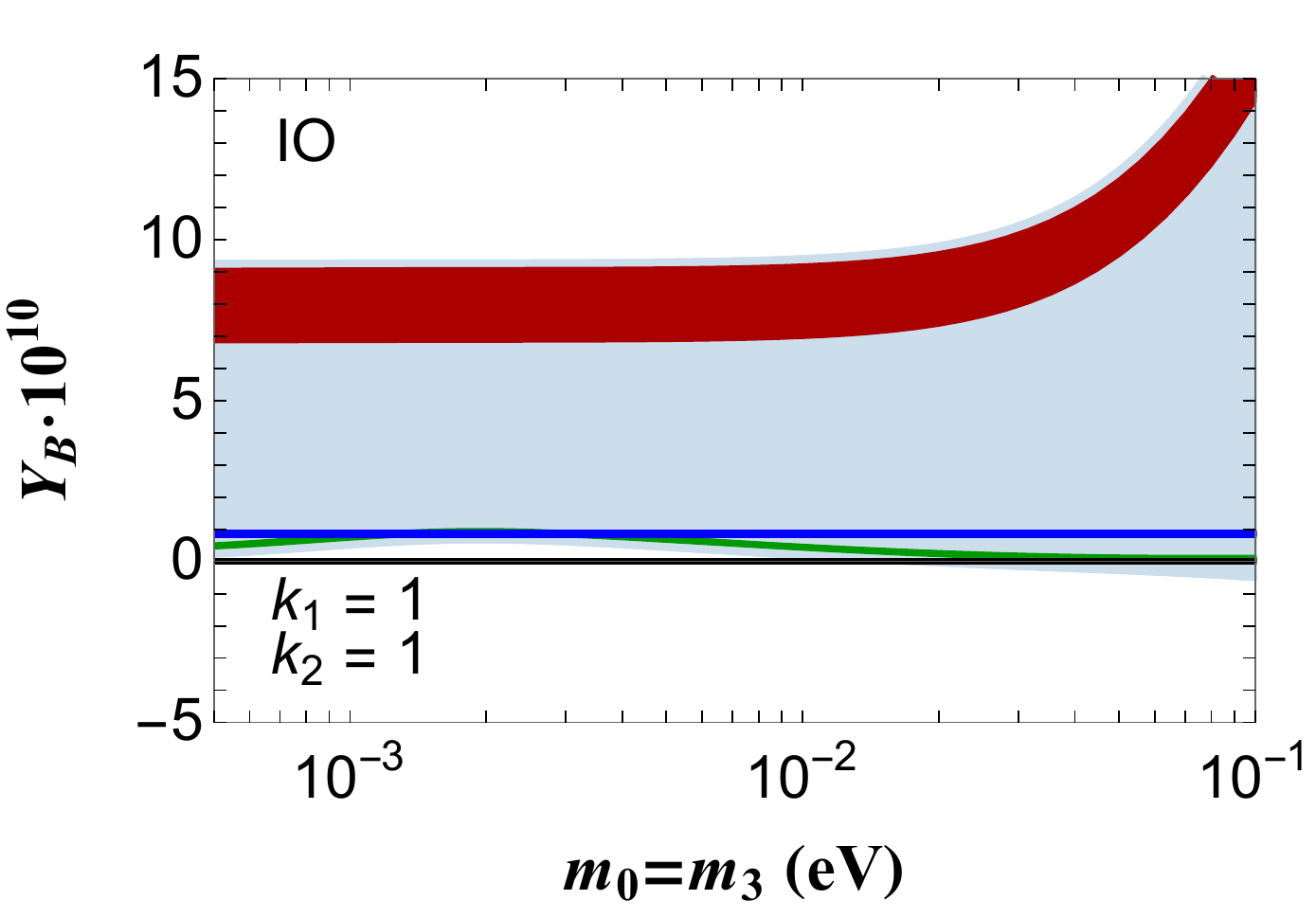}
      \caption*{}
    \end{subfigure}
    \begin{subfigure}[b]{0.48\textwidth}
      \includegraphics[width=\linewidth]{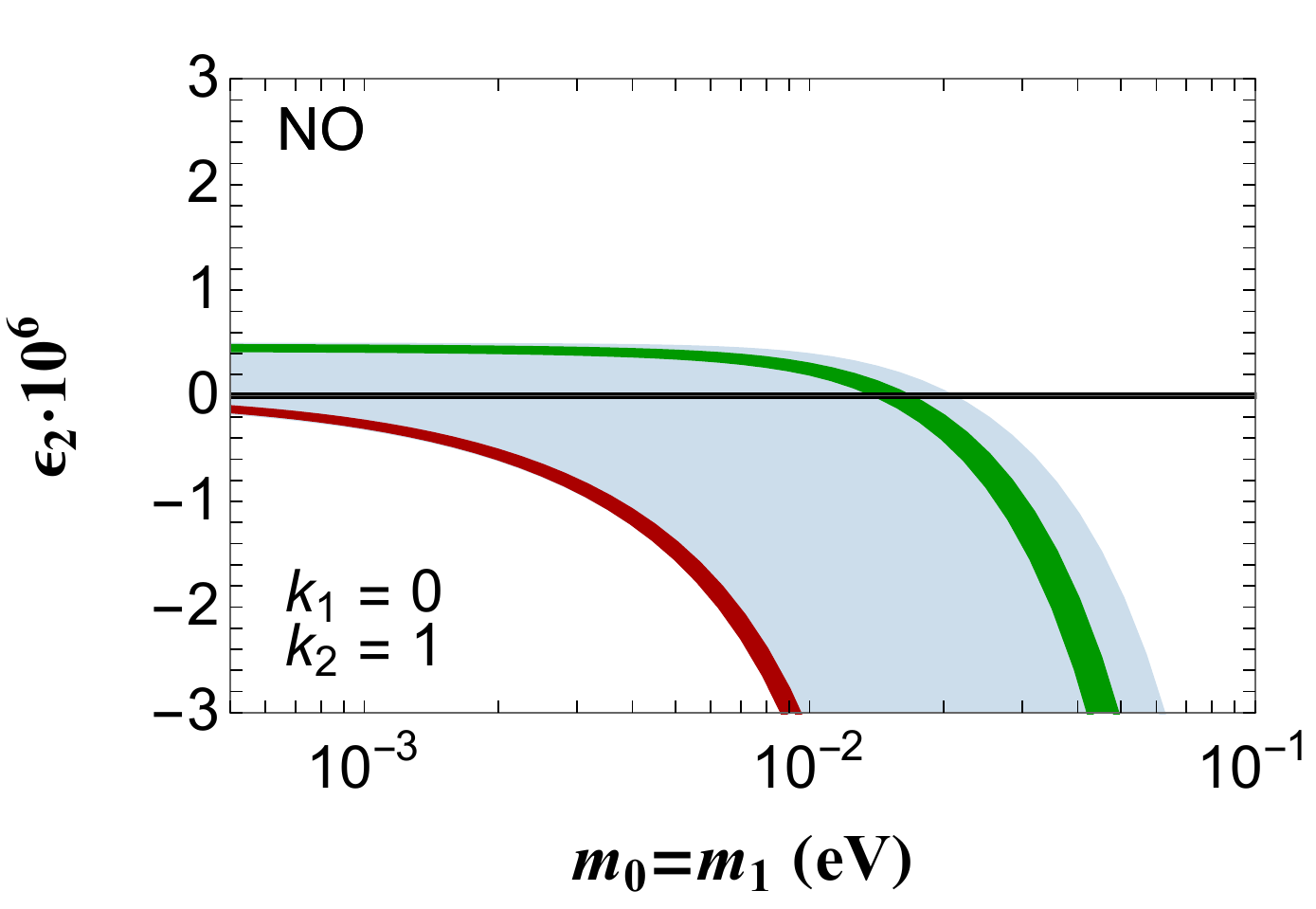}
      \caption*{}
    \end{subfigure}
        \hfill
    \begin{minipage}[b]{0.44\textwidth}
      \caption{\small{ \textbf{Case 1).} Upper panels: $Y_B$ versus $m_0$ for NO (left) and IO  (right) light neutrino mass spectrum.
      Lower panel: CP asymmetry $\epsilon_2$ versus $m_0=m_1$ in the case of NO. The group theoretical parameters are taken to be $n=4$ and $s=1$ and different choices of $k_1$ and $k_2$ are used. 
      All other parameters are the same as in figure~\ref{Fig:4}.
      The red, green and light-blue areas correspond to the values of $\zeta$ reported in (\ref{inzeta1}-\ref{zetacolor}).\vspace{1.55cm}}} 
      \label{Fig:5}
    \end{minipage}
  \end{figure}
%
In figure \ref{Fig:5}, we also consider
different choices of the parameters $k_1$ and $k_2$. In particular, we show in the plots on the left the effect of $k_2=1$. Most importantly, we see that for small values of $m_0$, 
$m_0 \lesssim 2.8 \times 10^{-3} \, \mathrm{eV}$ and NO light neutrino masses $Y_B$ is positive for all admitted values of $\zeta$, see (\ref{inzeta1}-\ref{inzeta2}). This leads to a unique prediction
of its sign compared to the case in which $k_2=0$ and thus the sign is correctly determined for all values of $\zeta$. It can be traced back to the change in the CP asymmetry
$\epsilon_2$, compare left panel in the middle of figure \ref{Fig:3} with the lower plot on the left of figure \ref{Fig:5}. At the same time, $Y_B$ is no longer positive for $m_0 \gtrsim 0.014 \, \mathrm{eV}$.
On the right of figure \ref{Fig:5} we see that for $k_1=k_2=1$ and IO light neutrino masses $Y_B$ behaves very similar for $s=1$ as for $s=3$ and $k_1=k_2=0$, see figure \ref{Fig:4}, 
showing clearly the importance of the choice of $k_{1,2}$. In particular, the choice $k_1=1$ is needed for achieving the correct sign of the baryon asymmetry. This value of $k_1$
and $s=1$ entail $\alpha=3 \pi/2$ which is also obtained for $k_1=0$ and $s=3$, compare (\ref{alphacase1}).

The approximate formulae given in subsection \ref{sec331} describe in most of the parameter space the results obtained with the 
formulae not expanded in $\tilde\kappa$ quite well. To quantify this statement better we first derive the NLO terms in $\tilde\kappa$
contributing to the three CP asymmetries $\epsilon_i$. The latter, expanded up to NLO in $\tilde\kappa$, can be written in the following compact form\footnote{The appearance
of $\sec\left(\theta+\zeta\right)$ and $\csc \left(\theta+\zeta\right)$ does not indicate any singular behavior of $\epsilon_i^{\text{NLO}}$, but arises only because we have extracted $\epsilon_i^{\text{LO}}$.
Indeed, the latter contain appropriate factors of $\cos\left(\theta+\zeta\right)$ and $\sin \left(\theta+\zeta\right)$ to cancel the former factors.}
\begin{eqnarray}
\epsilon_1^{\text{NLO}} & =&  \epsilon_1^{\text{LO}} \left[1 +\frac{\tilde\kappa}{\sqrt{3}\,y_0}\sec\left(\theta+\zeta\right)\left( -\cos3\,\theta+\frac{w^2}{z^2} \cos\left(\theta-2\,\psi\right)\right) \right]\,,
\\
\epsilon_2^{\text{NLO}} &= & -\epsilon_1^{\text{LO}} \, \left[1 +\frac{\tilde\kappa}{\sqrt{3}\,y_0}\sec\left(\theta+\zeta\right)\left( \cos\left(2\,\zeta-\theta\right)+\frac{w^2}{z^2} \cos\left(\theta-2\,\psi\right)\right) \right] \, \frac{f\left(m_2/m_1\right)}{f\left(m_1/m_2\right)}
\nonumber
\\
&& -  \epsilon_3^{\text{LO}} \,    \left[1 +\frac{\tilde\kappa}{\sqrt{3}\,y_0}\csc \left(\theta+\zeta\right)\left( -\sin\left(2\,\zeta-\theta\right)+\frac{w^2}{z^2} \sin\left(\theta-2\,\psi\right)\right) \right]
	                                \frac{f\left(m_2/m_3\right)}{f\left(m_3/m_2\right)} \,,
\\
\epsilon_3^{\text{NLO}} & = & \epsilon_3^{\text{LO}} \left[1 +\frac{\tilde\kappa}{\sqrt{3}\,y_0}\csc \left(\theta+\zeta\right)\left( \sin3\,\theta+\frac{w^2}{z^2} \sin\left(\theta-2\,\psi\right)\right) \right]
\end{eqnarray}
where we have defined 
\begin{equation}
\label{Imz12}
\mathrm{Im} (z_1) = w \, \cos \psi \;\;\; \mbox{and} \;\;\; \mathrm{Im} (z_2) = w \, \sin \psi
\end{equation}
with $w >0$ being of order one and $0 \leq \psi < 2 \, \pi$, since also the imaginary parts of the two complex parameters $z_{1,2}$ enter at this level. We have, furthermore, made use of the
LO expressions  $\epsilon_i^{\text{LO}}$ of the CP asymmetries that can be found in (\ref{eps1case1}-\ref{eps3case1}). As one can see, the NLO terms are suppressed by $\tilde\kappa/y_0$
relative to the LO term. Leaving aside for the moment the effect of the imaginary parts of $z_{1,2}$, by setting $w=0$, we can check that the NLO terms are small compared to the LO term, if 
\begin{equation}
\label{boundkappa}
\tilde\kappa \; \ll\; 0.07 \;\lesssim\;\sqrt{3}\, y_0  \, 
\end{equation}
assuming as allowed range of $y_0$ the one given in (\ref{rangey0}). With our choice of $\tilde\kappa$ in (\ref{kappafix}) this bound is clearly fulfilled. Consequently, we find in our numerical
analysis that for all admitted values of $\zeta$ the relative difference between the LO approximations in (\ref{eps1case1}-\ref{eps3case1}) and the results obtained with the formulae not expanded in 
$\tilde\kappa$ is less than $10\%$. If we consider the impact of the imaginary parts of $z_{1,2}$, we see that for $w/z \lesssim 1$ the LO approximations still describe the unexpanded result
very well. Indeed, in more than $90\%$ of the parameter space of $\zeta$, $\psi$ and $w$ with $w/z \lesssim 1$ the relative difference between the two is less than $10\%$. Only if $w$ is
taken to be larger than $z$, we find a considerable decrease in the goodness of the LO approximations, i.e. a relative deviation of the latter from the unexpanded results less than $10\%$ is only
given in about $60\% \div 70\%$ of the parameter space. Similar results are also obtained in the numerical analyses of case 2) and case 3 b.1) [as well as case 3 a)]. This shows that 
the simple LO expressions are rather powerful in describing the results for the CP asymmetries $\epsilon_i$ adequately.

\subsubsection*{Leptogenesis in case 2)}
\label{sec342}

As example of case 2), we perform a numerical calculation of $Y_B$ for the group theoretical parameters
\begin{equation}
\label{case2leptonum}
 n=10 \;\;\; \mbox{and} \;\;\; u=4 \; .
\end{equation}
It is important to note that in this case the matrix $U_R$ (and thus also the PMNS mixing matrix) is given by the first matrix in (\ref{shiftPcase2}), i.e.
in order to achieve compatibility with the measured values of the lepton mixing angles we have to apply a cyclic permutation to the rows of the original matrix
in (\ref{URcase2}). As a consequence, when using the formulae for the mixing angles, found in \cite{HMM}, the CP invariants given in (\ref{CPinvcase2}-\ref{I3case2}) and 
the LO approximations for the CP asymmetries $\epsilon_i$ in (\ref{e1case2}-\ref{e3case2}), we have to apply the set of transformations in (\ref{shift1case2}).
For completeness, we mention the results of the lepton mixing angles, the CP phases $\alpha$, $\beta$ and $\delta$, using $1.40\lesssim \theta\lesssim 1.44$, in table~\ref{tab:case2}.
This interval of $\theta$ is also used in our numerical analysis.

The example of case 2) has a much richer phenomenology than the one studied for case 1), since all three CP phases are in general non-trivial and, in addition,
the Majorana phase $\alpha$ assumes three considerably different values for the different choices of the group theoretical parameter $v$.
As shown in table \ref{tab:case2},  the Dirac phase $\delta$ as well as the Majorana phase $\beta$ are predicted to be almost maximal in this case, while
the other Majorana phase $\alpha$ can take three different values (small, almost maximal, intermediate), depending on $v$ which can take five different values
\begin{equation}
v=0, \, 6, \, 12, \, 18, \, 24 \; ,
\end{equation}
see table \ref{tab:case2}. These values of $v$ are admitted for $n=10$ and $u=4$. 
Let us mention again that the sign of $\sin\alpha$ depends as well on the choice of the parameter $k_1$ and all values displayed in table \ref{tab:case2} refer to $k_1=0$, compare (\ref{sinaapproxcase2}).
In the same vein, the sign of $\sin\beta$ given in this table holds for $k_2=0$ and changes for $k_2=1$, see (\ref{CPinvcase2}).
%
\begin{table}[t!]
\centering
\catcode`?=\active \def?{\hphantom{0}}
\begin{tabular}{!{\vrule width 1pt}@{\quad}>{\rule[-2mm]{0pt}{6mm}}l@{\quad}!{\vrule width 2pt}@{\quad\quad}c@{\quad\quad}|@{\quad\quad}c@{\quad\quad}|@{\quad\quad}c@{\quad\quad}!{\vrule width 1pt}}
\Xhline{2\arrayrulewidth}
$n$  &   \multicolumn{3}{c@{\quad}!{\vrule width 1pt}}{~10} \\[0.2mm]
\Xhline{3\arrayrulewidth}
$u$  &   \multicolumn{3}{c@{\quad}!{\vrule width 1pt}}{~4} \\[0.2mm]
\Xhline{3\arrayrulewidth}
 $\theta$  &  \multicolumn{3}{c@{\quad}!{\vrule width 1pt}}{1.40 $\div$ 1.44}   \\[0.2mm]
\Xhline{3\arrayrulewidth}
 $\sin^{2}\theta_{12}$ &  \multicolumn{3}{c!{\vrule width 1pt}}{0.340 $\div$ 0.342 }   \\[0.2mm]
\hline
 $\sin^{2}\theta_{13}$ &  \multicolumn{3}{c!{\vrule width 1pt}}{0.0187 $\div$ 0.0250} \\[0.2mm]
\hline
 $\sin^{2}\theta_{23}$ & \multicolumn{3}{c!{\vrule width 1pt}}{0.558 $\div$ 0.559} \\[0.2mm]
\Xhline{3\arrayrulewidth}
 $\sin\beta$  & \multicolumn{3}{c@{\quad}!{\vrule width 1pt}}{$0.83 \div 0.94 $}  \\[0.2mm]
 \hline
 $\sin\delta$  & \multicolumn{3}{c@{\quad}!{\vrule width 1pt}}{$-0.86 \div -0.80 $}  \\[0.2mm]
 \Xhline{3\arrayrulewidth}
  $v$ &  0  &  ~6 , 24 & 12 , 18   \\[0.2mm]
\hline
 $\sin\alpha$  &   $-0.035 \div -0.028 $   &  0.94 $\div$ 0.96 & $-0.62 \div -0.56$    \\[0.2mm]
\Xhline{2\arrayrulewidth}
\end{tabular}
\caption{\label{tab:case2}{\small \textbf{Case 2)}. Results for lepton mixing angles and CP phases $\alpha$, $\beta$ and $\delta$ for $n=10$, $u=4$.
All are compatible at the $3 \, \sigma$ level or better with the experimental data. The values of $\sin\alpha$, $\sin\beta$ and $\sin\delta$ refer to the choice $k_1=k_2=0$.
The parameter $v$ takes five different values: $v=0$, 6, 12, 18 and 24. 
The interval of $\sin\alpha$ corresponding to $v=18$ ($v=24$) is the same as for $v=12$ ($v=6$). 
An atmospheric mixing angle in the first octant,  $0.441\lesssim \sin\theta_{23} \lesssim 0.442$, is obtained by a permutation of the second and third rows of the PMNS mixing matrix \cite{HMM}. 
In this case $\sin\delta$ changes sign, while $\sin\alpha$ and $\sin\beta$ are invariant.
For values of the parameter $\theta$ in the second admitted interval,  $1.70 \lesssim \theta \lesssim 1.74$, compare \cite{HMM}, the mixing angles and $\sin\alpha$ are the same, but 
$\sin\beta$ and $\sin\delta$ change sign.}} 
\end{table}
%

\begin{figure}[t!]
    \begin{subfigure}[b]{0.5\textwidth}
      \includegraphics[width=\linewidth]{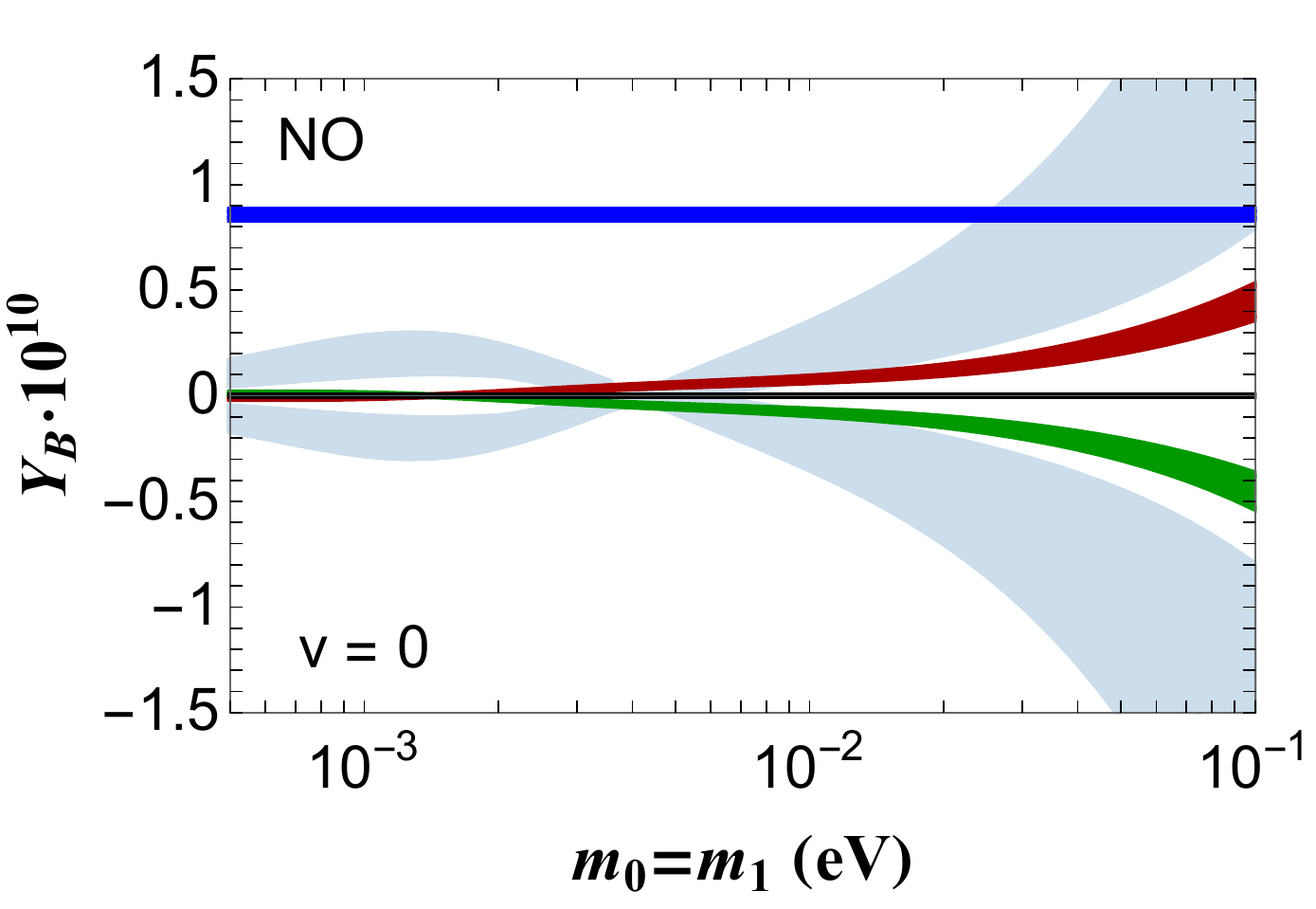}
      \caption*{}
    \end{subfigure}
    \begin{subfigure}[b]{0.5\textwidth}
     \includegraphics[width=\linewidth]{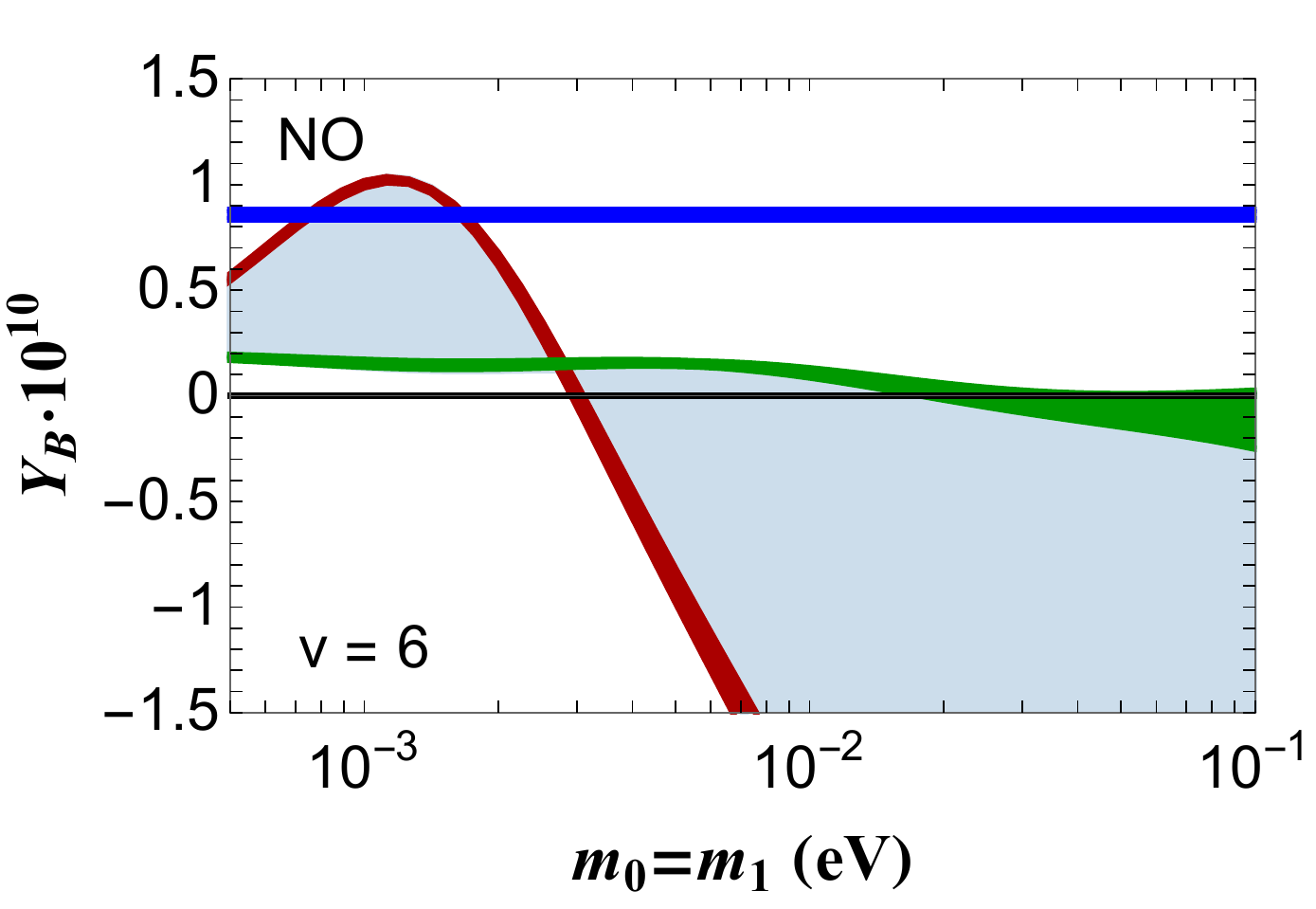}
      \caption*{}
    \end{subfigure}
    \begin{subfigure}[b]{0.5\textwidth}
      \includegraphics[width=\linewidth]{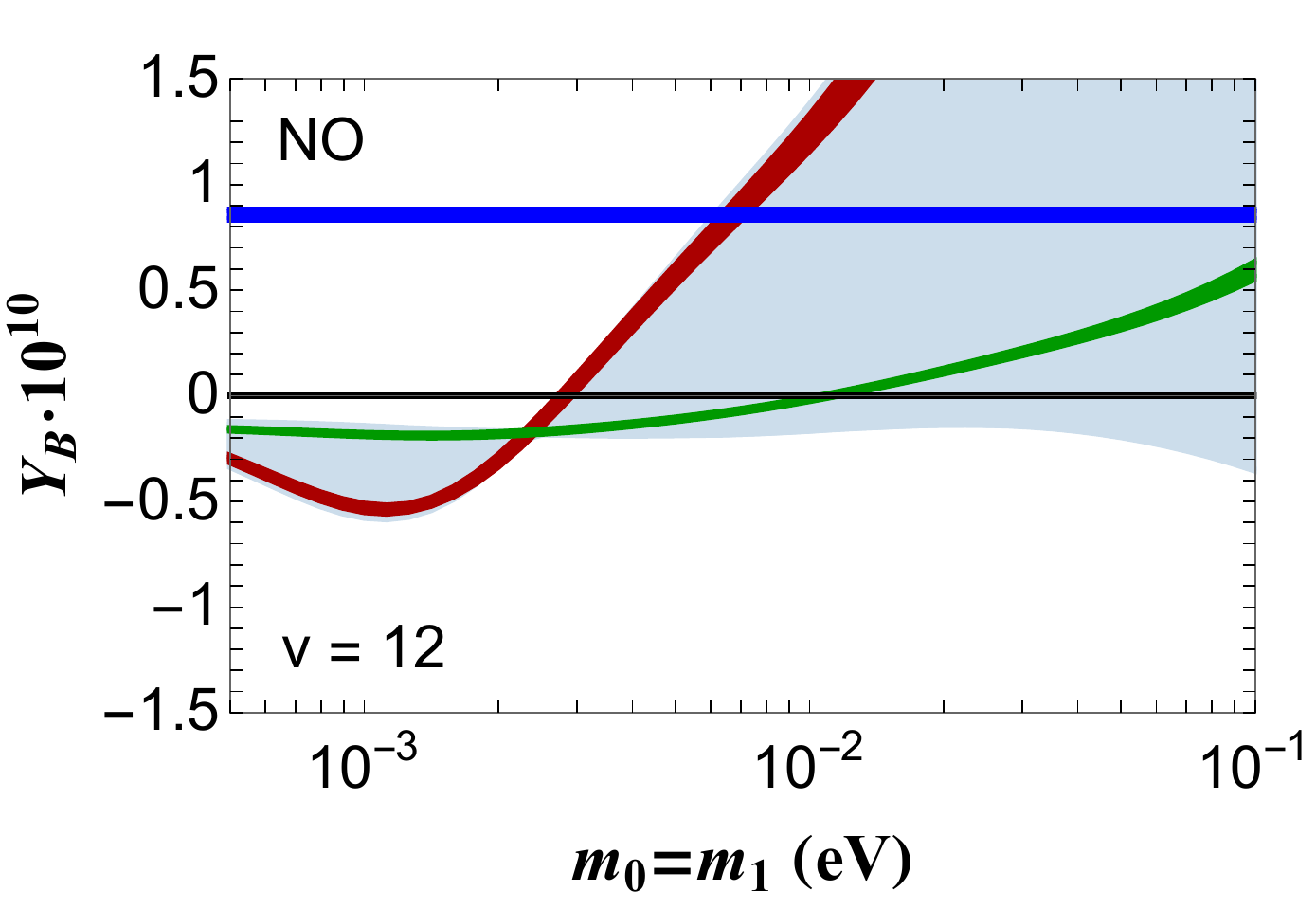}
      \caption*{}
    \end{subfigure}
        \hfill
    \begin{minipage}[b]{0.44\textwidth}
      \caption{\small {\textbf{Case 2).} $Y_B$ versus $m_0$ for $n=10$ and $u=4$ and 
      three choices of   $v$. Results for $v=18$ ($v=24$) are equal to the plot with $v=12$ ($v=6$). 
      In all the plots $1.40\lesssim \theta \lesssim 1.44$ and
      the neutrino mass squared differences are varied within their experimentally preferred $3 \, \sigma$ ranges. The horizontal blue band represents the measured value of $Y_B$
      at the experimentally preferred $3 \, \sigma$ level.
      The red, green and light-blue areas correspond to the choices of $\zeta$ reported in (\ref{inzeta1}-\ref{zetacolor}).\vspace{1.1cm}}} 
      \label{Fig:6}
    \end{minipage}
  \end{figure}

In figure~\ref{Fig:6} the variety of results for $Y_B$, its sign and its size, are shown for three different choices of the group theoretical parameter $v$.
As mentioned, in case 2) we only display plots for a light neutrino mass spectrum with NO and always set $k_1=k_2=0$.
The smallness of $\sin\alpha$ for $v=0$ entails a strong suppression of the otherwise dominant contribution to the CP asymmetries. A consequence
of this suppression is that the sign of $Y_B$ cannot be predicted in this case, but depends crucially on $\zeta$, see (\ref{epsisignundetcase2}) with $\phi_u$ being
appropriately replaced according to (\ref{shift1case2}). In this case, as can be checked by explicit computation,
 the Dirac phase $\delta$ provides the dominant source of CP violation in leptogenesis, yielding $|Y_B|\gtrsim 5\times10^{-11}$ for $m_0\gtrsim 0.02$ eV (see the light-blue area in the upper-left panel of figure~\ref{Fig:6}).
As in the numerical example of case 1) discussed before,
the red and green bands
represent the special choices
$\zeta=0, \, \pi$ and $\zeta= \pi/2, \, 3 \pi/2$, respectively. These confirm our analytic findings, see (\ref{epsisignundetcase2}),
that a shift in $\zeta$ by $\pi/2$ changes the sign of all the CP asymmetries and therefore of $Y_B$. Notice that 
the red and green bands do not need to overlap with the light-blue area, since the intervals of $\zeta$ in (\ref{inzeta2}) used to obtain the latter area
do not contain the special values $\zeta=0,\, \pi/2, \, \pi, \, 3\pi/2$. Indeed, for values of $m_0 \lesssim 3 \times 10^{-3} \, \mathrm{eV}$ not arbitrarily small values of $Y_B$
can be achieved. The minimum possible values of $|Y_B|$ are obtained for $\zeta \approx 0.75$ and $\zeta \approx 2.32$.
The vanishing of $Y_B$ for $m_0 \approx 3 \times 10^{-3} \, \mathrm{eV}$ is due to the vanishing of the loop function $f (m_1/m_2)$, see upper-left panel of figure \ref{Fig:1}.
Although for large enough $m_0$, $m_0 \gtrsim 2.7 \times 10^{-2} \, \mathrm{eV}$, $Y_B$ sufficiently large can be obtained, we do not consider this case
as a successful one, since the value of $Y_B$ is equally likely positive and negative.
 In contrast, for $v=6$ the sign of $Y_B$ can be univocally predicted for $m_0 \lesssim 3 \times 10^{-3} \, \mathrm{eV}$ and is in accordance with the experimental
observations. The dominant contribution to the baryon asymmetry (with the correct sign) for NO is produced by the decays of the heaviest RH neutrino, $N_1$, 
as expected from the general results given in subsection~\ref{sec32}.
 For our particular choice of $\tilde\kappa$ in (\ref{kappafix}) we see that in the interval $7\times 10^{-4}  \, \mathrm{eV} \lesssim m_0 \lesssim 2\times 10^{-3}  \, \mathrm{eV}$  the size of $Y_B$ can be correctly achieved.
We also see that for $m_0 \gtrsim 3 \times 10^{-3} \, \mathrm{eV}$ the sign of $Y_B$ is almost always negative. If we consider another choice of the parameters $k_{1,2}$
as in (\ref{k1k2num}), we can make $Y_B$ positive for $m_0 \gtrsim 3 \times 10^{-3} \, \mathrm{eV}$. The situation is reverse for the choice $v=12$, simply because the
sign of $\sin\alpha$ is opposite compared to the one in the case $v=6$. So, for $v=12$ the baryon asymmetry $Y_B$ is always negative for $m_0 \lesssim 2 \times 10^{-3} \, \mathrm{eV}$
and almost always positive for $m_0 \gtrsim 2 \times 10^{-3} \, \mathrm{eV}$. Consequently, only for values of $m_0$ larger than $6 \times 10^{-3} \, \mathrm{eV}$ also the
correct size of $Y_B$ can be obtained for a certain choice of $\zeta$, if we keep $\tilde\kappa$ fixed to the value in (\ref{kappafix}). In the range $6 \times 10^{-3} \, \mathrm{eV} \lesssim m_0
\lesssim 0.1 \, \mathrm{eV}$ positive $Y_B$ is obtained for more than $90\%$ of the choices of $\zeta$.
Again, like in the case $v=6$ the sign
of $Y_B$ can be changed by changing the values of the parameters $k_{1,2}$. Comparing the results for $v=6$ and $v=12$ we note in addition that
the size of $|Y_B|$ is a bit smaller for $v=12$ than for $v=6$, because $|\sin\alpha|$ is smaller for $v=12$ than for $v=6$, see table \ref{tab:case2}.

%

Concerning the IO light neutrino mass spectrum,  the results are very similar to the ones discussed in  case 1), that is
the dominant contribution to $Y_B$ comes from  $N_1$ and $N_2$ decays, which can generate large CP asymmetries 
of order few times $10^{-5}$ and with the correct sign. For this reason we do not show the corresponding plots of $Y_B$.
We only remark that, as for the NO light neutrino mass spectrum, in the case $v=0$, the sign of $Y_B$ is not fixed, see (\ref{epsisignundetcase2}), while
for other choices of $v$ the sign can be determined.

\subsubsection*{Leptogenesis in case 3)}
\label{sec343}

A representative example for case 3) is given by
\begin{equation} 
n=8 \;\;\; \mbox{and} \;\;\; m=4 
\end{equation}
which leads to good agreement with the experimental data on lepton mixing angles, if we consider case 3 b.1), as has been shown in \cite{HMM}.
The discussion of this case has two advantages: the value of the index $n$ of the group $\Delta (6 \, n^2)$ is still moderately small and thus the 
group itself is not too large (it has 384 elements) and, at the same time, several values of the group theoretical parameter $s$, characterizing the CP
transformation $X$, are admitted by the requirement to accommodate all three lepton mixing angles at the $3 \, \sigma$ level or better, so that different types of 
results for leptogenesis are achieved. Concretely, the viable choices of $s$ are \cite{HMM}
\begin{equation}   
s=1, \, 2, \, 4
\end{equation}
as well as the values of $s$ with $s>n/2=4$ that are related to the mentioned ones via $n-s=8-s$, i.e. $s=6$ and $s=7$. These are not discussed independently,
since results for these cases can be deduced from those for $s \leq n/2=4$, see table 6 and explanations in \cite{HMM} regarding the mixing parameters and
(\ref{symmetryepsicase3}) for the CP asymmetries $\epsilon_i$. The results of lepton mixing angles and CP phases for the different choices of $s$ are summarized in
table \ref{tab:case3b1}. As one can see, all values of $s$ give rise to a similar fit to the solar and the reactor mixing angles, while the results obtained for $\theta_{23}$
differ. The value of the latter also depends on the interval chosen for the parameter $\theta$, since replacing $\theta$ by $\pi-\theta$ changes $\sin^2 \theta_{23}$
into $\cos^2 \theta_{23}$. Consequently, for most of the choices of $s$ two intervals of $\theta$ are admitted that lead to different results not only for the atmospheric
mixing angle, but also to a different sign in $\sin\delta$. In our numerical analysis of leptogenesis, we always stick to choices of $\theta$ belonging to the intervals
displayed in table \ref{tab:case3b1} and only briefly comment on results originating from other choices of $\theta$.
A feature of this example is  $m=n/2=4$ which leads to a considerable simplification of the formulae
for the CP phases, see (\ref{phaseseasy3b1}). From there we see immediately that $\sin\alpha$ and $\sin\beta$ must be equal up to a sign (for the particular choice
$k_1=k_2=0$ they are equal) for all choices of $s$ and, moreover, that  for $s=n/2=4$ both Majorana phases are trivial, whereas the Dirac phase is maximal. 
As shown in table \ref{tab:case3b1} in our particular example the Dirac phase is large in all cases and also the Majorana phases are sizable, if they are not trivial. Furthermore,
we note that $\sin\alpha$ and $\sin\beta$ turn out to be negative for $s=1$, while they are positive for $s=2$. All these observations are relevant for the prediction
of the sign of the baryon asymmetry $Y_B$.
%
\begin{table}[t!]
\centering
\catcode`?=\active \def?{\hphantom{0}}
\begin{tabular}{!{\vrule width 1pt}@{\quad}>{\rule[-2mm]{0pt}{6mm}}l@{\quad}!{\vrule width 2pt}@{\quad\quad}c@{\quad\quad}|@{\quad\quad}c@{\quad\quad}|@{\quad\quad}c@{\quad\quad}!{\vrule width 1pt}}
\Xhline{2\arrayrulewidth}
$n$  &   \multicolumn{3}{c@{\quad}!{\vrule width 1pt}}{~8} \\[0.2mm]
\Xhline{3\arrayrulewidth}
$m$  &   \multicolumn{3}{c@{\quad}!{\vrule width 1pt}}{~4} \\[0.2mm]
\Xhline{3\arrayrulewidth}
 $\sin^{2}\theta_{12}$ &  \multicolumn{3}{c!{\vrule width 1pt}}{0.316 $\div$ 0.321}   \\[0.2mm]
\hline
 $\sin^{2}\theta_{13}$ &  \multicolumn{3}{c!{\vrule width 1pt}}{0.0186 $\div$ 0.0250} \\[0.2mm]
\Xhline{3\arrayrulewidth}
  $s$ &  1  &  2 & 4   \\[0.2mm]
\Xhline{3\arrayrulewidth}
 $\theta$  &  1.29 $\div$ 1.33  & 1.81 $\div$ 1.82 & 1.29 $\div$ 1.33  \\[0.2mm]
\Xhline{3\arrayrulewidth}
 $\sin^{2}\theta_{23}$ & 0.573 $\div$ 0.584 & 0.635 $\div$ 0.643 & 1/2  \\[0.2mm]
\Xhline{3\arrayrulewidth}
 $\sin\alpha=\sin\beta$  &  $-1/\sqrt{2}$ & 1  & 0      \\[0.2mm]
 \hline
 $\sin\delta$   &  0.934 $\div$ 0.937  & $-0.738 \div -0.734 $   & $-1$  \\[0.2mm]
\Xhline{2\arrayrulewidth}
\end{tabular}
\caption{\label{tab:case3b1}{\small \textbf{Case 3 b.1)}. Lepton mixing angles and CP phases $\alpha$, $\beta$ and $\delta$ for the choice $n=8$ and $m=4$.
The group theoretical parameter $s$, characterizing the CP transformation $X$, can take different values. Here we display those leading to good agreement (at the 
$3 \, \sigma$ level or better) with the experimental results on the lepton mixing angles and that fulfill $s \leq n/2=4$. Results for $s > n/2=4$ can be easily deduced
from the ones shown in this table using the transformation in (\ref{trafocase3}). The equality of the absolute values of $\sin\alpha$ and $\sin\beta$ originates from $m=n/2=4$, see (\ref{phaseseasy3b1}).
Setting $k_1=k_2=0$ leads to $\sin\alpha=\sin\beta$. Using (\ref{phaseseasy3b1}) we can also check that $s=n/2=4$ entails trivial Majorana phases and a maximal
Dirac phase. We display only one interval for $\theta$. However, in most cases $\theta$ can also be in a second admitted interval. This is related
to the one shown via the transformation of $\theta$ in $\pi-\theta$. In this case, $\sin^2\theta_{23}$ becomes $\cos^2\theta_{23}$ and $\sin\delta$ changes
sign. For further details see \cite{HMM}.}} 
\end{table}
%

\begin{figure}[t!]
    \begin{subfigure}[b]{0.48\textwidth}
      \includegraphics[width=\linewidth]{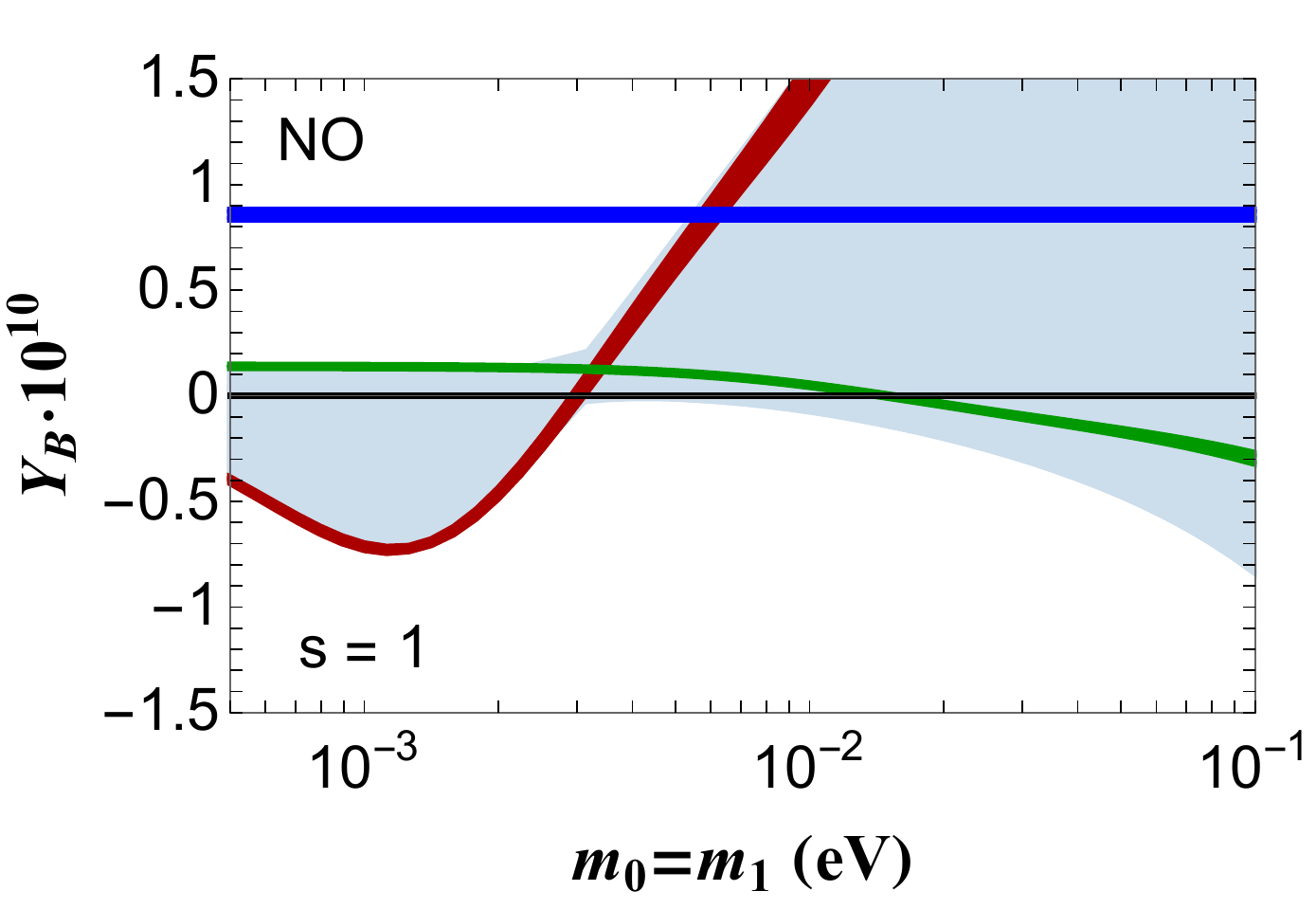}
      \caption*{}
    \end{subfigure}
    \begin{subfigure}[b]{0.48\textwidth}
     \includegraphics[width=\linewidth]{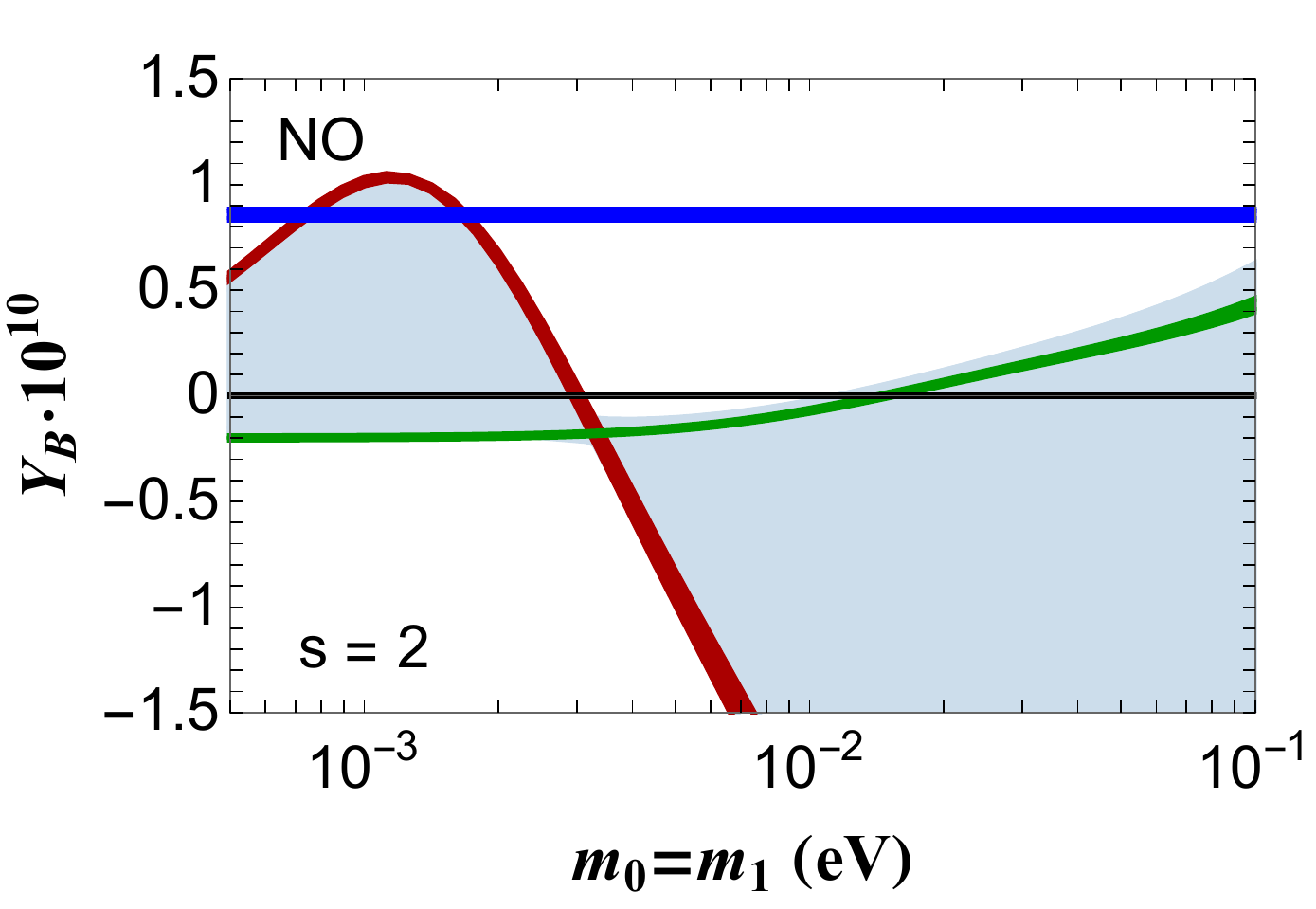}
      \caption*{}
    \end{subfigure}
    \begin{subfigure}[b]{0.48\textwidth}
      \includegraphics[width=\linewidth]{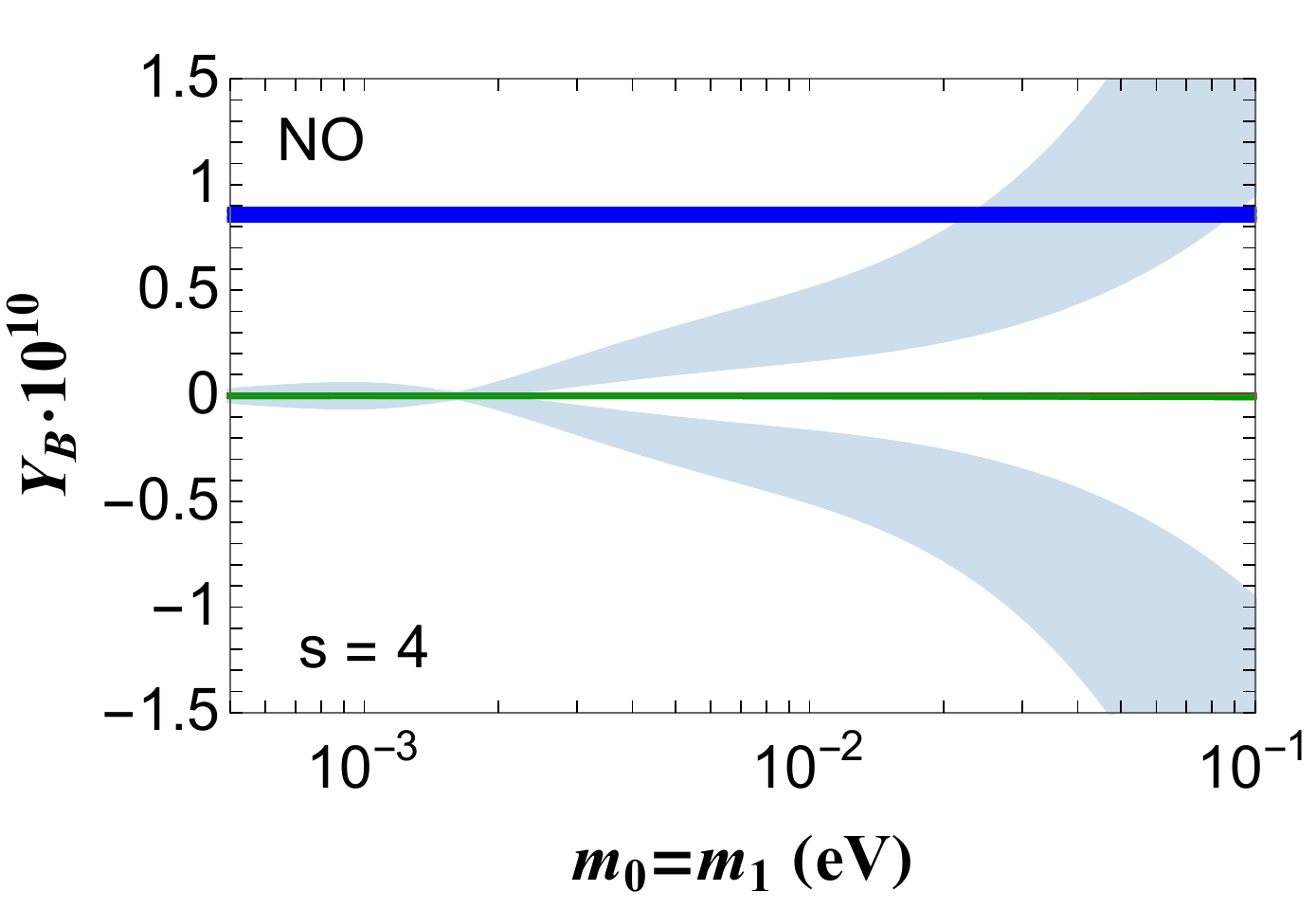}
      \caption*{}
    \end{subfigure}
        \hfill
    \begin{minipage}[b]{0.44\textwidth}
      \caption{\small{\textbf{Case 3 b.1).} $Y_B$ versus $m_0$ for $n=8$ and $m=4$. We display results for 
     $s=1, \, 2, \, 4$. Those for $s > n/2=4$ can be deduced from the ones presented here.  
     The parameter $\theta$ lies in the intervals given in table \ref{tab:case3b1}.    
      Differently colored areas indicate different choices of $\zeta$, see (\ref{inzeta1}-\ref{zetacolor}). 
      Light neutrino masses follow NO and their mass squared differences are varied within their experimentally 
      preferred $3 \, \sigma$ ranges. 
     The horizontal blue band corresponds to the experimentally preferred $3 \, \sigma$ range of $Y_B$ 
     \cite{YBexp}: $Y_B =(8.65\pm0.27)\times 10^{-11}$.  
         \vspace{0.5cm}}}
      \label{Fig:7}
    \end{minipage}
  \end{figure}

We display the results for $Y_B$ depending on the light neutrino mass $m_0$ and for the different choices $s=1$, $s=2$ and $s=4$ in figure \ref{Fig:7}. 
Like for case 2), we focus on a light neutrino mass spectrum with NO. Again, the areas in different colors represent different choices of the parameter $\zeta$, see (\ref{inzeta1}-\ref{zetacolor}).
As can be seen, the correct sign and size can be achieved for all three values of $s$. For $s=1$ the sign of $Y_B$ is mostly negative for small $m_0$, $m_0 \lesssim 3 \times 10^{-3} \, \mathrm{eV}$.
In fact, for such values of the lightest neutrino mass we have $\epsilon_1<0$ and $\epsilon_2>0$. Then, the dominant contribution to the baryon asymmetry (with a negative sign)  comes from the heaviest RH neutrino, $N_1$,
provided $|\epsilon_1|$ is not strongly suppressed (this suppression happens for $\zeta\approx \pi/2,~3\pi/2$).  In the interval $3\times 10^{-3}~\text{eV}\lesssim m_0\lesssim 10^{-2}~\text{eV}$ both $\epsilon_1$ and $\epsilon_2$ are positive and thus the overall sign of $Y_B$. This is also true for 
 $m_0\gtrsim 10^{-2}~\text{eV}$ as long as the contribution coming from $N_1$ is not suppressed due to the particular choice of $\zeta$. Thus, we correctly predict the 
 sign of $Y_B$ in  more than $90\%$ of the parameter space for $3\times 10^{-3}~\text{eV}\lesssim m_0\lesssim 0.1~\text{eV}$. Regarding the size of $Y_B$, this is correctly achieved 
 for $\tilde\kappa=4 \times 10^{-3}$, as long as $m_0\gtrsim 5\times 10^{-3}$ eV. 
The results for $s=2$ are similar to those for $s=1$, with the
sign of $Y_B$ reversed, because the dominant contribution to the CP asymmetries that is proportional to $\sin\alpha$ changes sign, as $\sin\alpha$ is negative for $s=1$ and 
positive for $s=2$, see table \ref{tab:case3b1}.
Consequently, only for small values of $m_0$ the sign is correctly reproduced. $Y_B$ compatible with its measured value at the 
$3 \, \sigma$ level or better prefers $m_0$ in the interval $6.8\times 10^{-4}~\text{eV}\lesssim m_0\lesssim 1.7 \times 10^{-3} \, \mathrm{eV}$. In this particular interval, the sign of $Y_B$ 
is positive in more than $80\%$ of the parameter space. 
The size of the maximally achievable value of $|Y_B|$ is slightly smaller for $s=1$ than for $s=2$, simply because the dominant term is proportional
to $\sin\alpha$ that is smaller by a factor $1/\sqrt{2} \approx 0.71$ for $s=1$ than for $s=2$, compare also table \ref{tab:case3b1}. The remaining choice $s=4$ allows to achieve the correct
sign and size of $Y_B$ in some parameter space for larger $m_0$, for which light neutrino masses become more and more degenerate. 
However, like for $v=0$ in case 2), we cannot call this a prediction of the sign of $Y_B$, since there is no preference 
for a certain sign observable in figure \ref{Fig:7}. This is explained by the absence of the leading term(s) in the CP asymmetries that are sourced by non-trivial Majorana phases. 
A way to see this is to use the formulae in (\ref{e1case3b1sim}-\ref{e3case3b1sim}) that are derived under the assumption that $m=n/2$ and $s=n/2$ in case 3 b.1). As one
can see, the resulting CP asymmetries are proportional to $\sin 2 \zeta$ as well as to $\sin 2\theta$. Thus, considering $\zeta$ in the intervals in (\ref{inzeta2}) leads equally
likely to $\sin 2\zeta >0$ as to $\sin 2\zeta <0$. Consequently, the sign of $Y_B$ cannot be predicted in this case, only by fixing the group theoretical parameters, $k_1$, $k_2$ and
the interval of $\theta$. In addition, we note that also a change in the interval of $\theta$ gives rise to a change in the sign of the CP asymmetries, since both intervals are connected
by replacing $\theta$ with $\pi-\theta$. According to the approximations in (\ref{e1case3b1sim}-\ref{e3case3b1sim}) the CP asymmetries are very small for the special
choices $\zeta=0, \, \pi$ and $\zeta= \pi/2, \, 3 \pi/2$, represented by the red and green bands in the figures. As already mentioned in subsection \ref{sec333},  for the first choice the CP asymmetries
vanish exactly to all orders in $\tilde\kappa$, whereas for the second choice this statement is only true at LO. Nevertheless, also in the latter case the CP asymmetries become way too small
for explaining the experimentally observed value of $Y_B$. The two light-blue areas are delimited by the curves defined by $\zeta=\pi/4$ ($\zeta=3\pi/4$)  and $\zeta\approx 0.24$ ($\zeta\approx 1.82$). 
The former two choices refer to the maximally achievable values of $Y_B$ (with positive and negative sign, respectively). This can be understood by using the formulae in 
(\ref{e1case3b1sim}-\ref{e3case3b1sim}) which show that the LO expressions of all CP asymmetries $\epsilon_i$ are maximized for maximal $|\sin 2\zeta|$ which occurs, for example, for $\zeta=\pi/4$
and $\zeta=3 \pi/4$. The latter two choices describe the boundaries of the light-blue areas that lead to the smallest absolute values for $Y_B$ for a certain value of $m_0$. They simply correspond
to two limiting values of the intervals ${\cal I}_i$ in which $\zeta$ can vary, if it does not take special values, compare (\ref{inzeta2}).

%
%
%
%
It is interesting to compare in more detail the results for $s=4$ in this case with the ones obtained for $v=0$ in case 2), see figure \ref{Fig:6}. As said, they share the feature that the sine of the 
Majorana phase $\alpha$ is small (or even exactly zero), see table \ref{tab:case2} and \ref{tab:case3b1}, and thus the otherwise dominant contribution to the CP asymmetries 
$\epsilon_i$ is absent. Furthermore, the Dirac phase is in both cases large, $|\sin\delta| \gtrsim 0.8$.  However, there is also a crucial difference between them, namely the fact that 
for $v=0$ in case 2) the Majorana phase $\beta$ is large, whereas it is trivial for $s=4$ in this case. Thus, comparing these two cases allows us to disentangle the effects of the CP phases $\beta$
and $\delta$ on the resulting baryon asymmetry. Indeed, we see that the two figures, upper-left plot of figure \ref{Fig:6} and lower-left plot of figure \ref{Fig:7}, are quite similar, with the 
striking difference that for $m_0 \lesssim 2 \times 10^{-3} \, \mathrm{eV}$ for $s=4$, $\sin\beta=0$, $Y_B$ is very small, while $Y_B$ can reach values up to $\pm 0.3 \times 10^{-11}$ for $v=0$
in case 2), i.e. $0.83 \lesssim |\sin\beta| \lesssim 0.94$.

Finally, we comment on some peculiarities of examples belonging to case 3 a): clearly, for the (always admitted) choice $s=0$ no non-vanishing
CP asymmetry can be obtained, because in this case an accidental CP symmetry, common to charged lepton and neutrino sectors, exists, for details
see \cite{HMM}. Furthermore, we note that in some cases such an accidental CP symmetry arises from a particular choice of the parameter $\theta$,
see e.g. $n=16$, $m=1$ and $s=8$ in which the best fitting point of $\theta$ is at $\theta_{\mathrm{bf}}=0$. Considering the whole experimentally preferred $3\, \sigma$ ranges
of the lepton mixing angles also non-zero values of $\theta$ are admitted. However, we expect then that CP phases are still
small and thus the size of $Y_B$ is suppressed. 
 Eventually, we notice
that there are cases in which $\sin\alpha=0$ is obtained at the best fitting point $\theta_{\mathrm{bf}}$ (in these cases the best fit value of the solar mixing angle
cannot be accommodated). These are cases in which 
a prediction of the sign of $Y_B$ is impossible, since we expect a similar behavior as for $s=4$ for case 3 b.1), discussed here, or for $v=0$ in case 2).

\section{Comments  on flavored leptogenesis}
\label{sec40}

In this section we briefly comment on the case of flavored leptogenesis which is realized if RH neutrino masses are smaller
than $10^{12} \, \mathrm{GeV}$ \cite{Abada:2006ea, Abada:2006fw, Nardi:2006fx} (and larger than $10^8 \, \mathrm{GeV}$ in order to correctly reproduce the light neutrino mass scale \cite{Davidson:2002qv}
for $y_0 \gtrsim 10^{-3}$).
In particular, for $10^9 \, \mathrm{GeV} \lesssim M_i \lesssim 10^{12} \, \mathrm{GeV}$ the Yukawa interactions of RH charged leptons of $\tau$-flavor are in thermal equilibrium
and hence leptogenesis occurs in the so-called two-flavor regime. For even lower RH neutrino masses, $M_i \lesssim 10^9 \, \mathrm{GeV}$, also the $\mu$-flavor becomes dynamical. 

The formula for the flavored CP asymmetry $\epsilon_i^{\alpha}$ ($\alpha=e,\,\mu,\,\tau$) reads \cite{review_lepto_2}
\begin{eqnarray}
\label{eCPfl}
\epsilon_i^\alpha & = &  -\frac{\Gamma (N_i \rightarrow H l_\alpha) - \Gamma (N_i \rightarrow H^\star \bar{l}_\alpha)}{\sum_\beta [\Gamma (N_i \rightarrow H l_\beta) + \Gamma (N_i \rightarrow H^\star \bar{l}_\beta)]}\nonumber\\
& = &  -\frac{1}{8 \pi\,(\hat{Y}_D^\dagger \hat{Y}_D)_{ii} } \sum_{j \neq i}\Big\{  \mathrm{Im} \left((\hat{Y}_D^\dagger \hat{Y}_D)_{ij}  (\hat{Y}_D)^\star_{\alpha i} (\hat{Y}_D)_{\alpha j} \right) \, f(M_j/M_i) \nonumber\\
&&+  \,
 \mathrm{Im} \left( (\hat{Y}_D^\dagger \hat{Y}_D)_{ji} (\hat{Y}_D)^\star_{\alpha i} (\hat{Y}_D)_{\alpha j}  \right) \, g\left(M_j/M_i \right) \Big\}\,,
\end{eqnarray}
where $f(x)$ in the SM is given in (\ref{floop}) and 
\begin{equation}
	g(x) \;=\; \frac{1}{1-x^2}\,.\label{gloop}
\end{equation}
Notice that the term in (\ref{eCPfl}) with $g(x)$ does not depend on Majorana phases, because it originates from
the lepton number conserving, but lepton flavor violating one-loop  contribution to the CP asymmetries. 
The analytic expression of the unflavored CP asymmetries $\epsilon_i$  in (\ref{epsi}) is recovered, if we sum over the flavor index $\alpha$
\begin{equation}
\label{releCPflunfl}
	\epsilon_i\; \;= \sum\limits_{\alpha=e,\,\mu,\,\tau}\epsilon_i^\alpha\,.
\end{equation}
The contribution to  $Y_B$ from the decay of each RH neutrino has a form similar to the expression in (\ref{YBi}), but with $\epsilon_i^\alpha$ instead of $\epsilon_i$ and 
efficiency factors with an additional index, depending on the dynamical flavor.

In analogy to the study of leptogenesis performed in the unflavored regime, we consider a neutrino Yukawa matrix $Y_D$ of the form  $Y_D=y_0\, 1 +\delta Y_D$, with 
the  correction $\delta Y_D$ defined as in  (\ref{deltaYD}) and proportional to the small expansion parameter $\kappa$ ($\tilde{\kappa}$).
We can first determine the allowed interval of the parameter $y_0$ using (\ref{miMi}) and taking into account $10^{9}~\text{GeV}\lesssim M_i\lesssim 10^{12}$~GeV
\begin{equation}
\label{y0FL}
	0.001\, \lesssim\, y_0\, \lesssim\, 0.04\,.
\end{equation}
Applying (\ref{eCPfl}) we can check that for our form of $Y_D$ the flavored CP asymmetries $\epsilon_i^\alpha$ are proportional to the product of $y_0$ and $\kappa$ ($\tilde{\kappa}$)
at LO. Thus, corrections $\delta Y_D$ are crucial also in this case in order to achieve non-vanishing CP asymmetries. However, the dependence on the parameters $y_0$ and $\kappa$
is different than in the unflavored case in which $\epsilon_i$ are independent of $y_0$  and  scale as $\kappa^2$. Since (\ref{releCPflunfl}) holds,
we see that the sum of $\epsilon_i^\alpha$ over $\alpha$ must add up to zero at LO. Hence, we can always express one of the flavored CP asymmetries as the negative of the
sum of the other two at this order. For $y_0$ varying in the interval in (\ref{y0FL})
the absolute value of the CP asymmetries $\epsilon_i^\alpha$ is larger than $10^{-6}$, if we choose $\tilde\kappa\gtrsim 10^{-3}$. This lower bound
coincides with the one necessary to obtain successful leptogenesis in the unflavored regime, see (\ref{estimatekappa}). As explained, this is also the expectation 
of the natural size of $\kappa$ from model building. 
Another important feature of the LO result of the flavored CP asymmetries, common to the case of unflavored CP asymmetries, is the fact that the only source of CP violation 
relevant  for leptogenesis is provided by the phases in the matrix $U_R$, coinciding with the CP phases of the PMNS mixing matrix. The reason for this to happen is the symmetric form of $\delta Y_D$ in flavor space
so that the quantity $(\hat{Y}_D^\dagger \hat{Y}_D)_{ij}$, $i \neq j$, in (\ref{YDYDexp}) only depends on the real part of $\delta Y_D$ at LO. 

We first derive the predictions for 
the flavored CP asymmetries, assuming that $U_R=U_{PMNS}$ with $U_{PMNS}$ being of its general form as in (\ref{UPMNSdef}) in appendix \ref{app11}.
In order to simplify the resulting formulae we assume in the following $\zeta= k_\zeta \, \pi$ with $k_\zeta=0, \, 1$ (equivalent to vanishing real part of $z_2$) 
and the low energy CP violation to be encoded only in the Dirac phase $\delta$, i.e.~
\begin{equation}
	\alpha\;=\;  k_\alpha\,\pi\quad\text{and}\quad \beta\,+\,2\,\delta\,=\,k_\beta\,\pi,\quad\text{with}\quad k_{\alpha,\beta}=0,\,1\,.
\end{equation}
At LO in $\tilde\kappa$, the flavored CP asymmetries $\epsilon_1^\alpha$ due to $N_1$ decays then read 
\begin{eqnarray}
\label{epsFLgen}
	\epsilon_1^e & \approx & (-1)^{k_\beta+ k_\zeta} \frac{\sqrt{3}\,y_0\,\tilde\kappa}{16\,\pi}\,\cos^2\theta_{12}\,\sin^2 2\theta_{13}\,\sin2\delta\,f\left(\frac{m_1}{m_3}\right)\,,\nonumber\\
	\epsilon_1^\mu & \approx &  (-1)^{k_\zeta+1} \, \frac{\sqrt{3} \,y_0\,\tilde\kappa}{4\,\pi}\, J_{CP}\,\left[(-1)^{k_\alpha}\,f\left(\frac{m_1}{m_2}\right)+(-1)^{k_\beta}\,f\left(\frac{m_1}{m_3}\right)+g\left(\frac{m_1}{m_2}\right)-
	g\left(\frac{m_1}{m_3}\right) \right] -\,\sin^2\theta_{23}\,\epsilon_1^e\,,\nonumber\\
	\epsilon_1^\tau & = & -\epsilon_1^e\,-\,\epsilon_1^\mu\,,
\end{eqnarray}
where the CP invariant $J_{CP}\propto\sin\delta$ is defined in (\ref{JCP}) and we have used  relation (\ref{miMi}). 
First of all, this example confirms that $\epsilon_i^\alpha$ are proportional to the product $y_0 \, \tilde\kappa$. It also 
reflects the property that the contribution containing $g(x)$ must be independent of the Majorana phases, in particular
of the low energy ones $\alpha$ and $\beta$, appearing in the PMNS mixing matrix. Furthermore, this example clearly shows
that, unlike in the case of unflavored CP asymmetries, the sign of the CP asymmetries $\epsilon_i^\alpha$ (and $Y_B$) cannot be predicted just from the knowledge of the sign of the sine of the CP phases, because it explicitly 
depends on $k_\zeta$ which encodes the sign of the real part of the correction $z_1$. 
Finally, we note that in this particular case $\epsilon_1^e\approx 0$ and $\epsilon_1^\tau\approx -\epsilon_1^\mu\propto J_{CP}$, if the CP phase $\delta$ is (close to) maximal, $\delta\approx \pi/2,\,3\pi/2$. 
The form of $\epsilon_2^\alpha$ and $\epsilon_3^\alpha$ is very similar to the one of  $\epsilon_1^\alpha$, in particular, regarding relations between the CP asymmetries of the different flavors $\alpha$
as well as the dependence on the CP phase $\delta$. 

Concerning the different scenarios with a flavor and a CP symmetry we focus on the flavored CP asymmetries $\epsilon_i^\alpha$ for case 1), since simple expressions can be derived. We obtain for $\epsilon^\alpha_{1,3}$ at LO
\begin{eqnarray}
	\epsilon_1^\alpha &  \approx & (-1)^{k_1+1}\, \frac{y_0\,\tilde\kappa}{6\,\sqrt 3\,\pi}\,\cos\left(\theta+\zeta\right)\,\cos\left(\theta+\rho_\alpha\,\frac\pi 3\right)\,\sin6\,\phi_s\, f\left(\frac{m_1}{m_2}\right)\, ,\nonumber\\
	\epsilon_3^\alpha &  \approx & (-1)^{k_1+k_2+1}\,\frac{y_0\,\tilde\kappa}{6\,\sqrt 3\,\pi}\,\sin\left(\theta+\zeta\right)\,\sin\left(\theta+\rho_\alpha\,\frac\pi 3\right)\,\sin6\,\phi_s\, f\left(\frac{m_3}{m_2}\right)\,,
	\label{rhoaindirect}
\end{eqnarray}
with $\rho_e=3$, $\rho_\mu=1$ and $\rho_\tau=-1$. The formulae for $\epsilon_2^\alpha$ are given by the linear combination in (\ref{e2bdtrivial}), 
replacing $\epsilon_{1,3}$ with their flavored counterparts $\epsilon_{1,3}^\alpha$. 
 Notice that the contribution to the CP asymmetries proportional to $g(x)$ is absent in this case. This is due to the fact that the
Dirac phase is trivial, see (\ref{CPinvcase1}). Most importantly, the sign of $\epsilon_i^\alpha$ (and $Y_B$) depends on the particular choice of the parameter $\zeta$ (the ratio between
the real parts of $z_1$ and $z_2$ that parametrize the correction $\delta Y_D$, see (\ref{Rez12}) and (\ref{deltaYD})). This is in contrast to the 
unflavored case where $\zeta$ appears only as argument of positive semi-definite functions, cf.~(\ref{eps1case1}-\ref{eps3case1}).

If $Y_D$ is only invariant under the residual symmetry $G_\nu$  instead of the full flavor group $G_f$ and the CP symmetry, $\epsilon_i^\alpha$ are in general non-zero even for vanishing corrections $\delta Y_D$.
For $Z$ and $X$ representing the residual $Z_2$ and the CP symmetry in ${\bf 3}$, respectively, $Y_D$ is in this case constrained by the conditions
\begin{equation}
	Z^\dagger\,Y_D\,Z\;=\;Y_D\quad\text{and}\quad X^\star\,Y_D\,X\;=\;Y_D^\star\,,
\end{equation}
such that it can be in general written as
\begin{equation}
\label{YDGnu}
Y_D = \Omega \, R_{ij} (\theta_L) \, \left( \begin{array}{ccc}
y_1 & 0 & 0\\
0 & y_2 & 0\\
0 & 0 & y_3
\end{array}
\right) \, R_{ij} (-\theta_R) \, \Omega^\dagger
\end{equation}
with $R_{ij} (\theta_{L,R})$ being rotations in the $(ij)$-plane around $\theta_{L,R}$, lying in the interval $[0, \pi)$, and $y_i$ real parameters.
All these five are not further restricted by $G_\nu$. As a consequence, the light neutrino masses as well as the lepton mixing angles and low
energy CP phases are less correlated with the parameters appearing in $M_R$ in (\ref{MR0}), i.e.~$U_\nu$ (and thus $U_{PMNS}$) 
does not only depend on $\theta$  and the relation between the 
light and heavy neutrino masses in (\ref{miMi})  does not hold except for the light neutrino mass $m_k$ with $k\neq i,j$. In particular, only for this generation
the CP parity of the light neutrino is given by the one of the RH one $N_k$, encoded in the matrix $K_\nu$.

Transforming $Y_D$ in (\ref{YDGnu}) to the mass basis of RH neutrinos using $U_R$ in (\ref{UR0}), we find
\begin{equation}
\label{YDhatGnu}
	\hat Y_D\,=\,\Omega\, R_{ij}(\theta_L)\,  \left( \begin{array}{ccc}
y_1 & 0 & 0\\
0 & y_2 & 0\\
0 & 0 & y_3
\end{array}
\right) \,R_{ij}(\theta-\theta_R)\,K_\nu\,.
\end{equation}
As can be checked, the flavored CP asymmetries, computed with (\ref{eCPfl}), are proportional to $\sin 2 (\theta-\theta_R)$
(but not $\theta_L$), the product $y_i \, y_j$ as well as the difference $y_i^2-y_j^2$. Thus, if any of these vanishes, i.e.~$\theta=\theta_R + p \, \pi/2$, $p$ integer,
or $y_i=0$ or $y_j=0$ or $y_i = \pm y_j$, the CP asymmetries all must vanish. Furthermore, $\epsilon^\alpha_k$ with $k \neq i,j$ is always zero.
The correct size $|\epsilon^\alpha_i| \gtrsim 10^{-6}$ can be achieved for $y_j \gtrsim 10^{-3}$. This constraint on the neutrino Yukawa couplings is consistent
with the requirement to reproduce light neutrino masses of order $0.1$ eV.
Interestingly enough, computing the unflavored CP asymmetries  $\epsilon_i$ we see that they vanish exactly in this scenario,
 unless a small correction $\delta Y_D$, which is not invariant under $G_\nu$, is introduced. 
 
 In addition to this general discussion, we have explicitly
 studied the predictions for $\epsilon_i^\alpha$ for all cases 1) to 3 b.1). In case 1) not only the unflavored CP asymmetries $\epsilon_i$, but also the flavored ones vanish identically. For this reason,
we present the exact expressions of $\epsilon^\alpha_{i}$ for case 2)  
\begin{eqnarray}
 	\epsilon_{1 \, (3)}^\alpha & = &  \frac{y_1\,y_3\,(y_{1 \, (3)}^2-y_{3 \, (1)}^2)\,\sin 2\left(\theta-\theta_R\right)\,\sin\left(\phi_u -\rho_\alpha\,\frac\pi 3\right)}{48\,\pi \left(y_{1 \, (3)}^2\,\cos^2\left(\theta-\theta_R\right)\,+\,y_{3 \, (1)}^2\,\sin^2\left(\theta-\theta_R\right) \right)}\left[(-1)^{k_2}\,f\left(\frac{M_{3 \, (1)}}{M_{1 \, (3)}}\right)+g\left(\frac{M_{3 \, (1)}}{M_{1 \, (3)}}\right)\right]\,,\nonumber\\ &&\label{epsFLcase2}\\
	\epsilon_2^\alpha & = & 0\, , \nonumber
 \end{eqnarray}
 with $\rho_\alpha$ defined as below (\ref{rhoaindirect}).
 All characteristics mentioned above can be verified: $\epsilon^\alpha_2=0$, since the rotation $R_{13} (\theta)$ acts in the $(13)$-plane;
furthermore, $\epsilon^\alpha_{1,3}$ are proportional to the product $y_1 \, y_3$, to the difference $y_1^2 - y_3^2$ as well as to $\sin 2 \, (\theta-\theta_R)$;
and the sum over all flavors $\alpha$, see (\ref{releCPflunfl}), vanishes. 
In addition, we observe that $\epsilon_{1,3}^\alpha$ are not sensitive to the parameter $\phi_v$, which determines the Majorana phase
$\alpha$, see  (\ref{sinaapproxcase2}).\footnote{Although the form of $Y_D$ is now more complicated than in the previous case and the PMNS mixing matrix does
not coincide anymore with $U_R$, it is still true that the Majorana phase $\alpha$ depends in this case (almost) only on $\phi_v$, as long as the 
measured lepton mixing angles are accommodated well.}  This result is in stark contrast with the ones of the unflavored CP asymmetries in (\ref{e1case2}-\ref{e3case2})
where the terms with $\phi_v$ usually dominate, if $v$ does not vanish.
Again, the sign of $Y_B$ is not determined, since $y_{1,3}$ can be both positive and negative. This observation is contrary to the findings in the case of unflavored leptogenesis,  see figure \ref{Fig:6}.
The formulae for $\epsilon^\alpha_i$ for case 3 a) and case 3 b.1) show a similar structure as those in (\ref{epsFLcase2}) and are thus not
explicitly discussed here.  

Finally, we remark that for the particular choice $\phi_u=0,~\pi$ ($u=0,\,n$), the flavored CP asymmetries in case 2) in (\ref{epsFLcase2}) fulfill
\begin{equation}
\label{epsflcase2_mutau}
 \epsilon_{1 \, (3)}^e\;=\;0\quad \text{and}\quad \epsilon_{1 \, (3)}^\tau\;=\;-\epsilon_{1 \, (3)}^\mu\,. 
 \end{equation}
This result is related to the presence of the $\mu\tau$ reflection symmetry \cite{mutaureflection}, since $X=P_{23}$ or $Z \, X=P_{23}$, 
if we set in addition $\phi_v=0$ ($v=0$), compare (\ref{possibleCPnu}).  
 A similar result can also be obtained for
 case 3 b.1): if we set $m=n/2$, we find $\epsilon^e_{2 \, (3)}=0$ and $\epsilon^\mu_{2 \, (3)}= -\epsilon^\tau_{2 \, (3)}$ (the CP asymmetries
 $\epsilon^\alpha_1$ vanish in all flavors due to the form of the rotation $R_{ij} (\theta)$). The $\mu\tau$ reflection symmetry is then recovered
 for $s=n/2$ \cite{HMM} and we find explicitly $Z X=P_{23}$.\footnote{We note that this choice is also possible for case 3 a), since the latter 
 arises from the same type of residual symmetries $Z$ and $X$, see (\ref{possibleCPnu}). However, it is clear that only for case 3 b.1) agreement with the 
 experimental data on lepton mixing angles can be achieved for the choice $m=n/2$ (and $s=n/2$) \cite{HMM}.} 
 
Flavored leptogenesis in a scenario with  $\mu\tau$ reflection symmetry has also been discussed in \cite{mutaureflectlepto}. In their analysis
the authors only impose this symmetry on the neutrino mass matrix.
A diagonal mass matrix for charged leptons is ensured by gauging $L_\mu-L_\tau$ symmetry. The authors stick to 
the two-flavor regime and discuss the case in which the RH neutrino mass 
spectrum is strongly hierarchical. Thus, only the out-of-equilibrium decays of the lightest RH neutrino are relevant for generating the baryon asymmetry of the Universe.
In \cite{GfCPflavorlepto} the authors focus on the flavor symmetries belonging to the series $\Delta (3 \, n^2)$ and $\Delta (6 \, n^2)$ 
combined with a CP symmetry, as in our analysis. 
They, however, concentrate on the case of flavored leptogenesis and a strongly hierarchical RH neutrino mass spectrum, which is similar to the study performed in \cite{mutaureflectlepto}.
Consequently, their results in general differ from ours.

\section{Comments  on SUSY framework}
\label{sec4}

Here we comment on the implementation of our scenario and results for leptogenesis in a SUSY framework, since most of the concrete
models with a flavor (and a CP) symmetry are formulated in a SUSY context, see e.g. \cite{AF2,A4S4CPmodels}.
The relevant superpotential of such a scenario reads
\begin{equation}
w_l= Y_l \, l \, h_d \, \alpha^c + Y_D \, l \, h_u \, \nu^c + M_R \, \nu^c \, \nu^c
\end{equation}
with $h_{u,d}$ denoting the two Higgs multiplets of the Minimal SUSY SM (MSSM). The latter are taken to transform trivially under all non-gauge symmetries.
 We assign the three generations of RH neutrinos $\nu^c$ to ${\bf 3}$ under $G_f$, while the fields $l$ 
now transform as ${\bf \bar{3}}$. In this way, the term $l \, h_u \, \nu^c$ is invariant under $G_f$ and thus arises at the renormalizable level. In addition,
the Yukawa coupling $Y_D$ is proportional to the identity matrix.
The RH charged leptons $\alpha^c$ are in the trivial representation of $G_f$. Like in the non-SUSY framework, also here we assume the existence of an auxiliary
symmetry $Z_3^{(\mathrm{aux})}$. Only the two fields $\mu^c$ and $\tau^c$ carry non-trivial phases under the latter symmetry: $\mu^c \sim \omega$ and $\tau^c \sim \omega^2$.

We note a few crucial differences relevant for our results of leptogenesis. First of all, the range of RH neutrino masses in which 
unflavored leptogenesis is the dominant generation mechanism of the lepton asymmetry is rescaled by the factor $1+\tan^2 \beta$
with $\tan\beta=\langle h_u \rangle/\langle h_d \rangle$ so that we have to require RH neutrino masses $M_i$ to lie in the interval
$10^{12} \, (1+\tan^2\beta) \, \mathrm{GeV} \lesssim M_i \lesssim 10^{14} \, (1+\tan^2\beta) \, \mathrm{GeV} $. We can estimate the allowed values of $\tan\beta$
in our scenario by considering the operator that gives rise to the tau lepton mass. This operator requires (at least) one insertion of a flavor symmetry
breaking field, since LH leptons transform as ${\bf \bar{3}}$ under $G_f$, whereas $\tau^c$ and $h_d$ are trivial singlets. Assuming that the 
size of the suppression due to the necessary insertion of one flavor symmetry breaking field is $\varepsilon \approx (0.01 \div 0.1)$, see (\ref{estimateeps})
in appendix \ref{app3}, and that the tau Yukawa coupling varies between $1/3$ and $3$, we get as range of $\tan\beta$ the following\footnote{For such values of $\tan\beta$
also the suppression of the bottom quark mass with respect to the top quark mass has to originate from the flavor symmetry. Thus, we expect in concrete models
that also the down type quark masses stem from non-renormalizable operators only.}
\begin{eqnarray}\nonumber 
\varepsilon=0.01 &:& \tan\beta \approx 3 \; ,
\\ \nonumber
\varepsilon=0.05 &:& 3 \lesssim \tan\beta \lesssim 15 \; ,
\\ \nonumber
\varepsilon=0.1 &:& 3 \lesssim \tan\beta \lesssim 30 \; .
\end{eqnarray}
Thus, RH neutrino masses are expected to be larger than in the SM case. In order to obtain the correct values for the light neutrino masses we 
have to rescale the coupling $y_0$ accordingly, see (\ref{miMi}).

On the other hand, the masses of RH neutrinos cannot be too large either 
in a SUSY framework, since larger values give rise to larger contributions to flavor non-universal soft terms of sleptons through renormalization group
running effects \cite{RGSUSY}. These flavor non-universal soft terms play a crucial role in charged 
lepton flavor violating processes \cite{BM1986} such as $\mu\rightarrow e\gamma$ and $\mu-e$ conversion that are strongly 
constrained experimentally  \cite{MEG,SINDRUM} (for a review see \cite{LFVreview}).

Furthermore, large RH neutrino masses also require a large reheating 
temperature $T_R \gtrsim 10^{12} \, \mathrm{GeV}$. This in turn gives rise to the well-known gravitino problem \cite{gravitinoproblem}, unless 
e.g.~the gravitino production is suppressed and/or it decays \cite{solutiongravitino} and/or there is an additional substantial contribution to the total energy
density of the Universe that dilutes the gravitino abundance, see discussion in \cite{review_lepto_1}. 
 
In a SUSY framework the CP asymmetries $\epsilon_i^{(\alpha)}$ not only arise from decays  of the RH neutrinos $\nu^c_i$  to SM particles, but also from decays of 
their SUSY partners and to SUSY particles, see e.g. \cite{review_lepto_1}. Thus, also the form of the loop functions $f(x)$ and $g(x)$ in the MSSM is different. In fact, the former now reads \cite{Covi:1996wh}
\begin{equation}
\label{floopMSSM}
f (x) \;=\; -x \left[ \frac{2}{x^2-1} + \ln \left( 1+ \frac{1}{x^2} \right) \right] \,.
\end{equation}
Comparing its behavior to the loop function in (\ref{floop}), relevant in the SM, we notice the following: both of them can be zero, however, the exact position in $x$ is 
slightly different ($x \approx 0.42$ for the SM loop function and $x\approx 0.34$ for $f(x)$ in (\ref{floopMSSM})); otherwise, their values are of the same order of 
magnitude for $x \lesssim 1$; for $x \gtrsim 1$ it holds to good approximation that $\left(f (x) \right)_{\footnotesize\mathrm{MSSM}} \approx 2 \, \left( f (x) \right)_{\footnotesize\mathrm{SM}}$.
They have a divergence at $x=1$ in common. Concerning the loop function $g(x)$ which enters in the flavored CP asymmetries, we have the exact relation $(g(x))_{\footnotesize\mathrm{MSSM}} = 2 \,(g(x))_{\rm SM}$,
where $(g(x))_{\rm SM}$ is given in (\ref{gloop}).

At the same time, additional washout effects are induced by the SUSY particles. The numerical factor in (\ref{YBi}) also slightly changes and reads $1.48  \times 10^{-3}$
due to the additional particles. Moreover, the sphaleron conversion factor is modified due to the presence of a second Higgs doublet.

Summarizing all the effects and assuming leptogenesis to be the dominant mechanism for generating a lepton (and thus the baryon) asymmetry, the results for the 
baryon asymmetry of the Universe $Y_{B, \footnotesize\mathrm{MSSM}}$  in the MSSM framework
are obtained by rescaling $Y_{B, \footnotesize\mathrm{SM}}$, computed in an SM context, in the following way \cite{review_lepto_2}
\begin{eqnarray}
\label{YBSUSYSM}
&&Y_{B, \footnotesize\mathrm{MSSM}} \approx \sqrt{2} \, Y_{B, \footnotesize\mathrm{SM}}  \;\;\; \mbox{in the regime of strong washout}
\\ \nonumber
\mbox{and}&&Y_{B, \footnotesize\mathrm{MSSM}} \approx 2 \, \sqrt{2} \, Y_{B, \footnotesize\mathrm{SM}}  \;\;\; \mbox{in the regime of weak washout.}
\end{eqnarray}

\mathversion{bold}
\section{$0\nu\beta\beta$ decay}
\mathversion{normal}
\label{sec5}

In this section we study $0\nu\beta\beta$ decay of even-even nuclei. This process, unobserved so far, is important for testing the Majorana nature
of  neutrinos and it explicitly depends on the values of the Majorana phases $\alpha$ and $\beta$
(see e.g. \cite{0nubbreview} for a recent review). Therefore, it is relevant in order to put constraints on the
scenarios introduced in subsection~\ref{sec22}.
Earlier studies of $0\nu\beta\beta$ decay in the context of models with combined flavor and CP symmetries can be found in the
first reference in \cite{D48CPmodel}, last reference in \cite{A4S4CPmodels},  in \cite{D6n2CPZ2Z2}, first reference in \cite{A5CP} as well as in \cite{DeltaCPothers}.
After a short subsection containing general information about the quantity measurable in $0\nu\beta\beta$ decay we discuss its predictions in different 
 scenarios of lepton mixing separately, scrutinizing, in particular, the examples for which leptogenesis has been analyzed.

\subsection{Preliminaries and general results}

The half-life $T_{1/2}^{0\nu\beta\beta}$ of a decaying nuclear isotope via this process is
\begin{equation}
	\frac{1}{T_{1/2}^{0\nu\beta\beta}} \;=\; G^{0\nu}\left|M^{0\nu}\right|^2\,\frac{m_{ee}^2}{m_e^2}\,,
\end{equation}
where $G^{0\nu}$ denotes the phase space factor, $M^{0\nu}$ the nuclear matrix element (NME) for a lepton number violating transition,
$m_e$ the electron mass and $m_{ee}$ the effective Majorana neutrino mass. The values of $G^{0\nu}$ and $M^{0\nu}$ can be computed and depend 
on the nuclear isotope, whereas $m_{ee}$ is expressed only in terms of neutrino masses and lepton mixing parameters,
\begin{equation}
\label{meedef}
m_{ee}= \left| U_{PMNS,11}^2 \, m_1 + U_{PMNS,12}^2 \, m_2 + U_{PMNS,13}^2 \, m_3  \right|\,,
\end{equation}
that, according to the parametrization of $U_{PMNS}$, given in appendix~\ref{app11}, reads
\begin{equation}\label{meedef2}
m_{ee}= \left| \cos^2 \theta_{12} \, \cos^2 \theta_{13} \, m_1 + \sin^2 \theta_{12} \, \cos^2 \theta_{13} \, e^{i \alpha} \, m_2 + \sin^2 \theta_{13} \, e^{i \beta} \, m_3  \right| \, .
\end{equation}
An upper bound on the effective Majorana neutrino mass has been set by several experiments, using different nuclear isotopes:
GERDA ($^{76}$Ge) \cite{GERDA}, KamLAND-Zen ($^{136}$Xe) \cite{KamlandZen}, EXO-200 ($^{136}$Xe) \cite{EXO200}, CUORE-0 ($^{130}$Te) \cite{CUORE0}, and 
NEMO 3 ($^{100}$Mo among others) \cite{NEMO3}. The strongest bound on $m_{ee}$ is given by the KamLAND-Zen experiment 
\begin{equation}
\label{meeBOUND}
	m_{ee}\;<\; \left(0.14\div0.28\right)\;\text{eV}\quad\text{at 90\% C.L.}
\end{equation}
with the largest uncertainty arising from the one of the associated NME.

For a  hierarchical light neutrino mass spectrum, i.e. for $m_0\approx 0$ in (\ref{massesNO}) and (\ref{massesIO}), the expected value of $m_{ee}$ strongly depends on the ordering of the
neutrino masses. In fact, in this limit we derive from (\ref{meedef2}), for NO
\begin{equation}
m_{ee}^\mathrm{NO} \approx \left|  \sin^2 \theta_{12} \, \cos^2 \theta_{13} \, e^{i \alpha} \, \sqrt{\Delta m_{\mathrm{sol}}^2} +  \sin^2 \theta_{13} \, e^{i \beta} \, \sqrt{\Delta m_{\mathrm{atm}}^2} \right| 
\end{equation}
and for IO
\begin{equation}
\label{meeIOgen}
m_{ee}^\mathrm{IO} \approx  \left| \cos^2 \theta_{12} + \sin^2 \theta_{12} \, e^{i \alpha} \right| \, \cos^2 \theta_{13} \, \sqrt{|\Delta m_{\mathrm{atm}}^2|}\, .
\end{equation}
In the last expression we have neglected the subdominant term proportional to  $\Delta m_{\mathrm{sol}}^2$.
For a given value of $\theta_{12}$ and $\theta_{13}$, the maximum and minimum  of $m_{ee}$ are obtained for trivial Majorana phases, see e.g.~the lower-left panel of figure \ref{Fig:9}. In particular, for  IO,
using that $\sin^2 \theta_{12} \approx 1/3$, we get
\begin{equation}
\label{regionIO}
\frac 13 \, \sqrt{|\Delta m_{\mathrm{atm}}^2|} \lesssim m_{ee}^\mathrm{IO} \lesssim \sqrt{|\Delta m_{\mathrm{atm}}^2|} \, .
\end{equation}
In the case of a QD light neutrino mass spectrum, we expect from (\ref{meedef2})
 $m_{ee}$ to be proportional to $m_0$, for both NO and IO. Indeed, neglecting $\sin\theta_{13}$, we have
 \begin{equation}
 \label{meeQDgen}
 		m_{ee}^\mathrm{QD} \approx \sqrt{1-\sin^2 \frac\alpha 2 \, \sin^2 2 \, \theta_{12}} \;\; m_0\,,
 \end{equation}
which, for $\sin^2 \theta_{12} \approx 1/3$ and a trivial Majorana phase $\alpha$, takes values in the interval 
\begin{equation}
\label{regionQD}
\frac 13 \, m_0 \lesssim m_{ee}^\mathrm{QD} \lesssim m_0 \, .
\end{equation}
 As is well-known, for NO $m_{ee}$ can be strongly suppressed due to cancellations for $m_0$ between $10^{-3}$ eV and $0.01$ eV. This can occur in principle, if both Majorana
 phases are trivial or both are non-trivial. In our numerical analysis we find examples for both situations, see e.g.~case 1) for the former one (compare (\ref{meecase1cancel})) and 
 case 3 a) for the latter one, see figure \ref{Fig:10}.
 
 In figures \ref{Fig:8}-\ref{Fig:10} we show in light-blue (orange)  the most general region in the $m_{0}-m_{ee}$  plane for NO (IO), 
 obtained by varying the lepton mixing angles within their experimentally preferred $3 \, \sigma$ ranges, see appendix~\ref{app12}, and all CP phases between $0$ and $2 \, \pi$.
 The boundaries of the areas are associated with trivial Majorana phases $\alpha$ and $\beta$ and  do depend on the lower and upper $3 \, \sigma$ limits
 of the solar mixing angle $\theta_{12}$. The experimental constraints in the $m_{0}-m_{ee}$ plane are set by the various $0\nu\beta\beta$ decay experiments, see (\ref{meeBOUND}),
 as well as by the Planck Collaboration that puts an upper limit on the lightest neutrino mass $m_0$, see (\ref{m0BOUND}).
The former is displayed as horizontal dashed line in figures \ref{Fig:8}-\ref{Fig:10}, while the latter as vertical line.

\mathversion{bold}
\subsection{Predictions of $m_{ee}$ in case 1)}
\mathversion{normal}

In this scenario the Majorana phase $\beta$ (as well as the Dirac phase $\delta$) is always trivial, for any choice of $\theta$
and the group theoretical parameters $n$ and $s$ (or their combination $\phi_s$), while $\alpha$ can take non-trivial values, see (\ref{alphacase1}). 
Then, for a  hierarchical neutrino mass spectrum ($m_0 \approx 0$), $m_{ee}$ reads 
\begin{eqnarray}
\label{meeNOcase1}
 m_{ee}^\mathrm{NO} &\approx& \frac 13 \, \left| \sqrt{\Delta m_{\mathrm{sol}}^2} + 2 \, (-1)^{k_1+k_2} \, \sin^2 \theta \, e^{6 \, i \, \phi_s} \, \sqrt{\Delta m_{\mathrm{atm}}^2} \right|\,,
\\
 m_{ee}^\mathrm{IO} &\approx& \frac 13 \, \left| 1 + 2 \, (-1)^{k_1} \, e^{6 \, i \, \phi_s} \, \cos^2 \theta \right| \, \sqrt{|\Delta m_{\mathrm{atm}}^2|}\; .
 \label{meeIOcase1}
\end{eqnarray}
Remembering that $\theta$ is close to $0$ or $\pi$, $m_{ee}^\mathrm{IO}$ can be simplified to
\begin{equation}
\label{meecase1IO}
	 m_{ee}^\mathrm{IO} \;\approx\;  \frac 13 \, \sqrt{5\pm4 \, \cos 6 \, \phi_s} \, \sqrt{|\Delta m_{\mathrm{atm}}^2|}\, ,
\end{equation}
with the plus (minus) sign corresponding to even (odd)  $k_1$.
 Similarly, we obtain for a QD light neutrino mass spectrum
\begin{equation}
m_{ee}^\text{QD}\; \approx\; \frac 13 \, \left| (-1)^{k_1} + 2 \, e^{6 \, i \, \phi_s} \, \left( \cos^2 \theta + (-1)^{k_2} \, \sin^2 \theta \right) \right| \, m_0\,.
\label{meeQDcase1}
\end{equation}
As one can see, for even $k_2$ the effective Majorana neutrino mass  is independent of $\theta$
\begin{equation}
\label{meecase1QD}
m_{ee}^\text{QD} \;\approx\; \frac 13 \, \sqrt{5\pm4 \, \cos 6 \, \phi_s} \;\; m_0 \, ,
\end{equation}
with the plus (minus) sign valid for  even (odd) $k_1$. Since $\theta \approx 0, \, \pi$, (\ref{meecase1QD}) is a good approximation
also in the case of $k_2$ being odd. This result is consistent with the one found in (\ref{meeQDgen}) for the generic case.

In the case $n=4$, that is the smallest value of the group theoretical parameter $n$ allowing for non-trivial $\alpha$ (see table~\ref{tab:case1}), we obtain
for $k_2$ even and $s=0$, $k_1$ even (odd) or $s=2$, $k_1$ odd (even)
\begin{equation}
m_{ee}^\text{NO}\approx 0.0040 \; (0.0018) \; \text{eV} \; ,\;\; m_{ee}^\text{IO} \approx 0.0479  \; (0.0149) \; \text{eV}  \;\;\; \mbox{and} \;\;\;  m_{ee}^\text{QD} \approx 0.11  \; (0.034) \; \text{eV}   \; ,
\end{equation}
while for $k_2$ odd $s=0$, $k_1$ even (odd) or $s=2$, $k_1$ odd (even) lead to 
\begin{equation}
m_{ee}^\text{NO}\approx 0.0019 \; (0.0039) \; \text{eV}  \; , \;\; m_{ee}^\text{IO} \;\; \mbox{like for $k_2$ even and} \;\;\;  m_{ee}^\text{QD} \approx 0.096  \; (0.029) \; \text{eV}  \; .
\end{equation} 
 Here we used the best fit values of the neutrino mass squared differences, see (\ref{dmbf}), $m_0=10^{-4}$ eV for NO and IO as well as $m_0=0.1$ eV for QD, 
  and $\theta \approx 0.18$ or $\theta \approx 2.96$ that lead to the best fitting of the experimental data on lepton mixing angles \cite{HMM}.
 If we choose $k_2$ even (odd) and instead $s=1$ or $s=3$ as well as any value of $k_1$, we find
 \begin{equation}
 m_{ee}^\text{NO}\approx 0.0031 \; (0.0031) \; \text{eV} \; , \;\; m_{ee}^\text{IO} \approx 0.0355 \; \text{eV for any $k_2$ and} \;\;\; m_{ee}^\text{QD} \approx 0.075  \; (0.071) \; \text{eV}  \; .
 \end{equation}
 These values of $m_{ee}$ agree well with the results from the approximate formulae given in (\ref{meeNOcase1}), (\ref{meecase1IO}) and (\ref{meecase1QD}).
 
 As in the generic case, for trivial $\alpha$, occurring for $s=0$ ($s=2$),  $m_{ee}$ is strongly suppressed for certain values of the lightest neutrino mass $m_0$ 
in the case of NO. In particular, $m_{ee}^\mathrm{NO}$ is smaller than $10^{-4}$ eV for 
\begin{equation}
\label{meecase1cancel}
0.0023\; \text{eV} \lesssim m_0 \lesssim 0.0037 \; \text{eV} \;\;\; \text{and} \;\;\;  0.0066 \; \text{eV} \lesssim m_0 \lesssim 0.0087\; \text{eV} 
\end{equation}
with the range of the intervals coming from the variation of the neutrino mass squared differences in their experimentally preferred $3 \, \sigma$ range and 
$0.169\lesssim \theta \lesssim 0.195$ or  $2.95 \lesssim \theta \lesssim 2.97$. For $m_0$ being in the first interval a cancellation requires
$k_1$ odd (even) and $k_2$ even, while for $m_0$ in the second interval $k_1$ and $k_2$ are required to be odd ($k_1$ is required to be even and $k_2$ odd) in order to find $m_{ee}^\mathrm{NO} \lesssim 10^{-4}$ eV.
We note that due to the constraints on lepton mixing angles and CP phases the range of $m_0$ in which $m_{ee}$ can be very small is 
considerably reduced with respect to the generic case. Especially, the smallest value of $m_0$ for which a cancellation can occur is larger than in the generic case because of 
 the lower bound on the solar mixing angle, $\sin^2\theta_{12} \gtrsim 1/3$.
 For $s=1, 3$ such a suppression is not possible and we, indeed, find a lower limit on $m_{ee}^{\mathrm{NO}}$ that is
 \begin{equation}
 m_{ee}^{\mathrm{NO}} \gtrsim 0.0029 \, \mathrm{eV} \; .
 \end{equation}

\mathversion{bold}
\subsection{Predictions of $m_{ee}$ in case 2)}
\mathversion{normal}

\begin{figure}[t!]
\begin{center}
\begin{tabular}{cc}
\includegraphics[width=0.48\textwidth]{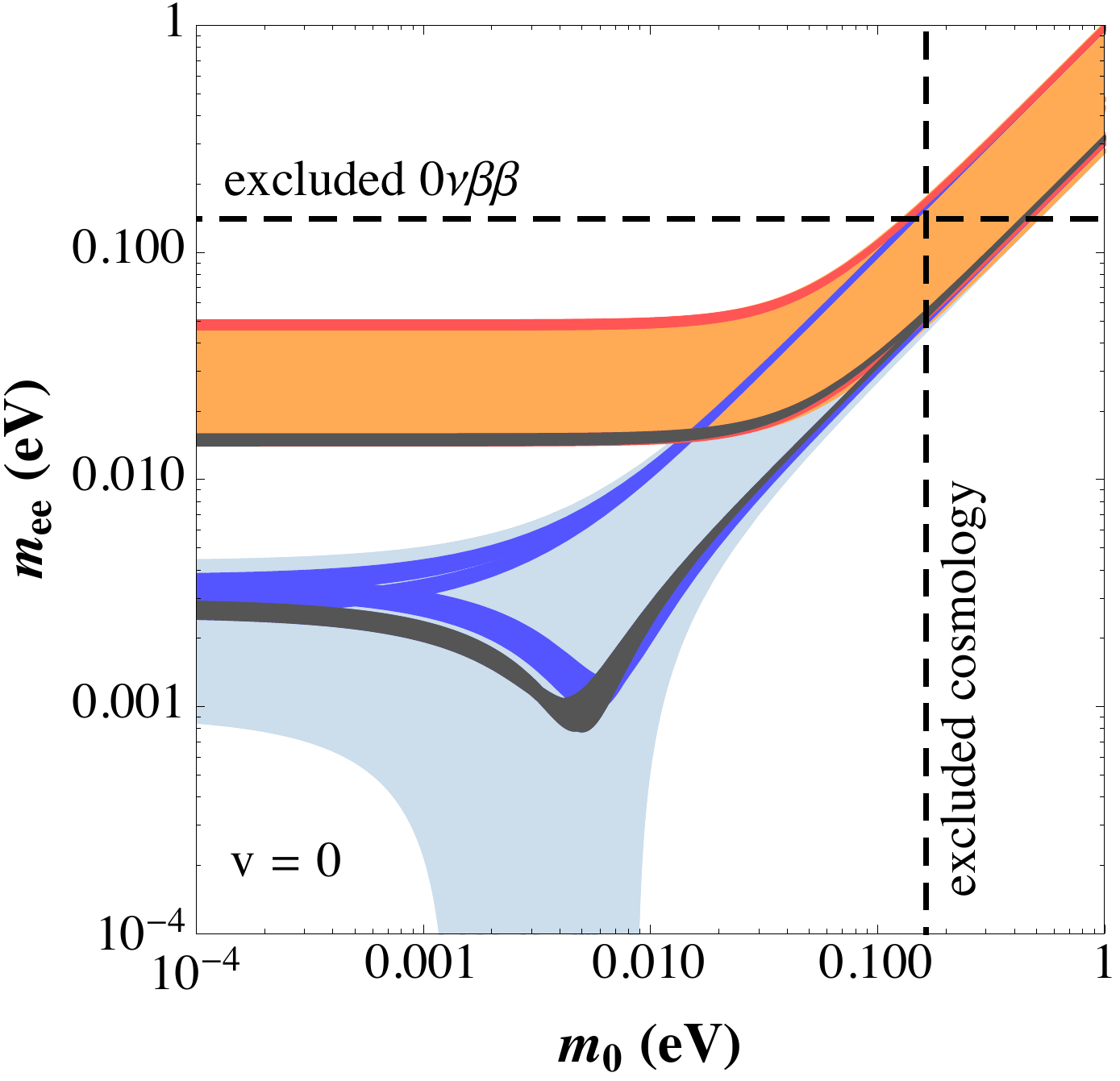} &
\includegraphics[width=0.48\textwidth]{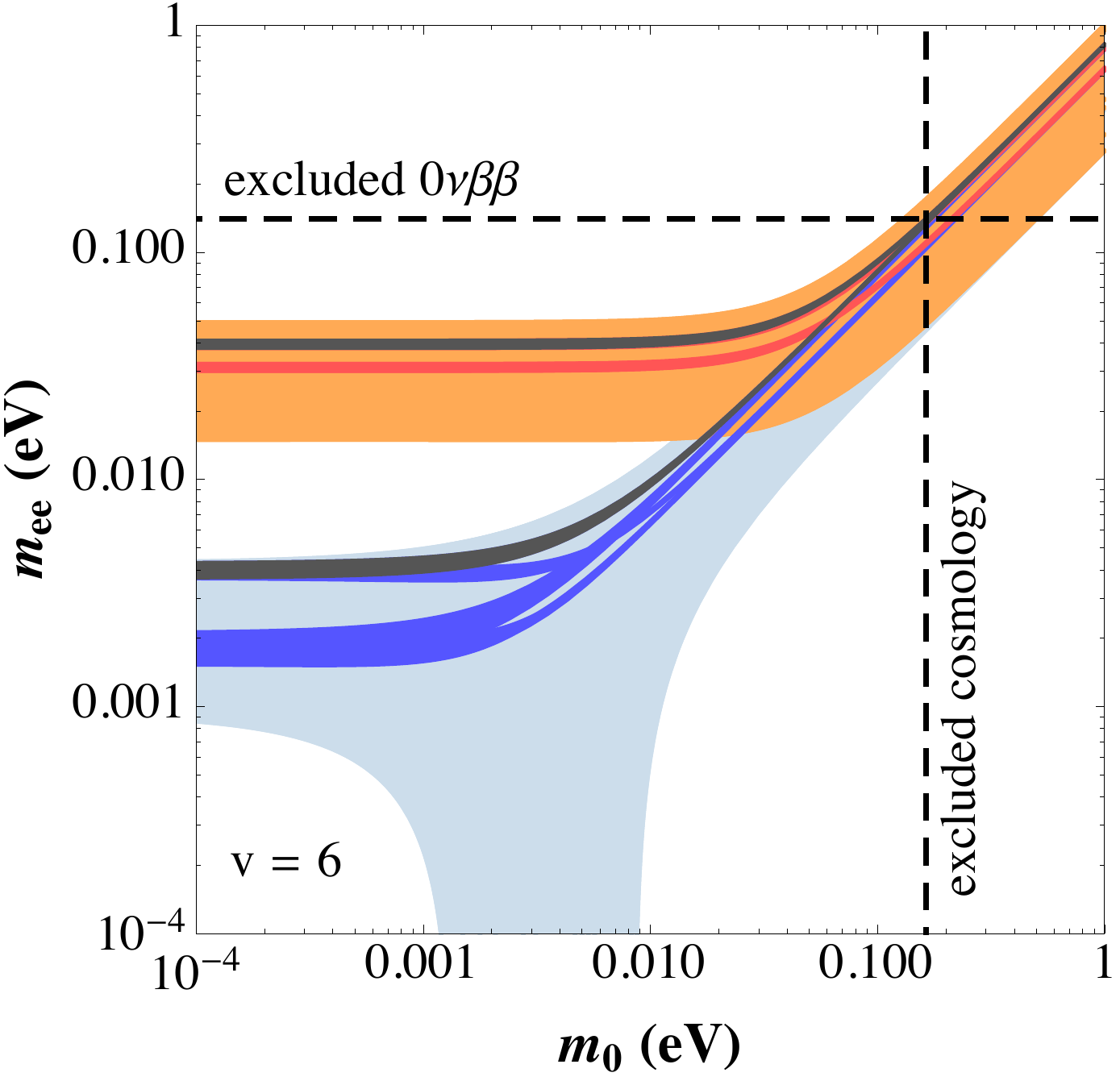} 
\end{tabular}
\caption{\small{\textbf{Case 2).} Effective Majorana neutrino mass $m_{ee}$ as function of the lightest neutrino mass $m_0$, for $n=10$, $u=4$ and two values of $v$. 
The parameter $\theta$ is varied in the
range $1.40\lesssim \theta \lesssim 1.44$. The neutrino mass squared differences are chosen within their experimentally preferred $3\, \sigma$ intervals. 
The blue (red) regions correspond to the variation of $m_{ee}$ in the case of NO (IO), for all possible combinations of  $k_{1,2}$. The particular choice $k_1=k_2=0$ is 
 highlighted with the dark-grey areas. The plot on the right is also obtained for $v=24$ and $1.70\lesssim \theta \lesssim 1.74$, with
 the dark-grey area corresponding to $k_1=1$ and $k_2=0$. The other allowed values of $v$ ($v=12$ and $v=18$) give predictions very similar to the ones shown here.
The light-blue and orange areas are the expectations obtained from the general expression of $m_{ee}$ in (\ref{meedef}) for NO and IO, respectively, 
where the neutrino oscillation parameters are taken within the experimentally preferred $3\, \sigma$ intervals given in appendix~\ref{app12}. The boundaries of these areas are obtained for CP conserving Majorana phases $\alpha$ and $\beta$ 
and depend on the experimentally preferred $3\, \sigma$ range of the solar mixing angle.}}
\label{Fig:8}
\end{center}
\end{figure}

In this case both Majorana phases $\alpha$ and $\beta$ can have in general non-trivial values, depending on $\theta$ and  the group theoretical parameters $n$, $u$ 
and $v$ (or their combinations $\phi_u$ and $\phi_v$). 
Taking the lepton mixing matrix $U_{PMNS,2}$ as defined in (\ref{URcase2}), the general expression (\ref{meedef}) for  a hierarchical neutrino mass spectrum can be approximated as 
\begin{eqnarray}
m_{ee}^\mathrm{NO} &\approx& \frac 13 \, \left| \sqrt{\Delta m_{\mathrm{sol}}^2} - 2 \, (-1)^{k_1+k_2} \, e^{i \, \phi_v} \, \left( \cos \theta \, \sin \frac{\phi_u}{2} - i \, \sin \theta \, \cos \frac{\phi_u}{2}  \right)^2  \, \sqrt{\Delta m_{\mathrm{atm}}^2} \right|\,,
\label{meecase2NO}
\\
m_{ee}^\mathrm{IO} &\approx &\frac 13 \, \left| 1 + (-1)^{k_1} \, e^{i\, \phi_v} \, \left( \cos \phi_u + \cos 2 \theta - i \, \sin 2 \theta \, \sin \phi_u \right) \right| \, \sqrt{|\Delta m_{\mathrm{atm}}^2|}  \, .
\label{meecase2IO}
\end{eqnarray}
For a QD light neutrino mass spectrum and $k_2$ even, as in case 1), the resulting $m_{ee}$ is actually independent of $\theta$
\begin{equation}
\label{meecase2QD}
m_{ee}^\text{QD} \approx \frac 13 \, \sqrt{3 + 2 \, \cos 2 \, \phi_u \pm 4 \, \cos \phi_u \, \cos \phi_v} \;\; m_0\,,
\end{equation}
with the plus (minus) sign referring to even (odd)  $k_1$. For $k_2$ odd instead one can show that $m_{ee}$  is independent of $\phi_u$, if
 $\theta\approx 0,~\pi$ or $\pi/2$ (these values are typically required for reproducing the observed lepton mixing angles see \cite{HMM}),
\begin{equation}
\label{meecase2QDb}
	m_{ee}^\text{QD} \;\approx\; \frac 13 \, \sqrt{5\pm4 \, \cos \phi_v} \;\; m_0 \, ,
\end{equation}
with again the plus (minus) sign referring to even (odd) $k_1$. This expression coincides with the one derived in the generic case in (\ref{meeQDgen}),
 if we use that in case 2) holds $\cos\alpha \approx (-1)^{k_1} \, \cos\phi_v$ (a relation similar to the approximate relation in (\ref{sinaapproxcase2})).

Expressions like in (\ref{meecase2NO}-\ref{meecase2QDb}) are also obtained, if we perform a permutation of the rows of $U_{PMNS,2}$, see (\ref{shiftPcase2}). In this case 
$m_{ee}$ can be computed by applying the transformations given in (\ref{shift1case2}-\ref{shift2case2}) to the approximations derived. 

We note that the quantity $m_{ee}$ is invariant under the set of transformations (see below (\ref{shift2case2}))
\begin{equation}
 \theta \rightarrow \pi-\theta\,, \quad v \rightarrow k \, n - v~(\text{$k$ odd})\quad\text{and}\quad \text{$k_1 \rightarrow k_1+1$}\,,\label{trtheta}
 \end{equation} 
which shows that results for different values of $v$ are related to each other, if we also take into account that the interval of $\theta$ has to be 
changed. 

In the explicit example, we choose for the flavor group the index $n=10$ and for the parameter characterizing the CP symmetry $u=4$, like in the numerical 
study of leptogenesis in subsection \ref{sec342}. We thus discuss a case in which we can use the formulae shown in (\ref{meecase2NO}-\ref{meecase2QD}), only after having applied the transformations in (\ref{shift1case2}).
This choice of parameters predicts $|\sin\beta|\approx 1$, while $|\sin\alpha|$ can take three different values, depending on the parameter $v$ (or  $\phi_v$), i.e. $v=0$, 6, 12, 18 and 24,
see table~\ref{tab:case2}.
In figure~\ref{Fig:8} we display the predictions of $m_{ee}$ as function of the lightest neutrino mass $m_0$ for two different choices of $v$:
$v=0$ in the left and $v=6$ in the right panel. These values lead either to $\alpha \approx 0, \, \pi$ ($v=0$) or  to almost maximal $\alpha$ ($v=6$).
The blue and red regions in each plot correspond to the predictions for NO and IO, respectively, in which we allow for all four possible combinations of 
$k_1$ and $k_2$.  Among these, we highlight the choice $k_1=k_2=0$ with the dark-grey area. 
Moreover, we vary $\theta$ in the range $1.40\lesssim \theta \lesssim 1.44$ and the neutrino mass squared 
differences within their experimentally preferred $3 \, \sigma$ intervals,  see appendix~\ref{app12}.  Using 
(\ref{trtheta}) we see that for $v=0$
values of $\theta$ in the second interval $1.70\lesssim \theta \lesssim 1.74$ lead to the same allowed areas, up to the exchange of $k_1=0$ with $k_1=1$.
For $v \neq 0$ instead, applying (\ref{trtheta}) shows, for example, that the plot on the right panel in figure~\ref{Fig:8} for $v=6$ and $1.40\lesssim \theta \lesssim 1.44$
is the same as the plot for $v=24$ and $\theta$ in the second interval $1.70\lesssim \theta \lesssim 1.74$. The only difference is that the dark-grey area corresponds
for $v=24$ to the choice $k_1=1$ and $k_2=0$ instead of $k_1=k_2=0$. The predictions of $m_{ee}$ for  $v=12$ and $v=18$ are related to each other in a similar way. Figures for 
these two values of $v$ resemble the ones displayed in figure~\ref{Fig:8}.

The most important feature of this case is the fact that there is no cancellation in $m_{ee}$ for any values of $v$ in the case of NO.
Furthermore, we note that for $v=0$ the predictions for IO coincide with the boundaries of the 
area allowed in the generic case. This happens, since the Majorana phase $\alpha$ is nearly trivial, see table \ref{tab:case2}, and the effect
of $\beta$ is suppressed by the reactor mixing angle as well as (the small mass) $m_3$, compare approximation in (\ref{meeIOgen}). 
 
Using  (\ref{meecase2QDb}) with $v=0$ shows that  $m_{ee}^{\text{QD}}$ either equals $m_{ee}^{\text{QD}} \approx m_0$ for $k_1$ odd
or $m_{ee}^{\text{QD}} \approx m_0/3$ for $k_1$ even (remember $k_1$ has to be replaced by $k_1+1$ in (\ref{meecase2QDb}) when applied to the case at hand).
For $m_0$ small and NO instead $m_{ee}$ takes values in the interval 
\begin{equation}
0.0026 \; \text{eV} \lesssim m_{ee}^{\text{NO}} \lesssim 0.0035 \; \text{eV} \; . 
\end{equation}
In contrast, for $v=6$ we see two different regimes realized 
\begin{equation}
m_{ee}^{\text{NO}} \approx 0.0018 \; \text{eV} \;\;\; \mbox{for} \;\;\; k_1+k_2 \;\; \mbox{odd and} \;\;\; m_{ee}^{\text{NO}} \approx 0.0039 \; \text{eV} \;\;\; \mbox{for} \;\;\; k_1+k_2 \;\; \mbox{even,}
\end{equation}
 respectively.
Since the Majorana phase $\alpha$ is non-trivial and not small for $v=6$, we find a non-trivial lower bound on $m_{ee}$ in the case of IO that is by a factor of two
larger than the generic lower bound, $m_{ee}^{\text{IO}} \approx 0.031$ eV. The other value of $m_{ee}$ is $m_{ee}^{\text{IO}} \approx 0.039$ eV, arising, if 
$k_1$ is even. Two different values are also obtained in  the regime of QD light neutrino masses in which we predict for $m_0=0.1$ eV, according to (\ref{meecase2QDb}), $m_{ee}^{\text{QD}} \approx 0.065$ eV or  $m_{ee}^{\text{QD}} \approx 0.083$ eV 
depending on the value of $k_1$.
Future experiments searching for  $0\nu\beta\beta$ decay   \cite{Artusa:2014lgv,Arnold:2010tu,Abgrall:2013rze} can probe almost the whole region for IO, down to $m_{ee}\approx 0.02$ eV, thus allowing 
for the possibility to distinguish between the different choices of the CP transformation $X$.

\mathversion{bold}
\subsection{Predictions of $m_{ee}$ in case 3)}
\mathversion{normal}

\begin{figure}[t!]
    \begin{subfigure}[b]{0.48\textwidth}
      \includegraphics[width=\linewidth]{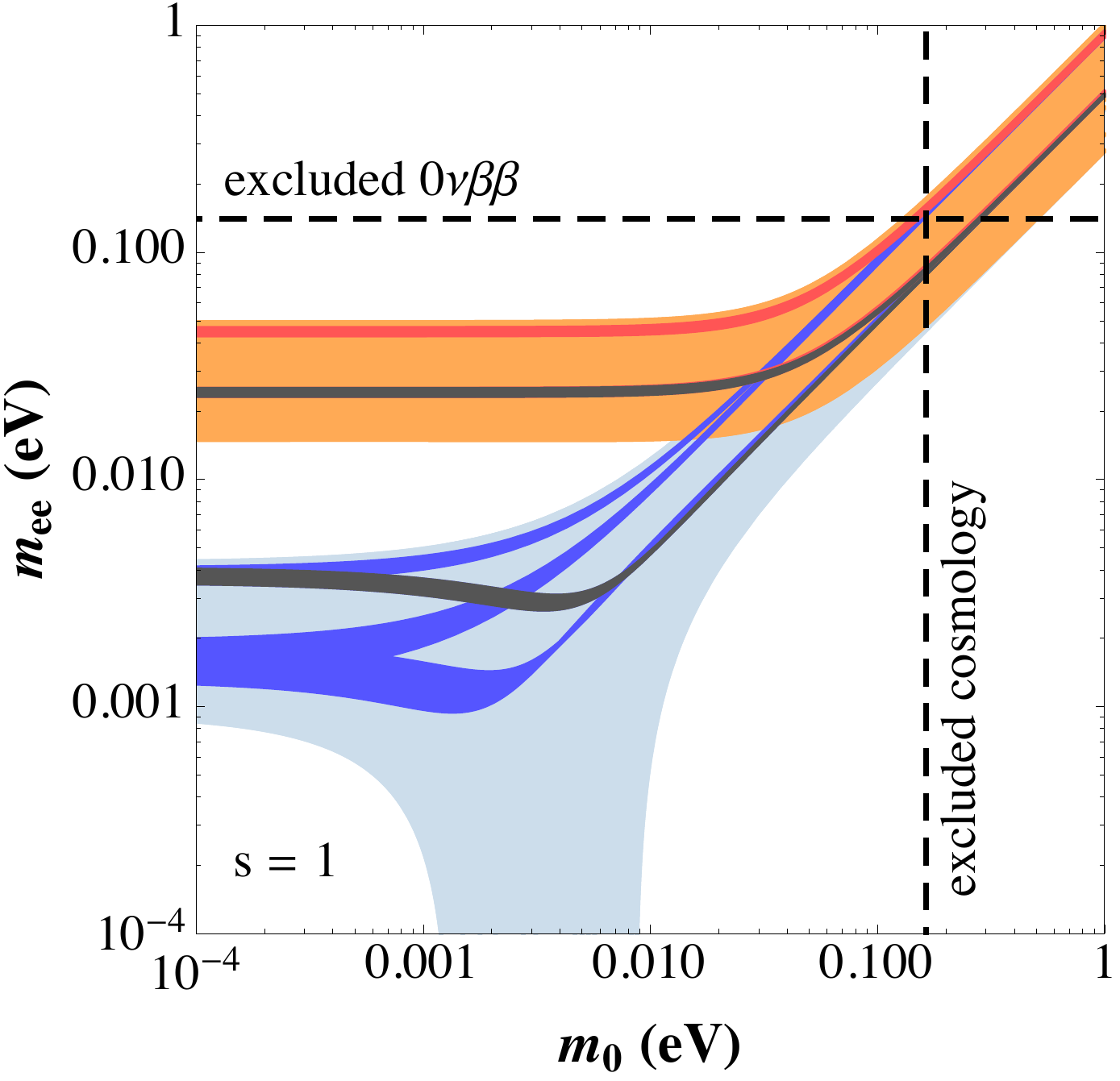}
      \caption*{}
    \end{subfigure}
    \begin{subfigure}[b]{0.48\textwidth}
     \includegraphics[width=\linewidth]{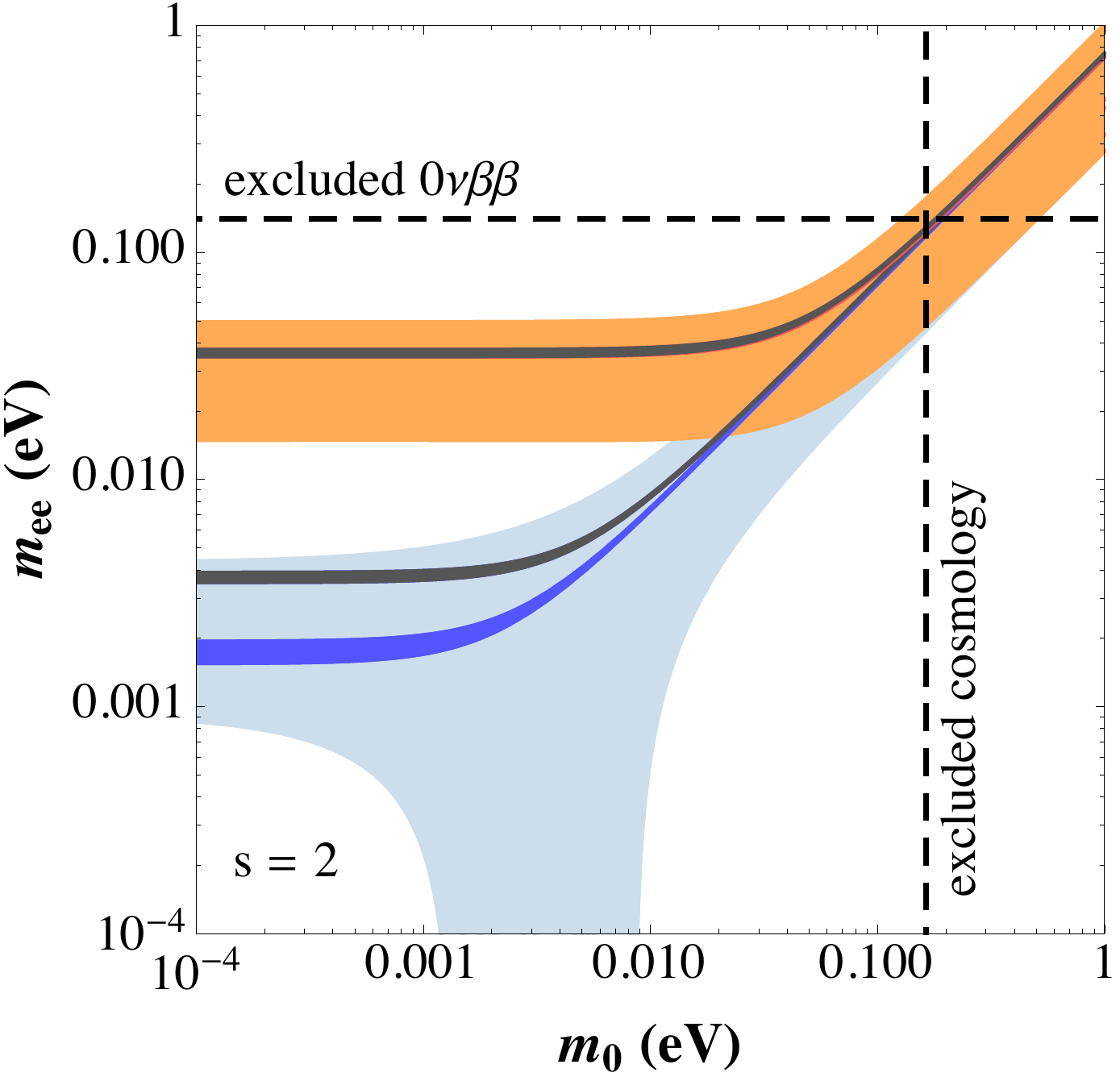}
      \caption*{}
    \end{subfigure}
    \begin{subfigure}[b]{0.48\textwidth}
      \includegraphics[width=\linewidth]{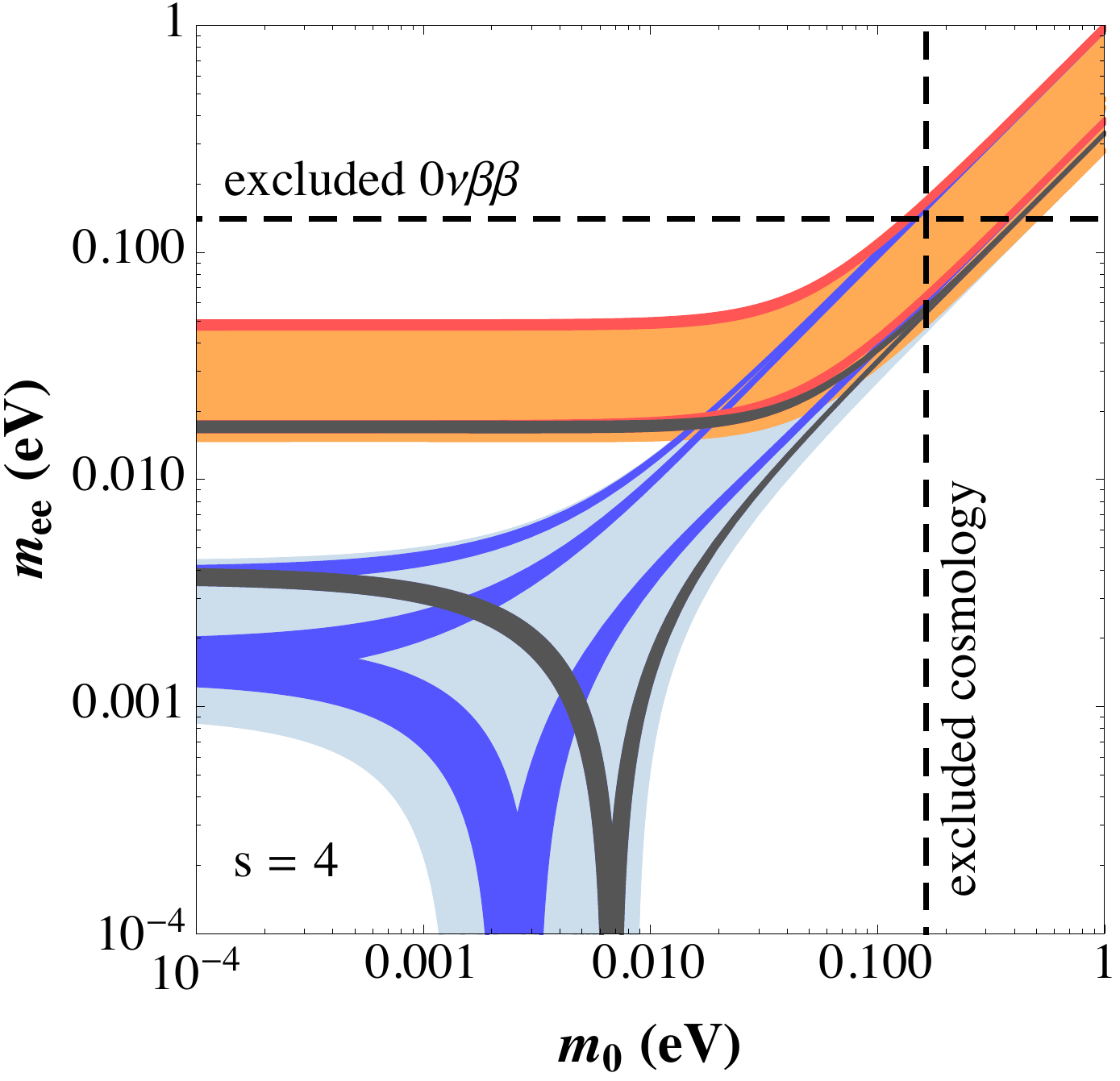}
      \caption*{}
    \end{subfigure}
        \hfill
    \begin{minipage}[b]{0.44\textwidth}
      \caption{\small{\textbf{Case 3 b.1).} $m_{ee}$ versus $m_0$ for $n=8$, $m=4$ and three possible choices of $s$. For a given $s$, $\theta$ takes values in the interval 
      shown in table~\ref{tab:case3b1}. In all the plots the mass squared differences are varied within  the experimentally preferred $3 \, \sigma$ ranges. The blue and red areas
      correspond to an NO  and IO light neutrino mass spectrum, respectively, for all combinations of $k_{1,2}$. 
      The region for $k_1=k_2=0$ is displayed in dark-grey.
      Results for $s > n/2$ are obtained from the ones presented by making use of the transformation in (\ref{trafocase3}).
         \vspace{2.3cm}}}
      \label{Fig:9}
    \end{minipage}
  \end{figure}

We also discuss  the effective Majorana neutrino mass given in  the last case introduced in subsection~\ref{sec22}.  We first consider case 3 b.1) that is characterized by the lepton mixing matrix in (\ref{URcase3}). We focus  on the choice $m=n/2$, as done in our numerical analysis of leptogenesis. In this case, it follows directly from  
 (\ref{trafo2case3}) that  $m_{ee}$ must be invariant under the replacement of $\theta$ with $\pi-\theta$.

Again, we can derive simple approximations that work well for hierarchical and QD light neutrino mass spectra. For vanishing $m_0$ we find
\begin{eqnarray}
&& m_{ee}^\mathrm{NO} \approx \frac 13 \, \left| \sin^2 \theta \,  \sqrt{\Delta m_{\mathrm{sol}}^2}   + (-1)^{k_1} \, \cos^2 \theta \, \sqrt{\Delta m_{\mathrm{atm}}^2} \right| \; ,
\label{meeNOcase3b1}
\\
&& m_{ee}^\mathrm{IO} \approx \frac 13 \, \left| 2  + (-1)^{k_2} \, e^{6 \, i \, \phi_s} \sin^2 \theta \right| \, \sqrt{|\Delta m_{\mathrm{atm}}^2|} \, .\label{meeIOcase3b1}
\end{eqnarray}
Interestingly enough, $m_{ee}^\mathrm{NO}$ is independent of $\phi_s$ (and thus of the chosen CP transformation) 
 and it takes  two distinct values for $k_1$ even and odd, respectively. 
Using for the neutrino mass squared differences the best fit values and choosing $\theta\approx 1.31$ 
(which sets the reactor mixing angle to its best fit value \cite{HMM}), we get
\begin{eqnarray}
\label{meeA}
&& m_{ee}^\mathrm{NO}\approx 0.0038~(0.0016)\;\;\text{eV} \;\;\; \text{\text{for} even (odd) } \; k_1 \; , \\
\label{meeB}
&& m_{ee}^\mathrm{IO}\approx \left|0.015\,e^{6\,i\,\phi_s}\,+\,(-1)^{k_2}\,0.033 \right|\;\text{eV}\, .
\end{eqnarray}
In contrast, for  the QD light neutrino mass spectrum we obtain 
\begin{equation}\label{meecase3b1QD}
m_{ee}^\text{QD} \approx \frac 13 \,\left|2\,+\,(-1)^{k_2}\,e^{6\,i\,\phi_s}\left( \sin^2\theta\,+\,(-1)^{k_1}\cos^2\theta \right)  \right|\,m_0 \, ,
\end{equation}
which becomes independent of $\theta$ for even  $k_1$. Using $\theta\approx \pi/2$ we can further simplify (\ref{meecase3b1QD}) and obtain the expression in (\ref{meecase1QD}), which is valid in case 1).

In figure~\ref{Fig:9} we show the quantity $m_{ee}$ versus $m_0$ for the group theoretical parameters $n=8$, $m=4$ and $s=1,2,4$ with $\theta$ chosen as in table~\ref{tab:case3b1}.
We, thus, consider the same example like in the numerical analysis of leptogenesis in subsection \ref{sec342}. 
The blue (red) areas correspond to NO (IO), for all combinations of $k_{1,2}$, with the particular case $k_1=k_2=0$ shown in dark-grey.

For $s=1$ and $s=2$, the Majorana phases are indeed non-trivial  (see table~\ref{tab:case3b1}) and $m_{ee}$ has a lower bound for a NO light neutrino mass spectrum.  This result is similar to what we found in the numerical example of case 2),
see figure~\ref{Fig:8}. Taking $m_0=10^{-4}~\text{eV}$ and $s=1$ we see that $m_{ee}^\text{NO}$ varies in the interval 
\begin{equation}
0.0034~\text{eV} \lesssim m_{ee}^\text{NO}\lesssim 0.0040~\text{eV} \;\;\; (0.0012~\text{eV} \lesssim m_{ee}^\text{NO}\lesssim 0.0020~\text{eV}) \;\;\; \mbox{for even (odd)} \;\; k_1 \; .
\end{equation}
For $s=2$ one can show that in the case of NO $m_{ee}$ only depends  on $k_1$ in all the range of $m_0$. This explains why for this choice of $s$ only two distinct areas are admitted.
The numerical results for $m_{ee}$ in the case $s=2$ are very similar to $s=1$ for $m_0=10^{-4}~\text{eV}$, i.e.
\begin{equation}
0.0035~\text{eV} \lesssim m_{ee}^\text{NO}\lesssim 0.0039~\text{eV} \;\;\; (0.0015~\text{eV} \lesssim m_{ee}^\text{NO}\lesssim 0.0019~\text{eV}) \;\;\; \mbox{ for even (odd)} \;\; k_1 \; . 
\end{equation}
In both cases the numerical values are in agreement with the analytic estimates for $m_{ee}^\text{NO}$ given in (\ref{meeA}).   
 In the case of IO  we find that for $s=2$ ($\phi_s=\pi/4$), $m_{ee}^\text{IO}$ is actually independent of $k_{1,2}$, see (\ref{meeIOcase3b1}), and thus only one narrow dark-grey shaded area exists in the right panel. It corresponds to $m_{ee}^\text{IO}\approx 0.036$ eV for small values of $m_0$. Instead, for $s=1$ $m_{ee}$ is given by $m_{ee}^\text{IO}\approx 0.024$ (0.045) eV  for even (odd) $k_2$ in the hierarchical regime. Again, these numerical results coincide with the analytic estimate in
 (\ref{meeB}). Similarly,  for a QD light neutrino mass spectrum two values are possible for $m_{ee}$, if $s=1$, namely $m_{ee}\approx 0.49\,m_0$ ($k_2=0$ and $k_1=0, 1$) and $m_{ee}\approx 0.93\,m_0$ ($k_2=1$ and $k_1=0, 1$), whereas we only find one value for $s=2$, i.e. $m_{ee}\approx 0.75\,m_0$.

On the contrary, in the case $s=4$ both Majorana phases $\alpha$ and $\beta$ are trivial and $m_{ee}$ can be 
strongly suppressed for a hierarchical NO light neutrino mass spectrum, as shown in the bottom-left panel of figure~\ref{Fig:9}. Since our approach constrains not only the CP phases, but also
the lepton mixing angles, see  table~\ref{tab:case3b1}, the suppression of $m_{ee}$ occurs only in two  small intervals of $m_0$, which depend on the integers  $k_{1,2}$, i.e. $m_{ee}^\text{NO}\lesssim 10^{-4} \, \mathrm{eV}$ 
is achieved for 
\begin{equation}
0.0019~\text{eV} \lesssim m_0 \lesssim 0.0033~\text{eV} \;\;\; (0.0059~\text{eV} \lesssim m_0 \lesssim 0.0074~\text{eV}) \;\;\; \mbox{and} \;\;\; k_1=k_2=1 \;\; (k_1=k_2=0) \; . 
\end{equation}
 In the strongly hierarchical regime, that is 
for $m_0\lesssim 10^{-4}~\text{eV}$, we find, instead,  
\begin{equation}
0.0035~\text{eV} \lesssim m_{ee}^\text{NO}\lesssim 0.0039~\text{eV} \;\;\; (0.0012~\text{eV} \lesssim m_{ee}^\text{NO}\lesssim 0.0020~\text{eV}) \;\;\; \mbox{for even (odd)} \;\; k_1 
\end{equation}
 in agreement with the analytic estimate, shown in (\ref{meeA}).
For light neutrino masses with IO or being QD, $m_{ee}$ is either close to its lower or its upper limit, as expected for trivial Majorana phases, compare figure~\ref{Fig:9}. 

%
\begin{figure}[t!]
\begin{center}
\begin{tabular}{cc}
\includegraphics[width=0.48\textwidth]{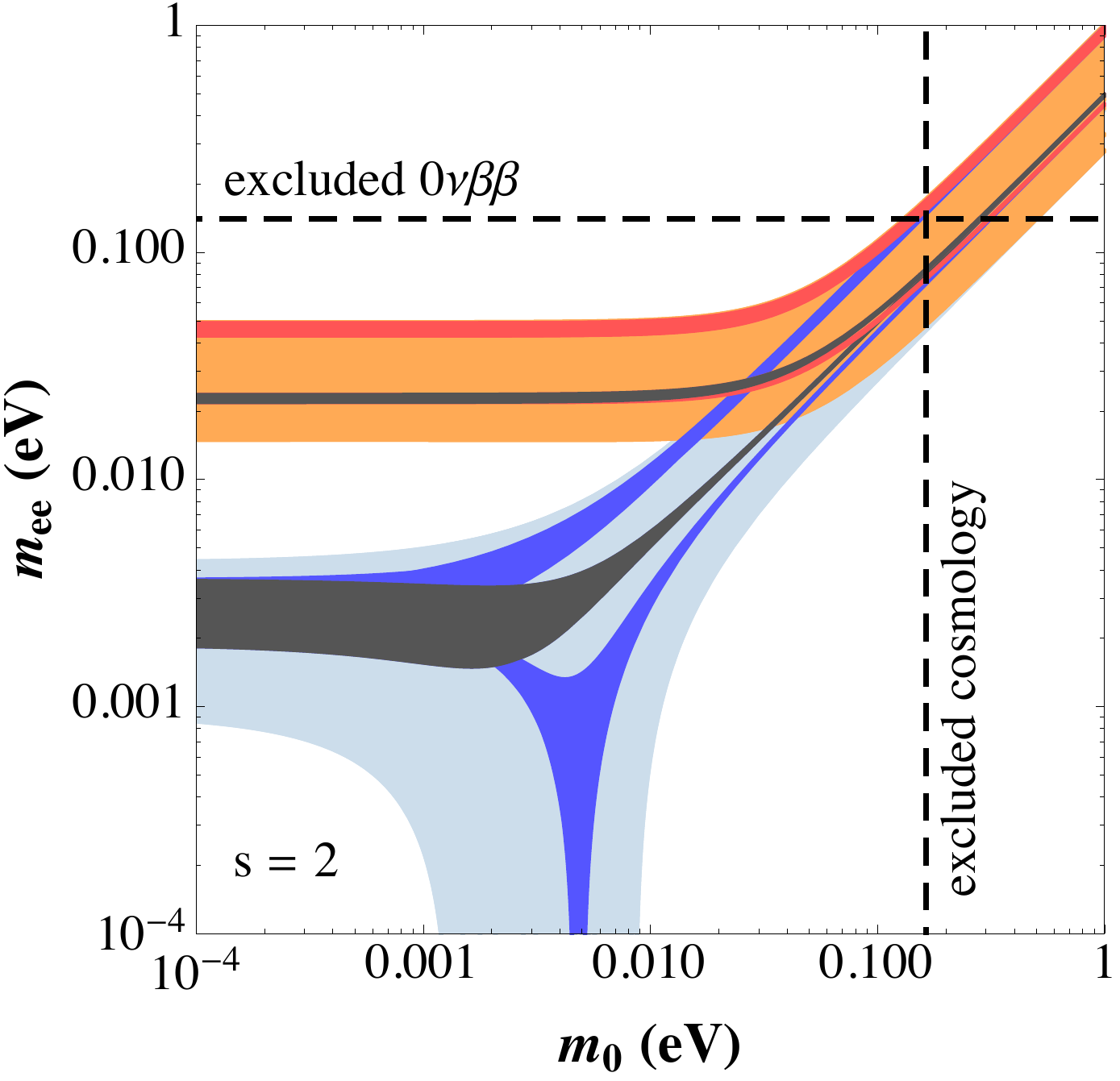} &
\includegraphics[width=0.48\textwidth]{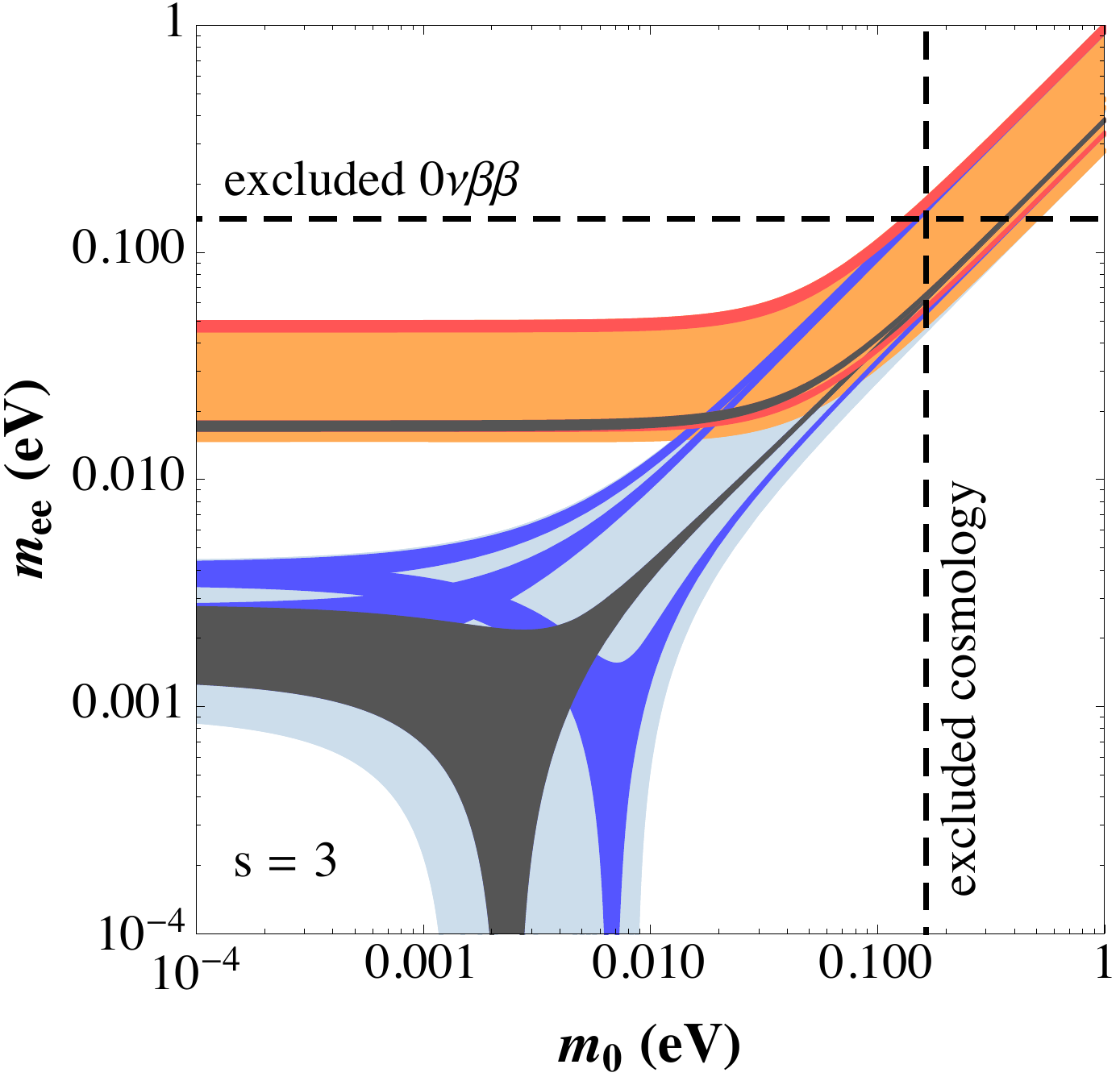} 
\end{tabular}
\caption{\small{\textbf{Case 3 a).} $m_{ee}$ versus $m_0$ for $n=16$, $m=1$ and two choices of $s$, $s=2$ and $s=3$. 
The parameter $\theta$  is varied in the intervals given in the text.
For these values of $s$ at least one of the Majorana phases is non-trivial, see (\ref{sinacase3as2s3}) and (\ref{sinbcase3as2s3}), and strong cancellations in $m_{ee}$ are realized in the case of NO.
The further parameters are chosen as in figure~\ref{Fig:9}.}}
\label{Fig:10}
\end{center}
\end{figure}
%
Finally, we remark that for case 3 a), which requires the group index $n$ to be larger than 10, 
 several choices of the CP transformation $X$ (or, equivalently, the parameter $s$) are admitted, see \cite{HMM}. 
  Some of these  allow to have strong cancellations in $m_{ee}$ for both Majorana phases being non-trivial. 
 Two examples are shown in figure~\ref{Fig:10}, in which the predictions of  $m_{ee}$
for $n=16$, $m=1$ and $s=2$ (left panel) and $s=3$ (right panel) are displayed.
The parameter $\theta$ has to lie in the interval  $2.28\lesssim \theta\lesssim 2.56$ ($0 \lesssim \theta\lesssim 0.540$) for $s=2$ ($s=3$) in order to accommodate the lepton mixing angles well \cite{HMM}.
For $k_1=k_2=0$ the  Majorana phases $\alpha$ and $\beta$ range in the intervals
\begin{equation}
\label{sinacase3as2s3}
0.330\lesssim \sin\alpha \lesssim 0.712 \;\;\; (-0.392\lesssim \sin\alpha \lesssim0.383) 
\end{equation}
and 
\begin{equation}
\label{sinbcase3as2s3}
0.654\lesssim \sin\beta \lesssim 0.872 \;\;\; (-0.675\lesssim \sin\beta \lesssim 0) 
\end{equation}
for $s=2$ ($s=3$). As a consequence, for NO light neutrino masses strong cancellations in $m_{ee}$ occur, if $s=2$ and
 \begin{eqnarray}
 	&& 0.0043~\text{eV}\lesssim m_0\lesssim 0.0052~\text{eV}\quad\text{and $k_1=k_2=1$,}
 \end{eqnarray}
 or $s=3$ and
 \begin{eqnarray}
 	&& 0.0019~\text{eV}\lesssim m_0\lesssim 0.0027~\text{eV} \quad\text{and $k_1=k_2=0$}\\
	\text{or} && 0.0062~\text{eV}\lesssim m_0\lesssim 0.0072~\text{eV} \quad\text{and $k_1=k_2=1$}\,,
 \end{eqnarray}
 as one can see in figure~\ref{Fig:10}. 
Such strong cancellations  
can also be achieved for $s=0$ and $s=8$. In these cases an accidental CP symmetry, common in the charged lepton and neutrino sectors, is present  \cite{HMM}.
For other values of $s$ with $s\leq 8$, not shown here, the resulting $m_{ee}$ has a lower bound, typically $m_{ee}\gtrsim 10^{-3}$ eV, within the range
$10^{-4}~\text{eV} \lesssim m_0 \lesssim 0.1~\text{eV}$.

In summary, we have shown that in the case of NO three different situations can be realized: no cancellations in $m_{ee}$ e.g.~for case 2), $n=10$ and $u=4$, strong suppression of $m_{ee}$
and trivial Majorana phases e.g.~for case 3 b.1), $n=8$, $m=4$ and $s=4$, as well as cancellations in the presence of two non-trivial Majorana phases e.g.~for case 3 a), $n=16$, $m=1$ and $s=2$ or $s=3$.
For light neutrino masses following IO or being QD we observe that $m_{ee}$ can only obtain values in a very limited range for all possible $m_0$. These values mainly depend on the chosen CP transformation and the parameters $k_1$ and $k_2$.
For this reason, at least part of these scenarios can be tested in future $0\nu\beta\beta$ decay experiments  \cite{Artusa:2014lgv,Arnold:2010tu,Abgrall:2013rze}.

\section{Summary}
\label{concl}

We have studied leptogenesis and $0\nu\beta\beta$ decay in a scenario with a flavor $G_f$ and a CP symmetry that are broken non-trivially in the charged lepton
and neutrino sectors. We have chosen as $G_f$ groups belonging to the series $\Delta (3 \, n^2)$ and $\Delta (6 \, n^2)$, $n \geq 2$, while
the CP symmetry is represented by a CP transformation $X$ that is a unitary and symmetric matrix, acting non-trivially on the flavor space. 
The residual symmetry $G_e$ in the charged lepton sector is fixed to a $Z_3$ subgroup of $G_f$ and $G_\nu$, the symmetry preserved
in the neutrino sector, is given by the direct product $Z_2 \times CP$ with $Z_2 \subset G_f$. In this way, the charged lepton mass matrix
can always be constrained to be diagonal, while the light neutrino mass matrix contains four independent parameters, corresponding 
to the three neutrino masses and $\theta \in [0, \pi)$, on which lepton mixing angles and CP phases in general depend.

Under the assumption of three RH neutrinos and a Dirac Yukawa coupling $Y_D$ that is invariant under $G_f$ and CP the CP asymmetries 
$\epsilon_i^{(\alpha)}$ vanish in the case of flavored as well as unflavored leptogenesis. If $Y_D$ is taken to be invariant
under the residual symmetry $G_\nu=Z_2 \times CP$ only, $\epsilon_i^{\alpha}$ become non-vanishing
and a non-zero value of $Y_B$ can be achieved via flavored leptogenesis. Still, in the case of unflavored
leptogenesis this is not sufficient and $\epsilon_i$ turn out to vanish. 

In our study of unflavored leptogenesis we introduce corrections $\delta Y_D$
to $Y_D$  in order to obtain $\epsilon_i$ non-zero. These corrections are expected to be in general proportional to the 
symmetry breaking parameter $\kappa$ of our scenario.
A particularly interesting case is to assume that $\delta Y_D$
is induced by the breaking of $G_f$ to $G_e$ and thus is invariant under $G_e$, the residual symmetry 
of  the charged lepton sector.
The two main consequences are: the suppression of the CP asymmetries $\epsilon_i \propto \kappa^2$ 
and the fact that the Dirac and Majorana phases, potentially measurable in terrestrial experiments, determine the sign of $Y_B$.
The first observation has already been made in approaches with a flavor symmetry only, whereas the second one, namely the
prediction of the sign of $Y_B$, is only possible thanks to the presence of a CP symmetry that controls the CP phases. Especially,
we have found the following: phases that are present in the correction $\delta Y_D$ are irrelevant for the sign of $Y_B$
at LO; if not suppressed (due to a special choice of the neutrino masses), we expect the terms involving the sine of the Majorana phase $\alpha$
to dominate the CP asymmetries and thus the sign of $Y_B$ (see figures~\ref{Fig:4} and \ref{Fig:5}, the upper-right and lower panel in figure~\ref{Fig:6} and the upper two panels in figure~\ref{Fig:7}); 
in case the dominant contribution to the CP asymmetries arises
from terms involving the Dirac phase the sign of $Y_B$ cannot be predicted and positive and negative values are equally possible (see the upper-left panel of figure~\ref{Fig:6} and the lower panel of figure~\ref{Fig:7}). These features are also confirmed by our analysis of the form of $Y_D$, needed for the CP asymmetries to only depend on the low energy CP phases, see subsection \ref{sec31add}, and by our study that assumes lepton mixing to be of a general form (and not constrained by symmetries), see subsection \ref{sec32}.

For flavored leptogenesis we have demonstrated with several examples that the sign of the CP asymmetries $\epsilon_i^\alpha$ depends on additional
input, e.g.~the relative size of the parameters of the correction $\delta Y_D$, and thus fixing the sign of $\epsilon_i^\alpha$
is in general impossible, having only information about the CP phases and the light neutrino mass ordering. 

We have argued that the symmetry breaking scenario considered by us is well-motivated and have presented
examples of non- as well as SUSY realizations of this scenario in appendix \ref{app3} where we show
that small corrections $\delta Y_D$ of the advocated form are naturally obtained.

We have also studied in detail the predictions for $m_{ee}$, accessible in experiments searching for $0\nu\beta\beta$ decay.
The following interesting properties are found: the constraints on CP phases and lepton mixing angles allow for cases
in which no cancellation in $m_{ee}$ occurs even for NO light neutrino masses and thus $m_{ee}^\mathrm{NO} \gtrsim 0.002 \, \mathrm{eV}$ can be achieved; 
in the case of an IO light neutrino mass spectrum such constraints can lead to values of $m_{ee}^\mathrm{IO} \gtrsim 0.05 \, \mathrm{eV}$ thus increasing the chances to measure this quantity
in the not-too-far future.

In summary, we have thoroughly analyzed a scenario in which flavor and CP symmetries can determine low and high energy CP phases
together with the lepton mixing angles. Thus, phenomena requiring CP violation can be related. In particular, we have shown that the 
knowledge of low energy CP phases (and the light neutrino mass spectrum) can be sufficient for fixing the sign of the baryon asymmetry $Y_B$ of the Universe, if the latter is generated via unflavored leptogenesis.

\section*{Acknowledgements}

C.H.~and E.M.~thank the Excellence Cluster `Universe' at the Technische
Universit\"{a}t M\"{u}nchen for partial support during the preparation of this work. We notice that
important parts of the final results of this work were presented by E.M.~at the workshop ``Nu@Fermilab", 21-25 July 2015, in Fermilab, Chicago, and by C.H.~at
the weekly theory seminar of the high energy physics group in Southampton on 29 January, 2016.

\appendix

\mathversion{bold}
\section{Conventions of mixing angles and CP invariants and global fit results}
\mathversion{normal}
\label{app1}

In this appendix we fix our conventions of mixing angles and of the CP invariants $J_{CP}$, $I_1$, $I_2$ and $I_3$ and list the global
fit results \cite{nufit} used here.

\subsection{Conventions of mixing angles and CP invariants}
\label{app11}

As parametrization of the PMNS mixing matrix we use  
\begin{equation}
\label{UPMNSdef}
U_{PMNS} = \tilde{U} \, {\rm diag}(1, e^{i \alpha/2}, e^{i (\beta/2 + \delta)}) \; , 
\end{equation}
with $\tilde{U}$ being of the form of the Cabibbo-Kobayashi-Maskawa (CKM) matrix $V_{CKM}$ \cite{pdg}
\be
\tilde{U} =
\begin{pmatrix}
c_{12} c_{13} & s_{12} c_{13} & s_{13} e^{- i \delta} \\
-s_{12} c_{23} - c_{12} s_{23} s_{13} e^{i \delta} & c_{12} c_{23} - s_{12} s_{23} s_{13} e^{i \delta} & s_{23} c_{13} \\
s_{12} s_{23} - c_{12} c_{23} s_{13} e^{i \delta} & -c_{12} s_{23} - s_{12} c_{23} s_{13} e^{i \delta} & c_{23} c_{13}
\end{pmatrix}
\ee 
and $s_{ij}=\sin\theta_{ij}$ and $c_{ij}=\cos\theta_{ij}$.  The mixing angles $\theta_{ij}$ range from $0$ to $\pi/2$, while the Majorana phases 
$\alpha, \beta$ as well as the Dirac phase $\delta$ take values between $0$ and $2 \pi$. The Jarlskog invariant 
$J_{CP}$ reads \cite{jcp}
\begin{eqnarray} \nonumber
J_{CP} &=&  {\rm Im} \left[ U_{PMNS,11} U_{PMNS,13}^\star U_{PMNS,31}^\star U_{PMNS,33}  \right] 
 \\ \label{JCP}
 &=& \frac 18 \sin 2 \theta_{12} \sin 2 \theta_{23} \sin 2 \theta_{13} \cos \theta_{13} \sin \delta \, .
\end{eqnarray}
Similar invariants, called $I_i$ with $i=1,2,3$, can be defined which depend on the Majorana phases $\alpha$ and $\beta$, see e.g. 
\cite{Jenkins_Manohar_invariants},
\begin{eqnarray}
&&I_1 = {\rm Im} [U_{PMNS,12}^2 (U_{PMNS,11}^\star)^2] = s^2_{12} c^2 _{12} c^4_{13} \sin \alpha \; ,
\\ \label{I1I2}
&&I_2 =  {\rm Im} [U_{PMNS,13}^2 (U_{PMNS,11}^\star)^2] = s^2 _{13} c^2 _{12} c^2_{13} \sin \beta \; ,
\end{eqnarray}
and
\begin{equation}
\label{I3def}
 I_3 = {\rm Im} [U_{PMNS,13}^2 (U_{PMNS,12}^\star)^2] = c^2_{13}  s^2 _{13} s^2 _{12} \sin \left(\beta-\alpha\right) \; .
\end{equation}
Notice that the Dirac phase has a physical meaning only if all mixing angles are different from $0$ 
and $\pi/2$, as indicated by the data. Analogously, the vanishing of the invariants $I_{1,2}$ only implies $\sin \alpha=0$, $\sin \beta=0$, if  solutions with $\sin2 \theta_{12} =0$, $\cos \theta_{13}= 0$ 
or $\sin 2 \theta_{13}=0$, $\cos \theta_{12}=0$ are discarded. 
In case of $I_3=0$ the vanishing of $\sin \left(\beta-\alpha\right)$ only follows, if $\sin 2 \theta_{13}=0$ and $\sin \theta_{12}=0$ are excluded. 
Furthermore, notice that one of the Majorana phases becomes unphysical, if the lightest neutrino mass vanishes.

\subsection{Global fit results}
\label{app12}

We use in our numerical analysis the results of mixing angles, the CP phase $\delta$ and the mass squared differences $\Delta m_{\mathrm{sol}}^2$ 
and $\Delta m_{\mathrm{atm}}^2$ taken from \cite{nufit}. For NO  the
best fit values of $\sin^2 \theta_{ij}$ and $\delta$ (given here in radian), the $1\, \sigma$ errors as well as $3\, \sigma$ ranges are 
\begin{eqnarray}\nonumber
&& \sin^2 \theta_{13} = 0.0218^{+ 0.0010} _{-0.0010} \;\;\;\;\;\;\;\;\,\, \mbox{and} \;\;\;\;\;\; 0.0186 \leq \sin^2 \theta_{13} \leq 0.0250 \; ,
\\ \label{anglesdeltabfappNO}
&& \sin^2 \theta_{12} = 0.304^{+ 0.013} _{-0.012} \;\;\;\;\;\;\;\;\;\;\;\;\,\mbox{and} \;\;\; \;\;\; 0.270 \leq \sin^2 \theta_{12} \leq 0.344 \; ,
\\ \nonumber
&& \sin^2 \theta_{23} = 0.452^{+ 0.052} _{-0.028} \;\;\;\;\;\;\; \;\;\;\;\;\; \mbox{and} \;\;\;\;\;\; 0.382 \leq \sin^2 \theta_{23} \leq 0.643 \; ,
\\ \nonumber
&& \delta=5.34^{+0.68}_{-1.22}    \;\;\;\;\;\;\; \;\;\;\;\;\;\; \;\;\;\;\; \;\;\;\;\;\; \mbox{and} \;\;\;\;\;\;\;\;\;\;\;\;\;\; 0 \leq \delta \leq 2 \, \pi 
\end{eqnarray}
as well as for the mass squared differences $\Delta m_{\mathrm{sol}}^2$ and $\Delta m_{\mathrm{atm}}^2$
\begin{equation}\label{massesbfappNO}
\Delta m_{\mathrm{sol}}^2 = \left( 7.50 _{-0.17}^{+0.19} \right) \, \times 10^{-5} \; \mathrm{eV}^2 \;\;\; \mbox{and} \;\;\; 
\Delta m_{\mathrm{atm}}^2 = \left( 2.457 _{-0.047}^{+0.047} \right) \, \times 10^{-3} \; \mathrm{eV}^2 
\end{equation}
and
\begin{eqnarray}\label{masses3sappNO}
7.02 \, \times 10^{-5} \; \mathrm{eV}^2 \leq  &\Delta m_{\mathrm{sol}}^2& \leq 8.09 \, \times 10^{-5} \; \mathrm{eV}^2 \; ,
\\ \nonumber
2.317 \, \times 10^{-3} \; \mathrm{eV}^2 \leq  &\Delta m_{\mathrm{atm}}^2& \leq 2.607 \, \times 10^{-3} \; \mathrm{eV}^2 \; .
\end{eqnarray}
For IO instead the global fit analysis in \cite{nufit} yields
\begin{eqnarray}\nonumber
&& \sin^2 \theta_{13} = 0.0219^{+ 0.0011} _{-0.0010} \;\;\;\;\;\;\;\;\,\, \mbox{and} \;\;\;\;\; 0.0188 \leq \sin^2 \theta_{13} \leq 0.0251 \; ,
\\ \label{anglesdeltabfappIO}
&& \sin^2 \theta_{12} = 0.304^{+ 0.013} _{-0.012} \;\;\;\;\;\;\;\;\;\;\;\;\,\mbox{and} \;\;\;\;\; 0.270 \leq \sin^2 \theta_{12} \leq 0.344 \; ,
\\ \nonumber
&& \sin^2 \theta_{23} = 0.579^{+ 0.025} _{-0.037} \;\;\;\;\;\;\; \;\;\;\;\;\; \mbox{and} \;\;\;\;\; 0.389 \leq \sin^2 \theta_{23} \leq 0.644 \; ,
\\ \nonumber
&& \delta=4.43^{+1.10}_{-1.08}    \;\;\;\;\;\; \;\;\;\;\;\;\; \;\;\;\;\;\;\; \;\;\;\;\ \mbox{and} \;\;\;\;\;\;\;\;\;\;\;\;\;\;  0 \leq \delta \leq 2 \, \pi
\end{eqnarray}
as well as the mass squared differences
\begin{equation}\label{massesbfappIO}
\Delta m_{\mathrm{sol}}^2 = \left( 7.50 _{-0.17}^{+0.19} \right) \, \times 10^{-5} \; \mathrm{eV}^2 \;\;\; \mbox{and} \;\;\; 
\Delta m_{\mathrm{atm}}^2 = \left(  -2.449 _{-0.047}^{+0.048} \right) \, \times 10^{-3} \; \mathrm{eV}^2 
\end{equation}
and
\begin{eqnarray}\label{masses3sappIO}
7.02 \, \times 10^{-5} \; \mathrm{eV}^2 \leq  &\Delta m_{\mathrm{sol}}^2& \leq 8.09 \, \times 10^{-5} \; \mathrm{eV}^2 \; ,
\\ \nonumber
-2.590 \, \times 10^{-3} \; \mathrm{eV}^2 \leq & \Delta m_{\mathrm{atm}}^2& \leq -2.307 \, \times 10^{-3} \; \mathrm{eV}^2 \; .
\end{eqnarray}

\mathversion{bold}
\section{Generators of $\Delta (3 \, n^2)$ and $\Delta (6 \, n^2)$}
\mathversion{normal}
\label{app2}

Here we present the generators of the groups $\Delta (3 \, n^2)$ and $\Delta (6 \, n^2)$ that have been used in \cite{HMM}. 
For further details on the groups $\Delta (3 \, n^2)$ and $\Delta (6 \, n^2)$ see \cite{Delta3n2} and \cite{Delta6n2}, respectively.
The generators $a$ and $c$ of $\Delta (3 \, n^2)$ for the representation ${\bf 3}$ that is faithful and irreducible for all groups
can be chosen as
\begin{eqnarray}
\label{genac}
&&a=\left( \begin{array}{ccc}
 1&0&0\\
 0&\omega&0\\
 0&0&\omega^2
\end{array}
\right) \; , 
\\
\nonumber
&&c=\frac 13 \left( \begin{array}{ccc}
 1+2 \cos \phi_n & 1-\cos \phi_n - \sqrt{3} \sin \phi_n & 1- \cos \phi_n + \sqrt{3} \sin \phi_n\\
 1-\cos \phi_n +\sqrt{3} \sin \phi_n & 1+ 2 \cos \phi_n & 1-\cos \phi_n - \sqrt{3} \sin \phi_n\\
 1-\cos \phi_n - \sqrt{3} \sin \phi_n & 1- \cos \phi_n +\sqrt{3} \sin \phi_n & 1+  2 \cos \phi_n
\end{array}
\right)
\end{eqnarray}
with $\omega=e^{\frac{2 \pi i}{3}}$ and  $\phi_n= \frac{2 \pi}{n}$.
They fulfill together with the generator $d= a^2 c \, a$ the relations
\begin{eqnarray}\nonumber
&& a^3=e \;\; , \;\;\; c^n=e \;\; , \;\;\; d^n=e \;\; , 
\\ \label{genD3n2}
&&c \, d=d \, c \;\; , \;\;\; a \, c \, a^{-1} = c^{-1} d^{-1} \;\; , \;\;\; a \, d \, a^{-1} = c 
\end{eqnarray}
with $e$ being the neutral element of the group. In order to arrive at the group $\Delta (6 \, n^2)$ we have to add 
another generator $b$, chosen for the representation ${\bf 3}$ as
\begin{equation} 
\label{genb}
b=\left( \begin{array}{ccc}
1&0&0\\
0&0&\omega^2\\
0&\omega&0
\end{array} \right) \; .
\end{equation}
It fulfills the following relations
\begin{equation} \label{genD6n2}
b^2=e \;\; , \;\;  (a \, b)^2=e \;\; , \;\;
b \, c \, b^{-1} = d^{-1} \;\; , \;\; b \, d \, b^{-1} = c^{-1} \; .
\end{equation}
We note that all groups have a trivial singlet ${\bf 1}$ for which all elements of the group are represented by unity.

\mathversion{bold}
\section{Results for CP asymmetries $\epsilon_i$ in the limit $z_1=0$}
\mathversion{normal}
\label{app2a}

For completeness, we report here the formulae for the CP asymmetries $\epsilon_i$ that are obtained in the limit $z_1=0$ (equivalent to $\zeta=\pi/2, \, 3 \pi/2$)
and for only one non-vanishing CP phase. 

First, we consider the case in which the Majorana phase $\alpha$ and the phase combination
$\beta +2 \, \delta$ are trivial, i.e.
\begin{equation}
\alpha= k_\alpha \, \pi \;\; , \;\; \beta+2 \delta= k_\beta \, \pi \;\;\; \mbox{with} \;\;\; k_{\alpha, \beta}=0,1 \; .
\end{equation}
This choice takes our non-standard definition of the second Majorana phase $\beta$ into account, see (\ref{UPMNSdef}) in appendix \ref{app11} 
and compare to the convention used by the Particle Data Group Collaboration \cite{pdg}. In this case the only source
of low energy CP violation (and thus in our scenario also of CP violation at high energies) is the Dirac phase $\delta$.
We find 

\begin{eqnarray}
\label{app2aspecialcasee1}
\epsilon_1 &\approx& \frac{\tilde\kappa^2}{\pi} \, \sin\delta \, \sin \theta_{13} \, \left( (-1)^{k_\alpha} \, \sin2\theta_{23} \, \left[ \cos\delta \, \sin\theta_{13} \,
\cos 2\theta_{12} \, \sin 2\theta_{23}  \phantom{ f \left( \frac{m_1}{m_2} \right)} \right.\right.
\\
\nonumber
&&\left.+ \frac 12 \, (1+\sin^2 \theta_{13}) \, \sin 2\theta_{12} \, \cos 2\theta_{23} \right] \, f \left( \frac{m_1}{m_2} \right) 
+ (-1)^{k_\beta} \, \cos\theta_{12} \, \cos^2\theta_{13} \, \cos 2\theta_{23}\, \times
\\ \nonumber
&& \left[ \cos\delta \, \sin\theta_{13}\, \cos\theta_{12} \, \cos 2\theta_{23} -\sin\theta_{12} \, \sin 2\theta_{23} \bigg]  \, f \left( \frac{m_1}{m_3} \right) \right)
\\
\label{app2aspecialcasee2}
\epsilon_2 &\approx& \frac{\tilde\kappa^2}{\pi} \, (-1)^{k_\alpha}\, \sin\delta \, \sin \theta_{13} \, \left( -\sin2\theta_{23} \, \bigg[ \cos\delta \, \sin\theta_{13} \,
\cos 2\theta_{12} \, \sin 2\theta_{23}   \right.
\\
\nonumber
&&+ \frac 12 \, (1+\sin^2 \theta_{13}) \, \sin 2\theta_{12} \, \cos 2\theta_{23} \bigg] \, f \left( \frac{m_2}{m_1} \right) 
+ (-1)^{k_\beta} \, \sin\theta_{12} \, \cos^2\theta_{13} \, \cos2\theta_{23}\, \times
\\ \nonumber
&&\left.\bigg[ \cos\delta \, \sin\theta_{13}\, \sin\theta_{12} \, \cos 2\theta_{23}  +\cos\theta_{12} \, \sin 2\theta_{23}\bigg]  \, f \left( \frac{m_2}{m_3} \right) \right)
\\ 
\label{app2aspecialcasee3}
\epsilon_3 &\approx& \frac{\tilde\kappa^2}{\pi} \, (-1)^{k_\beta}\, \sin\delta \, \sin\theta_{13}\, \cos^2\theta_{13}\, \cos 2\theta_{23} \, \left( \cos\theta_{12} \, \bigg[ 
 \sin\theta_{12} \, \sin 2\theta_{23} \phantom{ f \left( \frac{m_1}{m_2} \right)} \right.
\\
\nonumber
&&-\cos\delta \, \sin\theta_{13}\, \cos \theta_{12} \, \cos 2\theta_{23} \bigg] \, f \left( \frac{m_3}{m_1} \right) 
+ (-1)^{k_\alpha+1} \, \sin\theta_{12} \, \bigg[ \cos\delta \, \sin\theta_{13}\, \sin\theta_{12} \, \cos 2\theta_{23} 
\\ 
\nonumber
&&\left. +\cos\theta_{12} \, \sin 2\theta_{23}\bigg]  \, f \left( \frac{m_3}{m_2} \right) \right)
\end{eqnarray}

Note that none of the CP asymmetries $\epsilon_i$ can be written in a form similar to the one presented in (\ref{e1e3bdtrivial}). Thus, the sign of $\epsilon_i$
cannot be predicted with the knowledge of the sign of $\sin\delta$ (and of the loop function) only. For example, the sign of $\epsilon_3$ also depends on the octant of $\theta_{23}$,
$\theta_{23} \lessgtr \pi/4$, since $\epsilon_3$ is proportional to $\cos 2\theta_{23}$. These formulae are consistent with those derived for case 3 b.1), $m=n/2$ and $s=n/2$,
see (\ref{e1case3b1sim}-\ref{e3case3b1sim}), since both lead to vanishing CP asymmetries, if we take into account maximal atmospheric mixing and maximal Dirac phase for (\ref{app2aspecialcasee1}-\ref{app2aspecialcasee3})
and $\zeta=\pi/2, 3 \,\pi/2$ for  (\ref{e1case3b1sim}-\ref{e3case3b1sim}).

We also analyze the case in which the Majorana phase $\alpha$ and the Dirac phase $\delta$ are trivial
\begin{equation}
\alpha= k_\alpha \, \pi \;\; , \;\; \delta= k_\delta \, \pi \;\;\; \mbox{with} \;\;\; k_{\alpha, \delta}=0,1 \; .
\end{equation}
Like the case with trivial $\beta$ instead of $\alpha$, presented in subsection \ref{sec32},
the formulae of the CP asymmetries are compact and, especially, we can express $\epsilon_3$ in terms of the other two ones.
For $\epsilon_1$ and $\epsilon_2$ we find
\begin{eqnarray}
\epsilon_1 &\approx& -\frac{\tilde\kappa^2}{2 \, \pi} \, \sin\beta \, \cos^2 \theta_{13} \, \bigg[ \sin \theta_{12} \, \sin 2\theta_{23}
 - (-1)^{k_\delta} \, \sin \theta_{13} \, \cos \theta_{12} \, \cos 2\theta_{23} \bigg]^2 \, f \left( \frac{m_1}{m_3} \right) \; ,
\\ \nonumber
\epsilon_2 &\approx& \frac{\tilde\kappa^2}{2 \, \pi} \, (-1)^{k_\alpha+1} \, \sin\beta \, \cos^2 \theta_{13} \, \bigg[ \cos \theta_{12} \, \sin 2\theta_{23}
 + (-1)^{k_\delta} \, \sin \theta_{13} \, \sin\theta_{12} \, \cos 2\theta_{23} \bigg]^2 \, f \left( \frac{m_2}{m_3} \right) 
\end{eqnarray}
and $\epsilon_3$ reads
\begin{equation}
\epsilon_3 = - \epsilon_1 \, f \left( \frac{m_3}{m_1} \right) \left(f \left( \frac{m_1}{m_3} \right)\right)^{-1} 
 - \epsilon_2 \, f \left( \frac{m_3}{m_2} \right) \left(f \left( \frac{m_2}{m_3} \right)\right)^{-1}  \; .
\end{equation}

Here the CP asymmetries $\epsilon_{1,2}$ can be decomposed into three pieces: the sine
of the non-trivial Majorana phase $\beta$ (and $k_\alpha=0,1$), the loop function and a piece that can be written as a square. In this way it becomes evident what determines
the sign of these CP asymmetries.

\mathversion{bold}
\section{Sketch of models}
\mathversion{normal}
\label{app3}

In the following we present ideas for explicit realizations of our scenario in non-SUSY as well as SUSY contexts.
In particular, we motivate the size of the expansion parameter $\kappa$ and the flavor structure of the correction $\delta Y_D$.
For the sake of concreteness we stick to the flavor group $\Delta (6 \, n^2)$ with $n=4$ and choose one CP transformation $X$. 
We note that models with $\Delta (96)$ and the flavor group $\Delta (48)$ can also be found in the literature
\cite{D48CPmodel,D96modelnoCP}. 

\mathversion{bold}
\subsection{Non-SUSY setup}
\mathversion{normal}
\label{app31}

Here we discuss a sketch of a non-SUSY model.
Apart from the flavor group $\Delta (96)$ and the auxiliary symmetry
$Z_3^{(\mathrm{aux})}$ we also assume the presence of an additional $Z_{12}$
group that we use to segregate better the charged lepton and the neutrino sectors. LH leptons $l$ and RH neutrinos $N$ transform as ${\bf 3}$ that is a complex faithful representation of $\Delta (96)$, 
while RH charged leptons $\alpha_R$ are trivial singlets. The fields $l$ and $N$ are neutral under $Z_3^{(\mathrm{aux})}$, whereas $\alpha_R$
carry the charges $1$ for $\alpha=e$,  $\omega$ for $\alpha=\mu$ and  $\omega^2$ for $\alpha=\tau$, respectively. 
The fields $l$ and $N$ have both the charge $\omega_{12}^4$, $\omega_{12}=e^{2 \, \pi \, i/12}$, under the additional $Z_{12}$ group, while RH charged leptons are all assigned the phase $\omega_{12}^3$. As CP symmetry we impose the one generated by $X=a \, b \, c \, d^2 \, P_{23}$ in the representation ${\bf 3}$.\footnote{The explicit form of $X$ in the other representations of $\Delta (96)$ can be calculated using the information about the corresponding automorphism given in (\ref{XP23aut}).}

Charged lepton masses
are produced with the help of three different flavor symmetry breaking fields $\phi_\alpha$ that transform as
\begin{equation}
\phi_e \sim ({\bf 3}, 1, \omega_{12}) \;\;\; , \;\;\; \phi_\mu \sim ({\bf 3}, \omega^2, \omega_{12}) \;\;\; \mbox{and} \;\;\; \phi_\tau \sim ({\bf 3}, \omega, \omega_{12})  
\end{equation}
under $(\Delta (96), Z_3^{(\mathrm{aux})}, Z_{12})$ and are neutral under the gauge group. The relevant terms in the Lagrangian read
\begin{equation}
\label{Yukchl}
-\frac{y_e}{\Lambda} \, \bar{l} \, H \, \phi_e \, e_R - \frac{y_\mu}{\Lambda} \, \bar{l} \, H \, \phi_\mu \, \mu_R - \frac{y_\tau}{\Lambda} \, \bar{l} \, H \, \phi_\tau \, \tau_R
+ \mathrm{h.c.}
\end{equation}
with $\Lambda$ being the cutoff scale of the theory and $y_\alpha$ real couplings.
The VEVs of $\phi_\alpha$ are taken to be 
\begin{equation}
\label{VEVphie}
\langle \phi_e \rangle \propto \left( \begin{array}{c} 1\\0\\0 \end{array} \right) \;\;\; , \;\;\; 
\langle \phi_\mu \rangle \propto \left( \begin{array}{c} 0\\1\\0 \end{array} \right) \;\;\; \mbox{and} \;\;\;
\langle \phi_\tau \rangle \propto \left( \begin{array}{c} 0\\0\\1 \end{array} \right)  
\end{equation}
and thus break $G_f$ to the residual group $G_e=Z_3^{(D)}$, as desired. Since the phases of the VEVs are in general undetermined,
they break the CP symmetry imposed on the theory. The charged lepton mass matrix, computed from (\ref{Yukchl}) and (\ref{VEVphie}), is diagonal. For $y_\tau$ of order one 
the correct tau lepton mass is obtained for 
\begin{equation}
\label{estimateeps}
\left|\langle\phi_\tau\rangle\right|/\Lambda= \varepsilon \approx (0.01 \div 0.1) \; .
\end{equation}
Assuming all VEVs
$\langle \phi_\alpha \rangle$ to be of that order, the correct hierarchy among the charged lepton masses can be achieved with an additional 
Froggatt-Nielsen symmetry \cite{FN} under which RH charged leptons carry different charges, see e.g. \cite{AF2}.

Since RH neutrinos transform like LH leptons under the flavor and CP symmetry, the neutrino Yukawa coupling $Y_D$ arises from a renormalizable operator. Its flavor structure is trivial
and the coupling is real. The Majorana mass term of RH neutrinos originates from couplings to fields in two different triplets of $\Delta (96)$, one equivalent to ${\bf 3}$ and
another one to ${\bf 3^\prime}$ which is a real and unfaithful representation of $\Delta (96)$, 
\begin{equation}
\varphi_\nu \sim ({\bf 3}, 1, \omega_{12}^4) \;\;\; \mbox{and} \;\;\;  \psi_\nu \sim ({\bf 3^\prime}, 1, \omega_{12}^4) \; .
\end{equation}
As indicated, these fields are neutral under $Z_3^{(\mathrm{aux})}$, but carry the charge $\omega_{12}^4$ under $Z_{12}$, so that the Lagrangian contains the following terms
\begin{equation}
\label{couplingsMR}
-\frac 12 \, f_1 \, \overline{N^c} \, N \, \varphi_\nu - \frac 12 \, f_2 \, \overline{N^c} \, N \, \psi_\nu + \mathrm{h.c.}
\end{equation}
with $f_1$ and $f_2$ being real couplings. The VEVs $\langle\varphi_\nu\rangle$ and $\langle\psi_\nu\rangle$ are aligned as follows
\begin{equation}
\label{VEVfieldsnu}
\langle \varphi_\nu \rangle = i \, w \, \left( \begin{array}{c} 1\\1\\1 \end{array} \right)  \;\;\; , \;\;\; 
\langle \psi_\nu \rangle = \left( \begin{array}{c} v_1 + i \, v_2\\v_1 + i \, v_3\\v_1 - i \, (v_2+v_3) \end{array} \right)
\end{equation}
with $v_{1,2,3}$ and $w$ real parameters of order $\varepsilon^2 \, \Lambda$. Thus, RH neutrino masses between $10^{12} \, \text{GeV}$ and  $10^{14} \, \text{GeV}$
are achieved for $\Lambda$ close to the scale of grand unification. Notice that we have chosen the VEVs of $\varphi_\nu$ and $\psi_\nu$ to be smaller than those of the fields $\phi_\alpha$.
In this way the dominant correction to the Dirac neutrino mass matrix arises from the fields $\phi_\alpha$ only, see (\ref{corrphia}).
As one can check, $\langle \varphi_\nu\rangle$ and $\langle \psi_\nu \rangle$ leave $G_\nu$, generated by $Z=c^2$ and the CP transformation $X$,
invariant. This breaking pattern hence allows us to obtain the PMNS mixing matrix of case 1), see (\ref{URcase1}), for $n=4$ and $s=1$.
This is a choice of parameters also employed in our numerical discussion of unflavored leptogenesis in case 1), see subsection \ref{sec34}. The free parameter
$\theta$ depends on the VEV of the field $\psi_\nu$
\begin{equation}
\tan2\, \theta=- \frac{v_2 + 2 \, v_3}{\sqrt{3} \, v_2} \; .
\end{equation}
Its particular value, necessary for describing correctly the lepton mixing angles, should be explained in a more complete model. 
The three RH neutrino masses $M_i$ read\footnote{$\mathrm{sign} (x)$ stands for
the sign of the real parameter $x$.}
\begin{eqnarray}
\nonumber
&&M_1= \sqrt{3} \, \left|\sqrt{3} \, f_1 \, w + f_2 \, \mathrm{sign} (v_2) \, \sqrt{v_2^2+v_2 \, v_3+v_3^2} \right| \; ,
\\ \nonumber
&&M_2=3 \, \left| f_2 \, v_1 \right| \; ,
\\
&&M_3=  \sqrt{3} \, \left|\sqrt{3} \, f_1 \, w - f_2 \, \mathrm{sign} (v_2) \, \sqrt{v_2^2+v_2 \, v_3+v_3^2} \right|
\end{eqnarray}
and thus are functions of both couplings $f_{1,2}$ as well as all parameters of the VEVs of the fields $\varphi_\nu$ and $\psi_\nu$.
Using these we can also compute the masses of the light neutrinos. As one can see, we can accommodate in this way both mass orderings
as well as a QD light neutrino mass spectrum. 

The discussed operators necessary at LO in our scenario contain either no or one flavor symmetry breaking field.  
Each mass matrix, $m_l$, $m_D$ and $M_R$, receives corrections of relative order $\varepsilon^2$ with respect to the corresponding LO result.
Those to $m_l$ arise from insertions of three fields $\phi_\alpha$ and are of the generic form $\phi_\alpha \phi_\beta \phi^\dagger_\gamma$
with $\alpha$, $\beta$, $\gamma$ being $e$, $\mu$ or $\tau$. Clearly, these do not change the form of $m_l$ and thus the charged lepton
mass matrix is still diagonal. The dominant corrections to $m_D$ instead change the form of the latter and hence
constitute the leading form of $\delta Y_D$. They stem from the terms
\begin{equation}
\label{corrphia}
-\sum \limits_{\alpha=e, \mu, \tau} \sum\limits_{r=1,2,6} \frac{y^\nu_{\alpha,r}}{\Lambda^2} \, \bar{l} \, H^c \, \phi^\dagger_\alpha \, \phi_\alpha \, N + \mathrm{h.c.} \; .
\end{equation}
The index $\alpha$ indicates which field $\phi_\alpha$ is coupled, while the index $r$ takes into account the different possible contractions
via a one-, two- or six-dimensional representation of $\Delta (96)$. The contribution arising from the contraction to a singlet can be absorbed into the LO term, since it 
is always real and proportional to the identity matrix in flavor space. The one coming from the contraction to a doublet, indeed, is not there, since the residual symmetry $G_e$
that is left invariant by the VEVs of the fields in (\ref{VEVphie}) forces it to vanish. Consequently, the correction $\delta Y_D$ 
is generated via the terms with $y^\nu_{\alpha,6}$ in (\ref{corrphia}). Matching the form of $\delta Y_D$ given in (\ref{deltaYD}) the two couplings $z_1$ and $z_2$ 
turn out to be
 \begin{equation}
 \label{z1z2D}
 z_1= \frac{\sqrt{3}}{2} \, \left( 2\, y^\nu_{e,6} - y^\nu_{\mu,6} - y^\nu_{\tau,6} \right) \;\;\; \mbox{and} \;\;\; z_2 = \frac{3}{2} \, \left( y^\nu_{\tau,6} - y^\nu_{\mu,6} \right) \; ,
 \end{equation}
if we set $\langle \phi^\dagger_\alpha \, \phi_\alpha \rangle$ to $\varepsilon^2 \, \Lambda^2$ for $\alpha=e, \, \mu, \, \tau$ for simplicity.
The parameter $\kappa$ is thus of the order
\begin{equation}
\kappa \approx \varepsilon^2 \;\;\; \mbox{meaning} \;\;\; 10^{-4} \lesssim \kappa \lesssim 10^{-2} \; .
\end{equation}
We note that in this particular case $z_{1,2}$ in (\ref{z1z2D}) turn out to be real. 
Subleading corrections to the Dirac neutrino mass matrix arise from two types of terms: terms with two flavor symmetry breaking fields ($\varphi_\nu$
and $\psi_\nu$ and the conjugate fields) and terms with four flavor symmetry breaking fields of the type $\phi_\alpha^\dagger \, \phi_\beta^\dagger \, \phi_\gamma \, \phi_\delta$
with $\alpha$, $\beta$, $\gamma$, $\delta= e, \, \mu, \, \tau$. Both lead to corrections relatively suppressed by $\varepsilon^4$ with respect to the LO term.\footnote{We note
that the first type of terms is invariant under $G_\nu$, while the latter is invariant under $G_e$. We also note that these terms do not affect our LO results for the CP asymmetries $\epsilon_i$. We have checked that the latter statement is even correct in a variant of the model, in which the VEVs $\langle \varphi_\nu \rangle$ and 
$\langle \psi_\nu\rangle$ are of order $\varepsilon \, \Lambda$ so that the corrections to the Dirac neutrino mass matrix due to terms invariant under $G_\nu$
arise at the order $\varepsilon^2$ and thus at the same order as the desired corrections, stemming from the terms in (\ref{corrphia}).}
The RH neutrino mass matrix, being at LO of order $\varepsilon^2 \, \Lambda$, also receives corrections. The dominant ones are of order $\varepsilon^4 \, \Lambda$
and arise from three types of terms: terms with two conjugate fields $\varphi^\dagger_\nu$ and $\psi_\nu^\dagger$, terms with three fields of the form $\phi^\dagger_\alpha \, \phi_\alpha \, \varphi_\nu$
or $\phi^\dagger_\alpha \, \phi_\alpha \, \psi_\nu$ for $\alpha=e, \, \mu, \, \tau$ as well as terms with four fields of the type $\phi_\alpha$ that are in general different.
Clearly, the latter two types of terms break the residual symmetry $G_\nu$, if the VEVs of the flavor symmetry breaking fields are inserted.
These together with the correction $\delta Y_D$ that is also of relative order $\varepsilon^2$
with respect to the LO term induce small corrections to the LO results for the lepton mixing parameters. However, these are expected to
be suppressed by $\varepsilon^2$ and thus are at most at the percent level.

\mathversion{bold}
\subsection{SUSY setup}
\mathversion{normal}
\label{app32}

If we consider instead a SUSY framework, we can also construct a model of this type. Apart from the fact that $l$ and $\nu^c$ transform in complex conjugated three-dimensional
representations, $l \sim {\bf\bar{3}}$ and $\nu^c \sim {\bf 3}$, see section \ref{sec4}, the three main differences are: 
$a)$ we slightly change the additional symmetry and we use a $Z_5$ instead of a $Z_{12}$ group. The transformation properties of the fields are
\begin{equation}
l \sim \omega_5^3 \;\; , \;\; \alpha^c \sim \omega_5^2 \;\; , \;\; \nu^c \sim \omega_5^2 \;\; , \;\; \varphi_\nu \, , \, \psi_\nu \sim \omega_5 \;\;\; \mbox{(}\, \omega_5=e^{2 \, \pi \, i/5} \, \mbox{)}
\end{equation}
and the Higgs multiplets $h_u$ and $h_d$ are neutral;
$b)$ we use less fields in the charged lepton sector
\begin{equation}
\phi_\tau \sim ({\bf 3}, \omega, 1) \;\;\; \mbox{and} \;\;\; \chi \sim ({\bf 2}, \omega, 1) 
\end{equation}
under $(\Delta (96), Z_3^{(\mathrm{aux})}, Z_5)$. The VEVs of these fields leave $G_e=Z_3^{(D)}$ invariant, if they are chosen to be of the form
\begin{equation}
\label{VEVphitauchi}
\langle \phi_\tau \rangle \propto \left( \begin{array}{c} 0\\0\\1 \end{array} \right)\;\;\; \mbox{and} \;\;\;
  \langle \chi \rangle \propto \left( \begin{array}{c} 0\\1 \end{array} \right) \; .
\end{equation}
The terms in the superpotential contributing at lowest orders to the charged lepton mass matrix are
\begin{equation}
\frac{y_e}{\Lambda^3} \, l \, h_d \, \phi_\tau \, \chi^2 \, e^c +  \frac{y_\mu}{\Lambda^2} \, l \, h_d \, \phi_\tau \, \chi \, \mu^c + \frac{y_\tau}{\Lambda} \, l \, h_d \, \phi_\tau \, \tau^c \; .
\end{equation}
In this way, we can generate the mass hierarchy among the charged leptons with the help of insertions of several flavor symmetry breaking fields. Furthermore, the correct
ratio between muon and tau lepton masses and electron and tau lepton masses is achieved, if we assume that the field $\chi$ acquires a VEV of the order\footnote{In the following we treat 
$\varepsilon$ and $\lambda^2$ as expansion parameters of the same order of magnitude.} 
\begin{equation}
\left|\langle \chi \rangle\right|/\Lambda \approx \lambda^2 \approx 0.04 \, ,
\end{equation}
while the VEV of $\phi_\tau$ is chosen like in (\ref{estimateeps}) and its actual value depends on the size of $\tan\beta$, see details in section \ref{sec4}. 
We note also that the VEVs of $\varphi_\nu$ and $\psi_\nu$ still have the form as in (\ref{VEVfieldsnu}), but we now choose their size to be $\varepsilon \, \Lambda$;
$c)$ the lowest order correction to the Dirac mass matrix of the neutrinos and thus the source of $\delta Y_D$ are two operators with three insertions
of the field $\phi_\tau \, $\footnote{These two operators correspond to two independent contractions of the flavor indices.
Furthermore, we note that the operator with three fields $\chi$ gives a vanishing contribution to $\delta Y_D$, if the VEV of $\chi$, see (\ref{VEVphitauchi}), is inserted.}
\begin{equation}
\sum \limits_{i=1,2} \frac{y^\nu_{\tau, i}}{\Lambda^3} \, l \, h_u \, \phi_\tau^3 \, \nu^c 
\end{equation}
with $y^\nu_{\tau, i}$ being both real.
Thus, the size of the small parameter $\kappa$ is estimated as 
\begin{equation}
\kappa \approx \varepsilon^3 \;\;\; \mbox{meaning} \;\;\;  10^{-6} \lesssim \kappa \lesssim 10^{-3} \; .
\end{equation}
This has to be compared with the value given in (\ref{estimatekappa}) showing that the expansion parameter $\varepsilon$ should be in this SUSY realization
 $\varepsilon \approx 0.1$. The parameters $z_1$ and $z_2$ of $\delta Y_D$ in (\ref{deltaYD}) can be written as
\begin{equation}
z_1 = -\sqrt{3} \, \left( y^\nu_{\tau,1} + y^\nu_{\tau, 2} \right)  \;\;\; , \;\;\; z_2 = 3 \, \left( y^\nu_{\tau,1} + y^\nu_{\tau, 2} \right) 
\end{equation}
for $\langle \phi_\tau \rangle$ fixed to $\varepsilon \, \Lambda$.

As mentioned, the lowest order correction to $m_D$ is of the relative order $\varepsilon^3$ with respect to the LO term.
The dominant subleading corrections to the charged lepton mass matrix $m_l$ which involve the fields $\varphi_\nu$ and $\psi_\nu$
(and thus break the residual group $Z_3^{(D)}$) contribute to the first column of $m_l$, since they arise from operators 
with the field $e^c$ and five fields $\varphi_\nu$ and $\psi_\nu$. Their size relative to the LO term (i.e. the electron mass) is $\varepsilon^2$.
Corrections contributing to the third and second column of $m_l$ come from operators with 
more flavor symmetry breaking fields, because the requirement of invariance under the symmetry $Z_3^{(\mathrm{aux})}$ necessitates insertions
of  one and two fields $\phi_\tau$ and $\chi$, respectively. Eventually, the most relevant corrections to the RH neutrino mass matrix $M_R$
which break the residual symmetry $G_\nu$ originate from operators with one of the fields $\varphi_\nu$ and $\psi_\nu$ and three fields
$\phi_\tau$ and $\chi$, since these have to be invariant under the symmetries $Z_5$ and $Z_3^{(\mathrm{aux})}$. Their relative  
suppression with respect to the LO term is of the order $\varepsilon^3$. Hence, we expect corrections to the LO results for lepton mixing
to be at maximum at the percent level.

\mathversion{bold}
\subsection{General source of correction $\delta Y_D$}
\mathversion{normal}
\label{app33}

Lastly, we can also consider the general case in which a gauge singlet $\Phi$ in the 
six-dimensional irreducible
representation of the flavor group $\Delta (96)$ and 
uncharged under all auxiliary symmetries couples to LH leptons, RH neutrinos and the Higgs field $H$ in a non-SUSY context
\begin{equation}
-\frac{y^\nu_\Phi}{\Lambda} \, \bar{l} \, H^c \, \Phi \, N + \mathrm{h.c.} \; .
\end{equation}
In a SUSY context the corresponding term in the superpotential reads
\begin{equation}
\frac{y^\nu_\Phi}{\Lambda} \, l \, h_u \, \Phi \, \nu^c \; .
\end{equation}
In both cases $y^\nu_{\Phi}$ is a real coupling.
The most general form of the VEV of $\Phi$ that leaves the residual group $G_e=Z_3^{(D)}$ invariant is
\begin{equation}
\langle \Phi \rangle = \left( \begin{array}{c} x_1 \\0\\0\\x_2\\0\\0 \end{array}\right) \, \kappa \, \Lambda \;\;\; \mbox{with} \;\;\; x_i \;\;\; \mbox{complex} 
\end{equation}
and assuming $\kappa$ like requested in (\ref{estimatekappa}).
Thus, $z_1$ and $z_2$, parametrizing the correction $\delta Y_D$ in (\ref{deltaYD}), read 
\begin{equation}
z_1 = \frac{\sqrt{3}}{2} \, y^\nu_\Phi \, \left( x_1 + x_2 \right) \;\;\; \mbox{and}  \;\;\; z_2 = \frac{\sqrt{3}}{2} \, i \, y^\nu_\Phi \, \left( x_1 - x_2 \right) \; .
\end{equation}
This shows that the special cases, $z_1=0$ or $z_2=0$, discussed in section \ref{sec3}, can be achieved with a particular form
of the VEV of the field $\Phi$. Clearly, the latter can also arise from some combination of flavor symmetry breaking fields.

 

\end{document}